\documentclass[a4paper,11pt]{article}
\usepackage{jheppub}
\usepackage{amssymb}
\usepackage{enumitem}
\usepackage{axodraw4j}
\usepackage{color}
\usepackage{longtable}
\usepackage{appendix}
\usepackage{graphicx}
\usepackage{amsthm}
\usepackage{amsmath}
\usepackage{booktabs}
\usepackage{tabularx}
\usepackage{float}
\usepackage[dvipsnames]{xcolor}
\usepackage{empheq}
\usepackage{tikz}
\usetikzlibrary{arrows.meta}

\newtheorem{theorem}{Theorem}[section]

\newtheorem{lemma}[theorem]{Lemma}

\newcommand{\Fboot}{\mathbb{F}}

\title{\boldmath New analytic constraints for the soft anomalous dimension bootstrap at three loops}

\author[a,b]{Einan Gardi,}
\author[a]{Luke Lippstreu,}
\author[a]{Andrew J. McLeod,}
\author[a]{Laura Walsh,}
\author[a,c]{Zehao Zhu,}
\author[a]{and Enrico Zunino}
\affiliation[a]{Higgs Centre for Theoretical Physics, School of Physics and Astronomy,\\
                    The University of Edinburgh, Edinburgh EH9 3FD, Scotland, UK}
\affiliation[b]{CERN, Theoretical Physics Department, CH-1211 Geneva 23, Switzerland}
\affiliation[c]{PRISMA$^{++}$ Cluster of Excellence \& Institute of Physics,\\
Johannes Gutenberg University Mainz, 
Staudingerweg 7, 55128 Mainz, Germany}

\abstract{The soft anomalous dimension governs the universal long-distance singularities of gauge theory amplitudes.  It exhibits a highly constrained analytic structure and has been shown to be amenable to bootstrap methods. In this paper we revisit the relation between unitarity cuts and sequential discontinuities of Feynman integrals and use it to derive new constraints on the soft anomalous dimension. At three loops, we show that certain sequential discontinuities in non-overlapping two-particle channels vanish for every web contributing to the quadrupole function that correlates four coloured partons. These constraints hold for arbitrary parton masses, and we explicitly verify them for the known massless and one-mass cases. We then demonstrate their constraining power by substantially strengthening the massless bootstrap and carrying out the first bootstrap construction of the one-mass quadrupole function. The new constraints are largely complementary to those arising from the absence of unphysical branch points, symmetries, and collinear and Regge limits. 
With their help, we determine each
quadrupole function within the assumed function space
up to a single coefficient, which we fix by matching
to existing diagrammatic results.
These results open the way to bootstrapping the three-loop soft anomalous dimension for amplitudes with several massive partons.}
\begin{document}

\maketitle
\flushbottom

\section{Introduction}
\label{sec: intro}
The infrared (IR) structure of gauge theories has been a subject of focused study since the early days of quantum field theory~\cite{Bloch:1937pw}.
This long history of studying the IR  properties of amplitudes and cross sections has both phenomenological and theoretical motivations. The cancellation of IR divergences is one of the main technical challenges when calculating observables. This is especially true in the non-Abelian case, where -- although guaranteed by the Kinoshita--Lee--Nauenberg (KLN) theorem for appropriate energy-degenerate initial and final states~\cite{Kinoshita:1962ur,Lee:1964is} -- actually realising this cancellation in fixed-order perturbative calculations is extremely difficult. Engineering the cancellation for generic observables now constitutes a branch of research in its own right \cite{Giele:1991vf, Giele:1993dj, Giele:1994xd, Frixione:1995ms,Catani:1996vz, Catani:1996jh, Catani:2002hc, Nagy:2003qn, Frederix:2009yq, DelDuca:2024ovc, Alioli:2025hpa, Currie:2013vh, Bertolotti:2022aih, Magnea:2024jqg, Devoto:2023rpv, Devoto:2025kin}. At the same time, the IR sector of gauge theories has been a source of insights into the all-order structure of perturbative quantum field theory, 
factorisation and exponentiation properties of amplitudes and cross
sections~\cite{Sen:1982bt,Sterman:1986aj,Collins:1989gx,Catani:1998bh,Sterman:2002qn,Aybat:2006mz,Gardi:2009qi, Gardi:2009zv, Becher:2009cu, Becher:2009qa,Becher_2015,falcioni_scattering_2022} and their relations with
Wilson-line correlators~\cite{Gatheral:1983cz, Frenkel:1984pz, Mitov:2010rp, Gardi:2010rn, Gardi:2011yz, Gardi:2013ita,Gardi:2021gzz,Henn:2019swt,Grozin:2015kna,Grozin:2014hna,Gardi:2025ule},
properties of appropriate asymptotic states in perturbation theory (for unbroken gauge theories) \cite{Kulish:1970ut,CIAFALONI1985379,CATANI1985301,ciafaloni}, and conjectural connections with holographic conformal duals defined on the celestial sphere \cite{strominger_asymptotic_2014,he20162d, strominger2018lectures, Pasterski_2021, Magnea:2021fvy, Magnea:2025zut}. A recent review of some of the progress that has been made in this field can be found in Ref.~\cite{Agarwal:2021ais}. 

In this paper, we restrict our attention to soft divergences appearing in fixed-angle QCD scattering amplitudes, namely when all external invariants are parametrically of the same magnitude ($s_{ij}=(p_i+p_j)^2\sim Q^2, \forall i,j$) and are all large compared to the QCD scale. Under this assumption, the dimensionally regularised  amplitude is known to factorise into a singular, universal  colour operator acting on a finite, process-dependent hard scattering function, which is a vector in colour-flow space~\cite{Sen:1982bt,Sterman:1986aj,Collins:1989gx,Catani:1998bh,Sterman:2002qn,Aybat:2006mz,Gardi:2009qi, Gardi:2009zv, Becher:2009cu, Becher:2009qa}. 
Moreover, it has been shown that the divergences that arise from the exchange of soft gluons between external partons are encoded by correlators of Wilson lines, which exponentiate as a result of renormalisation~\cite{Gatheral:1983cz, Frenkel:1984pz, Mitov:2010rp, Gardi:2010rn, Gardi:2011yz, Gardi:2013ita}. The finite quantity governing this exponentiation is the {\emph{soft anomalous dimension}}; integrating it over the factorisation scale down to the infrared generates the soft divergences. To calculate the soft anomalous dimension, one generally exploits the multiplicative renormalisability of correlators of timelike Wilson lines~\cite{Brandt:1981kf, Korchemsky:1987wg}, before performing an asymptotic expansion near the lightcone of each massless parton~\cite{Almelid:2015jia,Gardi:2025ule,Gardi:2025lws}. 

For amplitudes involving only massless external partons, the soft anomalous
dimension takes the dipole form, involving only pairwise colour correlations, through two loops~\cite{Catani:1998bh,Sterman:2002qn,Aybat:2006mz,Gardi:2009qi, Gardi:2009zv, Becher:2009cu, Becher:2009qa}.
This structure follows from soft--collinear factorisation and rescaling
symmetry, which also constrain the possible corrections at higher
orders~\cite{Gardi:2009qi,Becher:2009qa}.
The first non-dipole corrections in massless scattering arise at three loops, when correlations involving four partons first become possible. These quadrupole corrections depend on conformally invariant cross ratios and have been
computed at three loops for any multileg amplitude~\cite{Almelid:2015jia}. 
When massive partons are present,  correlations between three coloured particles already
appear at two loops, provided at least two of the partons are massive;
these contributions to the soft anomalous dimension have been computed in Refs.~\cite{Ferroglia:2009ep,Ferroglia:2009ii}.
At three loops, the complete soft anomalous dimension is also known
for amplitudes involving one massive parton and any number of massless
partons~\cite{Liu:2022elt,Gardi:2025lws}. For amplitudes involving several massive particles such three-loop corrections   remain unknown.  

While the explicit results cited above have all been determined via direct diagrammatic computations,  the number of physical and mathematical constraints that the soft anomalous dimension satisfies suggests that it may also be amenable to bootstrap methods~\cite{Dixon:2009ur,Almelid:2017qju,Liu:2022elt,Gardi:2025lws}. This strategy generally involves constructing an ansatz for the functional form within an assumed function space, and  subsequently constraining its free coefficients  using symmetries and the expected behaviour of the target function in special kinematic limits. Ref.~\cite{Almelid:2017qju} has developed such a strategy and successfully applied it to the massless soft anomalous dimension at three loops.
More broadly, similar bootstrap methods have been shown to be highly effective in the planar limit of ${\mathcal N}=4$~Super Yang--Mills theory (which enjoys a dual conformal symmetry similar to the rescaling symmetry of the soft anomalous dimension), where they have been used to deduce amplitudes and form factors with nontrivial kinematic dependence up to eight\footnote{Recently, Anthropic announced that Claude Fable 5.1 bootstrapped the six-point amplitude at nine loops, \emph{in a week}. Their result is available \href{https://smsharma.io/cosmic-nine-loops/}{here}. A (human) collaboration independently made the symbol of the solution available \href{https://doi.org/10.5281/zenodo.22800070}{here}.} loops~\cite{Dixon:2016nkn,Drummond:2018caf,Caron-Huot:2018dsv,Dixon:2022rse,Dixon:2022xqh,Dixon:2023kop,Basso:2024hlx}. 

The bootstrap strategy is particularly well suited to function spaces
that can be systematically enumerated and manipulated, such as generalised polylogarithms (GPLs), which come equipped with a well-understood coaction and symbol~\cite{goncharov2001multiple,Goncharov:2010jf,Duhr:2011zq,Duhr:2012fh}.
In the successful applications discussed above, the target functions
are taken to be pure polylogarithmic functions of fixed transcendental
weight, with a symbol alphabet assumed or known at the
outset~\cite{Dixon:2013eka,Almelid:2017qju}.
The symbol alphabet specifies the allowed logarithmic integration kernels.
Within this framework, the absence of unphysical branch points restricts
the first entries of the
symbol~\cite{Gaiotto:2011dt,Abreu:2014cla,Abreu:2015zaa,Abreu:2017mtm};
at function level, regularity at unphysical branch points also
constrains terms that are invisible to the symbol~\cite{Dixon:2020bbt}. An important example of functions that satisfy these types of constraints is provided by the single-valued harmonic
polylogarithms (SVHPLs), which combine holomorphic and antiholomorphic
harmonic polylogarithms so that their monodromies
cancel~\cite{Brown2004,Dixon_2012}.
Choosing such a space incorporates single-valuedness in the Euclidean region directly into
the bootstrap ansatz.

For the soft anomalous dimension, analytic constraints combine with
permutation symmetries and factorisation properties in special
kinematic limits. In the massless case, these ingredients were used
in Ref.~\cite{Almelid:2017qju} to bootstrap the three-loop correction
to the dipole formula.
In timelike collinear limits, strict factorisation requires the
splitting operator to depend only on the momenta and quantum numbers
of the collinear partons~\cite{Kosower:1999xi,Feige_2014}.
Consistency with infrared factorisation therefore restricts correlations
involving the non-collinear
partons~\cite{Becher:2009qa,Dixon:2009ur}.\footnote{These constraints also apply to timelike collinear limits of massless partons in amplitudes containing massive non-collinear partons~\cite{Duhr:2025cye}.}
Similarly, the logarithmic structure of
massless amplitudes in the Regge
limit is constrained by 
both infrared factorisation and Regge factorization and their compatibility is highly restrictive~\cite{DelDuca:2011wkl,DelDuca:2011ae,Caron-Huot:2013fea,Caron-Huot:2017fxr}. 
This has been an area of significant progress in recent years, leading to predictions for the soft anomalous
dimension to all loop orders at next-to-leading logarithmic
accuracy~\cite{Caron-Huot:2017zfo}, and explicit four-loop constraints
at next-to-next-to-leading logarithmic
accuracy~\cite{Falcioni:2020lvv,falcioni_scattering_2022}.

A further source of analytic information comes from the relation between
unitarity cuts and discontinuities of Feynman integrals. Unitarity relates
a discontinuity in a physical momentum channel to a sum over cut diagrams,
whose cut lines represent on-shell intermediate
states~\cite{Cutkosky:1960sp,tHooft:1973wag}.
Such \emph{unitarity cuts} in specified external channels partition the
vertices in the graph into two sets on the two sides of the cut, and impose the corresponding
energy-flow conditions across the cut~\cite{tHooft:1973wag,Abreu:2014cla}.
Cuts and discontinuities of Wilson-line correlators have also been
studied in Refs.~\cite{andries,Laenen:2015jia}.

A broader notion is provided by \emph{generalised cuts}, which need not
correspond to such a vertex partition or satisfy the associated
energy-flow conditions. Their relation to discontinuities has been
formulated at one loop using residues and integration
contours~\cite{Abreu:2017ptx}.\footnote{This broader notion of cutting also underlies diagrammatic representations
of the coaction in terms of integrals and their generalised cuts,
developed at one loop~\cite{Abreu:2017enx,Abreu:2017mtm}, with extensions
to examples beyond one loop~\cite{Abreu:2021vhb}.}
For functions expressible in terms of multiple polylogarithms,
discontinuities act on the first factor of the
coaction~\cite{Duhr:2012fh}, linking this algebraic structure to the
analytic properties of the integrals.
In the present work, we restrict attention to physical-channel unitarity
cuts as defined above. Relations between cuts in multiple channels,
sequential discontinuities and entries of the
coaction~\cite{Abreu:2014cla,Bourjaily:2020wvq} provide the framework
for the constraints derived below.

A prominent example of constraints on sequential discontinuities
is provided by the Steinmann relations, which originate in causality
and forbid double discontinuities in partially overlapping momentum
channels within their domain of
applicability~\cite{Steinmann1960a,Steinmann1960b,Cahill:1973qp}.\footnote{
The physical-region qualifications and subtleties associated with
massless two-particle channels are discussed in
Refs.~\cite{Bourjaily:2020wvq,Hannesdottir:2025bss}.}
These relations have been successfully exploited in bootstraps
of suitably normalised planar ${\cal N}=4$ super-Yang--Mills
amplitudes~\cite{Caron-Huot:2016owq,Dixon:2016nkn}.
At symbol level, the ordinary relations restrict the first two
entries, while the stronger extended Steinmann relations constrain
adjacent entries throughout the
symbol~\cite{Caron-Huot:2019bsq}, with close connections to
cluster adjacency~\cite{Drummond:2017ssj}.

In the present paper, we develop bootstrap methods for the three-loop
quadrupole functions of the soft anomalous dimension, retaining the
full colour dependence of gauge-theory amplitudes. 
Our analysis applies to general gauge theories, including QCD and
\hbox{${\cal N}=4$} super-Yang--Mills theory, in the presence of
external coloured probes represented by lightlike or timelike Wilson lines.
We derive a new set of analytic constraints from an analysis of the
cut diagrams contributing to these functions.
These constraints restrict sequential discontinuities in non-overlapping two-particle channels of the quadrupole functions and are therefore distinct from the Steinmann relations. Notably, our argument only invokes the topology of the contributing webs (sets of Wilson-line diagrams), so these constraints are expected to be valid for any configuration of external masses. Correspondingly, we expect them to prove useful for bootstrapping the soft anomalous dimension involving two or more massive partons. As a proof of concept, we use the new sequential discontinuity constraints to re-bootstrap the massless case, requiring less kinematic boundary data than in Ref.~\cite{Almelid:2017qju}.  
Finally, we carry out the first bootstrap construction of the one-mass
quadrupole function. Within the assumed function space, the combined 
constraints leave just one undetermined coefficient before any explicit
diagrammatic input is used; this last coefficient can then be determined from a diagrammatic calculation in a suitable kinematic limit.

The paper is organised as follows. Section~\ref{sec: cutsanddiscsW} provides a review of the relevant known results on cuts and discontinuities. We start by illustrating unitarity cuts in a single channel and then discuss  cuts in multiple channels, before exploring the connections between cuts, discontinuities, and entries of the coaction. Section~\ref{sec:struct} provides a minimal primer on the structure of the soft anomalous dimension, and introduces cuts of diagrams with Wilson lines, setting the stage for the derivation of the constraint. In Section~\ref{sec: new_constr} we analyse the cuts of the webs contributing to the soft anomalous dimension at three loops, and we show that they imply vanishing of certain double discontinuities on appropriate combinations of functions that appear in the soft anomalous dimension in completely general kinematics. Section~\ref{sec: implement} provides tools to bridge the gap between the general constraints derived in the previous section and the bootstrap of the soft anomalous dimension in given configurations of massive and massless particle scattering. In Section~\ref{sec: 4LLbootstrap} we re-bootstrap the fully-massless case, providing a first demonstration of the effectiveness of our new  constraint. Section~\ref{sec: 1m3ll-bootstrap} then presents the first bootstrap of the soft anomalous dimension for amplitudes with a massive external leg. This result would not have been achievable without the use of our new  constraint. In Section~\ref{sec: concl}, we summarise our findings and discuss future directions of research. Appendices~\ref{sec:alpha_details} and \ref{app:color} provide additional details on the chosen parametrisation and a graphical representation of the colour structures involved in the derivation.

\section{Cuts and discontinuities of Feynman integrals}
\label{sec: cutsanddiscsW}

In this section, we review the relation between unitarity cuts
and discontinuities of Feynman integrals. We begin with the
single-channel cutting rules in the formulation of
't Hooft and Veltman~\cite{tHooft:1973wag}, before reviewing
the prescription for cuts in multiple channels and their
relation to sequential discontinuities introduced in
Ref.~\cite{Abreu:2014cla}. We recall the vertex-decoration
prescription for cuts in multiple channels, the compatibility
conditions that exclude crossed cuts, and the kinematic
conditions entering the relation to sequential discontinuities.
We also discuss the complementary formulation in terms of
time-ordered perturbation theory~\cite{Bourjaily:2020wvq}
and recall the connection between sequential discontinuities
and the coaction. This provides the framework for our subsequent
application to the webs contributing to the soft anomalous
dimension.

\subsection{Review of unitarity cuts}
\label{sec:unitarity_cuts}
In Ref.~\cite{tHooft:1973wag}, 't Hooft and Veltman explore the consequences of the unitarity of the $S$-matrix on Feynman diagrams. The starting point of their analysis is the decomposition of the Feynman propagator into a positive- and a negative-energy component
\begin{equation}
    \Delta_F(x_i-x_j)=\theta(x^0_i-x^0_j)\Delta^+_{ij}+\theta(x^0_j-x^0_i)\Delta^-_{ij},
\end{equation}
where $\Delta^\pm_{ij}\equiv\Delta^\pm(x_i-x_j)$ are defined as
\begin{equation}
    \Delta^\pm_{ij}=\frac{1}{(2\pi)^{3}}\int d^4ke^{ik(x_i-x_j)}\theta(\pm k^0)\rho(k^2)\, .
    \label{eq:pos/neg_comps}
\end{equation}
In view of the reality of the spectral function $\rho(k^2)$ (which for a free propagator is just $\rho_{\text{free}}(k^2)=\delta(k^2-m^2)$), we notice that $\Delta^+_{ij}=(\Delta^-_{ij})^*$. We also highlight that the components $\Delta^\pm_{ij}$ each only have support on either positive or negative values of the energy flowing across the cut (oriented from $x_j$ to $x_i$), as enforced by the theta functions.

Working in configuration space, we denote the Feynman integral associated with a diagram with $n$ vertices by $F(x_1, x_2, \dots, x_n)$. The cuts of this diagram, which separate these vertices into disjoint sets, can then be denoted by $F(x_1, x_2, \dots, \underline{x_i}, \dots, \underline{x_j}, \dots, x_n)$, where we underline all of the vertices that appear on one side of the cut.  To each such cut, we associate an integral according to the following prescription:
\begin{enumerate}[label=(\roman*)]
    \item\label{unchanged} propagators between vertices that have not been underlined retain the usual propagator $\Delta_F$,
    \item \label{conjugate_propagator} propagators between vertices that have both been underlined are replaced by their complex conjugate $\Delta_F^*$,
    \item \label{energy_flow1} propagators originating from vertices that have not been underlined ($x_k$) and ending on vertices that have been underlined ($\underline{x_i}$) are replaced by their positive-energy component $\Delta_{ik}^+$,
    \item \label{energy_flow2} propagators originating from vertices that have been underlined ($\underline{x_j}$) and ending on vertices that have not been underlined ($x_l$) are replaced by their negative-energy component $\Delta_{lj}^-$,
    \item \label{minus} the factor assigned an underlined vertex is $-1$ times the original vertex factor.
\end{enumerate}
With these conventions in place, we can derive the largest-time equation~\cite{tHooft:1973wag}. Namely, consider a vertex $x_i$ such that its time component $x_i^0$ is larger than those of all other vertices. Then, rules \ref{unchanged}--\ref{minus}, together with the definition of positive and negative energy components of the propagators, imply that
\begin{equation}
\label{eq: LTE}
        F(x_1, x_2, \dots, \underline{x_i}, \dots, \underline{x_j}, \dots, x_n)=-F(x_1, x_2, \dots, x_i, \dots, \underline{x_j}, \dots, x_n)\, ,
\end{equation}
where only the vertex with the largest time component $x_i$ has been changed from underlined to not underlined. 
Since this relation holds for any choice of underlinings of the remaining vertices, summing over all possible configurations yields
\begin{equation}
\label{eq: improvedLTE}
    \sum_{\text{\{underlinings\}}}F(x_1, x_2, \dots, \underline{x_i}, \dots, \underline{x_j}, \dots, x_n)=0\,.
\end{equation}
Many of the terms in this sum are in fact zero due to incompatible theta functions; the diagrams that actually contribute are those in which the sets of underlined and non-underlined vertices each form a single connected subgraph~\cite{tHooft:1973wag}.

While Eq.~\eqref{eq: LTE} cannot be transformed into a valid statement in momentum space (because it relies on $x^0_i$ being the largest time component), Eq.~\eqref{eq: improvedLTE} can be Fourier transformed. Since the resulting expression no longer depends directly on the vertices $x_i$, we keep track of which side of the cut each vertex appears on by decorating underlined vertices with $\{1\}$, and vertices that are underlined with $\{0\}$. In momentum space, our conventions for the (unitarity) cuts of a diagram can then be summarised as follows:
\begin{enumerate}[label=(\alph*)]
    \item \label{decorated1} to each propagator connecting two vertices decorated with $\{0\}$, assign the usual propagator,
    \item \label{decorated2} to each propagator connecting two vertices decorated with $\{1\}$, assign a complex conjugate propagator (a propagator with the sign of $i\varepsilon$ reversed),
    \item \label{decorated3} to each propagator connecting vertices decorated with $\{0\}$ and $\{1\}$, assign a factor $2\pi\delta(D)\theta(E)$, where $D$ is the denominator of the cut propagator (without the $i\varepsilon$ prescription) and $E$ is the energy component of the momentum flowing from the non-underlined vertex to the underlined vertex, 
    \item \label{decorated4} associate the usual vertex factor to each vertex decorated with $\{0\}$, and minus the usual vertex factor to each vertex decorated with $\{1\}$.
\end{enumerate}

For each cut, we define the \emph{cut channel} $s$ as the square
of the total momentum flowing across it. Grouping all cut diagrams
according to their cut channels, we can rewrite
Eq.~\eqref{eq: improvedLTE} in momentum space as~\cite{tHooft:1973wag}
\begin{equation}
\label{eq: Cutkosky}
    \Tilde{F}(k_1, k_2, \dots, k_n)+\Tilde{F}^*(k_1, k_2, \dots, k_n)=-\sum_{s}\sum_{\mathcal{C}_s}\text{Cut}_s\Tilde{F}(k_1, k_2, \dots, k_n),
\end{equation}
where we have used the fact that $\Tilde{F}(\underline{k_1}, \underline{k_2}, \dots, \underline{k_n})=\Tilde{F}^*(k_1, k_2, \dots, k_n)$ and we have introduced the notation $\sum_{\mathcal{C}_s}\text{Cut}_s\Tilde{F}$ to denote the sum over all decorated graphs with cut channel $s$. This equation is known as Cutkosky's cutting rule~\cite{Cutkosky:1960sp}. Its significance becomes apparent when expressed in terms of the transition amplitude $T$, defined by $S=1+iT$. Since the standard Feynman rules yield contributions to $iT$, we introduce the notation $\mathcal{T}=-i\Tilde{F}$ for the contribution of a given diagram to the transition amplitude; then, using the same notation for cut diagrams, Eq.~\eqref{eq: Cutkosky} becomes
\begin{equation} \label{eq:transfer_amplitude}
    \mathcal{T}(k_1, k_2, \dots, k_n)-\mathcal{T}^*(k_1, k_2, \dots, k_n)=-\sum_{s}\sum_{\mathcal{C}_s}\text{Cut}_s\mathcal{T}(k_1, k_2, \dots, k_n).
\end{equation}
We see that the left side of the equation can be interpreted as the difference between the original integral and the same integral evaluated with the opposite $i\varepsilon$ prescription. This motivates us to define the discontinuity of a function $\mathcal{T}(s)$ that has a branch cut starting at $s=0$ by
\begin{equation}
    \text{Disc}_s \mathcal{T} (s\pm i0) = \lim_{\epsilon\rightarrow0} \left[\mathcal{T} (s \pm i\epsilon) - \mathcal{T} (s \mp i\epsilon)\right].
    \label{eq:def_disc}
\end{equation}
Since $\text{Disc}_s \mathcal{T} (s\pm i0)$ can in principle be computed in any kinematic region, it can be useful to specify the region $R$ in which it has been computed. For this purpose, we adopt the notation
\begin{equation}
    \text{Disc}_{s} \mathcal{T}(s)\Big|_{R}.
    \label{eq:discR}
\end{equation} 
If there is no branch cut in the region $R$, or if $\mathcal{T}(s)$ does not in fact depend on $s$, the value of this discontinuity will be zero. 

To isolate the discontinuity in a specified channel from
Eq.~\eqref{eq:transfer_amplitude}, we choose a region in which
only that channel is above threshold. Specifically, we define
the region in which $s_{IJ}$ is above its threshold $s^{\text{th.}}_{IJ}=(m_I+m_J)^2$ while all other Mandelstam invariants $s_{KL}$  are between their pseudo-thresholds  $s^{\text{ps.}}_{KL}=(m_K-m_L)^2$ and their thresholds: 
\begin{equation}
\label{eq:kinematic_region}
    R_{s_{\scalebox{0.7}{$\scriptscriptstyle IJ$}}}\equiv \left\{s_{IJ}>s^{\text{th.}}_{IJ},\, \,s^{\text{ps.}}_{KL}<s_{KL}<s^{\text{th.}}_{KL}, \,\,\forall \,\,\,KL \neq IJ\right\}.
\end{equation}
There, $\mathcal{T}(\{s_{AB}\})$ will only have a branch cut in the $s_{IJ}$ plane. Eq.~\eqref{eq:transfer_amplitude} will therefore become
\begin{equation}
    \text{Disc}_{s_{IJ}} \mathcal{T}(\{s_{AB}\})\Big|_{R_{s_{\scalebox{0.7}{$\scriptscriptstyle IJ$}}}}=-\sum_{\mathcal{C}_{s_{\scalebox{0.7}{$\scriptscriptstyle IJ$}}}}\text{Cut}_{s_{\scalebox{0.7}{$\scriptscriptstyle IJ$}}}\mathcal{T}(\{s_{AB}\}),
    \label{eq:disc=-cut}
\end{equation}
where the sum on the right has collapsed to just terms that are non-zero in this kinematic region, namely the cuts in the $s_{IJ}$ channel \cite{tHooft:1973wag,Abreu:2014cla, Bourjaily:2020wvq}. In other words, in this kinematic region, Eq.~\eqref{eq:disc=-cut} relates the $s_{IJ}$ discontinuity of $\mathcal{T}(\{s_{AB}\})$ to a sum of cut diagrams. 

\subsection{Sequential discontinuities and cuts in multiple channels}
\label{sec:iterateddiscsandcuts}
It is natural to wonder whether Cutkosky's cutting rule~\eqref{eq: Cutkosky} can be generalised to a relation between sequential discontinuities of Feynman integrals and sums of (suitably-defined) cuts. Ref.~\cite{Abreu:2014cla} conjectured such a generalisation for sequential discontinuities with respect to distinct channels that do not partially overlap (meaning the sets of momenta appearing in these cuts are either disjoint or one is a proper subset of the other), which they related to sums over suitably compatible cuts evaluated in a specific kinematic region (see below for more details).
Similar relations were also considered in Ref.~\cite{Bourjaily:2020wvq}, which made use of Time-Ordered Perturbation Theory (TOPT) and considered the case of sequential discontinuities with respect to a single channel. In the present work, we focus on sequential discontinuities with respect to non-overlapping channels, and we verify the equivalence between these two prescriptions under appropriate conditions.

\begin{figure}
    \centering
    \scalebox{0.4}{
    \begin{picture}(276,276) (177,-147)
    \SetWidth{3.0}
    \SetColor{Black}
    \Line[arrow,arrowpos=0.5,arrowlength=8.333,arrowwidth=3.333,arrowinset=0.2](180,126)(240,66)
    \SetWidth{1.0}
    \Line(240,66)(240,-84)
    \Line(240,-84)(390,-84)
    \Line(240,66)(390,66)
    \Line(390,66)(390,-84)
    \SetWidth{3.0}
    \Line[arrow,arrowpos=0.5,arrowlength=8.333,arrowwidth=3.333,arrowinset=0.2](390,66)(450,126)
    \Line[arrow,arrowpos=0.5,arrowlength=8.333,arrowwidth=3.333,arrowinset=0.2](180,-144)(240,-84)
    \Line[arrow,arrowpos=0.5,arrowlength=8.333,arrowwidth=3.333,arrowinset=0.2](390,-84)(450,-144)
    \SetColor{Red}
    \Arc[dash,dashsize=10](410,-100)(100,90,185)
    \SetColor{ForestGreen}
    \Arc[dash,dashsize=10](220,80)(100,-90, 5)
    \SetColor{ForestGreen}
    \Line[arrow,arrowpos=1,arrowlength=8.333,arrowwidth=3.333,arrowinset=0.2](299,81)(333,81)
    \Line[arrow,arrowpos=1,arrowlength=8.333,arrowwidth=3.333,arrowinset=0.2](227,9)(227,-28)
    \SetColor{Red}
    \Line[arrow,arrowpos=1,arrowlength=8.333,arrowwidth=3.333,arrowinset=0.2](301,-102)(331,-102)
    \Line[arrow,arrowpos=1,arrowlength=8.333,arrowwidth=3.333,arrowinset=0.2](403,9)(403,-28)
    \Text(238,74)[lb]{\LARGE{\Black{$\{{\color{red} 0}, {\color{ForestGreen} 0}\}$}}}
    \Text(395,52)[lb]{\LARGE{\Black{$\{{\color{red} 0},{\color{ForestGreen} 1}\}$}}}
    \Text(195,-85)[lb]{\LARGE{\Black{$\{{\color{red} 0},{\color{ForestGreen} 1}\}$}}}
    \Text(358,-110)[lb]{\LARGE{\Black{$\{{\color{red} 1},{\color{ForestGreen} 1}\}$}}}
    \end{picture}}\hspace{-30pt}
    \scalebox{0.4}{\begin{picture}(487,316) (148,-129)
    \SetWidth{3.0}
    \SetColor{Black}
    \Line[arrow,arrowpos=0.5,arrowlength=8.333,arrowwidth=3.333,arrowinset=0.2](415,106)(530,136)
    \Line[arrow,arrowpos=0.5,arrowlength=8.333,arrowwidth=3.333,arrowinset=0.2](270,-50)(360,-15)
    \Line[arrow,arrowpos=0.5,arrowlength=8.333,arrowwidth=3.333,arrowinset=0.2](510,46)(540,46)
    \Line[arrow,arrowpos=0.5,arrowlength=8.333,arrowwidth=3.333,arrowinset=0.2](250,80)(290,76)
    \SetWidth{1.0}
    \Line(290,76)(360,-15)
    \Line(360,-15)(510,46)
    \Line(290,76)(415,105)
    \Line(415,105)(510,46)
    \SetWidth{3.0}
    \SetColor{Red}
    \Line[dash,dashsize=10](450,-74)(450,136)
    \SetColor{ForestGreen}
    \Line[dash,dashsize=12.2](330,136)(330,-74)
    \Text(270,88)[lb]{\LARGE{\Black{$\{{\color{red} 0}, {\textcolor{ForestGreen}{ 0}}\}$}}}
    \Text(400,115)[lb]{\LARGE{\Black{$\{{\color{red} 0},{\textcolor{ForestGreen}{ 1}}\}$}}}
    \Text(350,-40)[lb]{\LARGE{\Black{$\{{\color{red} 0},{\textcolor{ForestGreen}{ 1}}\}$}}}
    \Text(508,20)[lb]{\LARGE{\Black{$\{{\color{red} 1},{\textcolor{ForestGreen}{ 1}}\}$}}}
    \SetWidth{1.5}
    \SetColor{Black}
    \Line[arrow,arrowpos=1,arrowlength=6.667,arrowwidth=2.667,arrowinset=0.2](235,166)(560,166)
    \Text(550,140)[lb]{\Huge{\Black{$t$}}}
    \end{picture}}\hspace{-55pt}
    \scalebox{0.4}{\begin{picture}(487,316) (148,-129)
    \SetWidth{3.0}
    \SetColor{Black}
    \Line[arrow,arrowpos=0.5,arrowlength=8.333,arrowwidth=3.333,arrowinset=0.2](360,106)(530,136)
    \Line[arrow,arrowpos=0.5,arrowlength=8.333,arrowwidth=3.333,arrowinset=0.2](270,-50)(415,-15)
    \Line[arrow,arrowpos=0.5,arrowlength=8.333,arrowwidth=3.333,arrowinset=0.2](510,46)(540,46)
    \Line[arrow,arrowpos=0.5,arrowlength=8.333,arrowwidth=3.333,arrowinset=0.2](250,80)(290,76)
    \SetWidth{1.0}
    \Line(290,76)(415,-15)
    \Line(415,-15)(510,46)
    \Line(290,76)(360,105)
    \Line(360,105)(510,46)
    \SetWidth{3.0}
    \SetColor{Red}
    \Line[dash,dashsize=10](450,-74)(450,136)
    \SetColor{ForestGreen}
    \Line[dash,dashsize=12.2](330,136)(330,-74)
    \Text(270,88)[lb]{\LARGE{\Black{$\{{\color{red} 0}, {\textcolor{ForestGreen}{ 0}}\}$}}}
    \Text(345,115)[lb]{\LARGE{\Black{$\{{\color{red} 0},{\textcolor{ForestGreen}{ 1}}\}$}}}
    \Text(395,-40)[lb]{\LARGE{\Black{$\{{\color{red} 0},{\textcolor{ForestGreen}{ 1}}\}$}}}
    \Text(508,20)[lb]{\LARGE{\Black{$\{{\color{red} 1},{\textcolor{ForestGreen}{ 1}}\}$}}}
    \SetWidth{1.5}
    \SetColor{Black}
    \Line[arrow,arrowpos=1,arrowlength=6.667,arrowwidth=2.667,arrowinset=0.2](235,166)(560,166)
    \Text(550,140)[lb]{\Huge{\Black{$t$}}}
    \end{picture}}\\
    \scalebox{0.4}{
    \begin{picture}(276,276) (177,-147)
    \SetWidth{3.0}
    \SetColor{Black}
    \Line[arrow,arrowpos=0.5,arrowlength=8.333,arrowwidth=3.333,arrowinset=0.2](180,126)(240,66)
    \SetWidth{1.0}
    \Line(240,66)(240,-84)
    \Line(240,-84)(390,-84)
    \Line(240,66)(390,66)
    \Line(390,66)(390,-84)
    \SetWidth{3.0}
    \Line[arrow,arrowpos=0.5,arrowlength=8.333,arrowwidth=3.333,arrowinset=0.2](390,66)(450,126)
    \Line[arrow,arrowpos=0.5,arrowlength=8.333,arrowwidth=3.333,arrowinset=0.2](180,-144)(240,-84)
    \Line[arrow,arrowpos=0.5,arrowlength=8.333,arrowwidth=3.333,arrowinset=0.2](390,-84)(450,-144)
    \SetColor{Red}
    \Line[dash,dashsize=12.2](314,116)(314,-134)
    \SetColor{ForestGreen}
    \Line[dash,dashsize=10](190,-9)(430,-9)
    \SetColor{ForestGreen}
    \Line[arrow,arrowpos=1,arrowlength=8.333,arrowwidth=3.333,arrowinset=0.2](227,9)(227,-28)
    \SetColor{ForestGreen}
    \Line[arrow,arrowpos=1,arrowlength=8.333,arrowwidth=3.333,arrowinset=0.2](403,9)(403,-28)
    \SetColor{Red}
    \Line[arrow,arrowpos=1,arrowlength=8.333,arrowwidth=3.333,arrowinset=0.2](299,81)(333,81)
    \Line[arrow,arrowpos=1,arrowlength=8.333,arrowwidth=3.333,arrowinset=0.2](301,-102)(331,-102)
    \Text(238,74)[lb]{\Large{\Black{$\{{\color{red} 0}, {\color{ForestGreen} 0}\}$}}}
    \Text(395,52)[lb]{\Large{\Black{$\{{\color{red} 1},{\color{ForestGreen} 0}\}$}}}
    \Text(203,-82)[lb]{\Large{\Black{$\{{\color{red} 0},{\color{ForestGreen} 1}\}$}}}
    \Text(358,-105)[lb]{\Large{\Black{$\{{\color{red} 1},{\color{ForestGreen} 1}\}$}}}
  \end{picture}}\hspace{35pt}
    \scalebox{0.4}{\begin{picture}(487,316) (148,-129)
    \SetWidth{3.0}
    \SetColor{Black}
    \Line[arrow,arrowpos=0.5,arrowlength=8.333,arrowwidth=3.333,arrowinset=0.2](150,106)(270,76)
    \SetWidth{1.0}
    \Line(270,76)(390,-104)
    \Line(390,-104)(510,46)
    \Line(270,76)(390,106)
    \Line(390,106)(510,46)
    \SetWidth{3.0}
    \Line[arrow,arrowpos=0.5,arrowlength=8.333,arrowwidth=3.333,arrowinset=0.2](390,106)(540,136)
    \Line[arrow,arrowpos=0.5,arrowlength=8.333,arrowwidth=3.333,arrowinset=0.2](180,-104)(390,-104)
    \Line[arrow,arrowpos=0.5,arrowlength=8.333,arrowwidth=3.333,arrowinset=0.2](510,46)(600,46)
    \SetColor{Red}
    \Line[dash,dashsize=12.2](330,136)(330,-74)
    \SetColor{ForestGreen}
    \Line[dash,dashsize=10](450,-74)(450,136)
    \Text(255,88)[lb]{\LARGE{\Black{$\{{\color{red} 0}, {\textcolor{ForestGreen}{ 0}}\}$}}}
    \Text(365,115)[lb]{\LARGE{\Black{$\{{\color{red} 1},{\textcolor{ForestGreen}{ 0}}\}$}}}
    \Text(364,-132)[lb]{\LARGE{\Black{$\{{\color{red} 0},{\textcolor{ForestGreen}{ 1}}\}$}}?}
    \Text(508,20)[lb]{\LARGE{\Black{$\{{\color{red} 1},{\textcolor{ForestGreen}{ 1}}\}$}}}
    \SetWidth{1.5}
    \SetColor{Black}
    \Line[arrow,arrowpos=1,arrowlength=6.667,arrowwidth=2.667,arrowinset=0.2](150,166)(600,166)
    \Text(590,145)[lb]{\Huge{\Black{$t$}}}
  \end{picture}}
    \caption{The equivalence between the crossed cut condition and the TOPT perspective. In the first line, we depict a compatible cut and the pair of TOPT diagrams to which it can be mapped. In the second line, we illustrate why crossed cuts cannot be represented in TOPT; namely, we see that (with this choice of ordering of the red and green cuts) there exists no time at which the vertex $\{{\color{red} 0},{\textcolor{ForestGreen}{ 1}}\}$ can exist, because it is supposed to remain to the right of the green cut and to the left of the red one.}
    \label{fig:TOPT}
\end{figure}
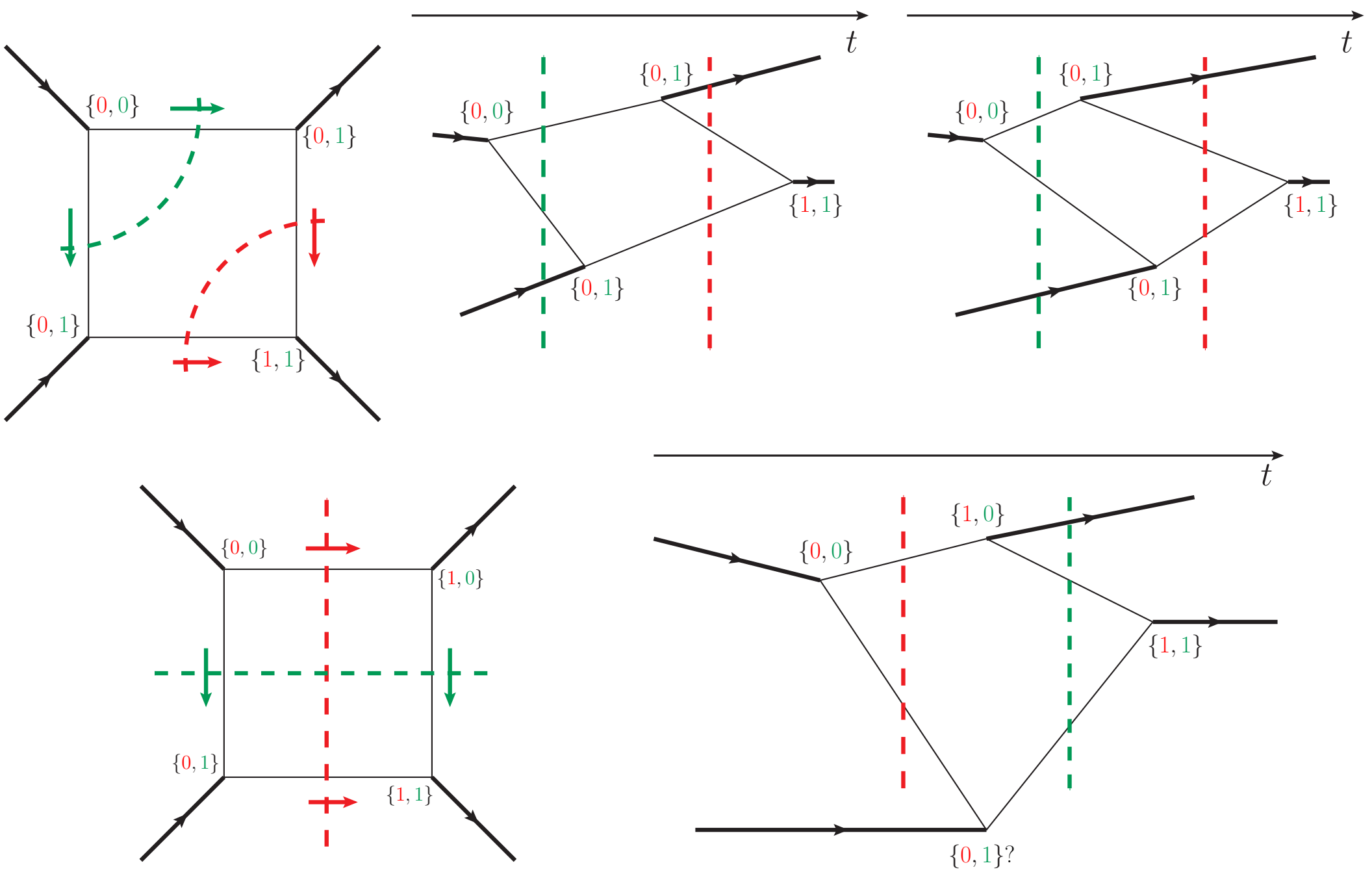

To extend Eq.~\eqref{eq:disc=-cut} to several channels, we first need to say what a cut in multiple channels is, and which integral it corresponds to. A cut in a channel $s_i$ splits the vertices of the graph into two sets, so a candidate configuration of $k$ cuts in distinct channels $\{s_1,\dots,s_k\}$ can be specified by giving each vertex $v$ a sequence of labels $(c_1(v),\dots,c_k(v))$, with $c_i(v)\in\{0,1\}$ recording which side of cut $i$ the vertex lies on. The individual labels keep track of each partition separately, while their parity $c(v)\equiv\sum_{i=1}^k c_i(v)\pmod{2}$ fixes the final decoration of the vertex. Following Ref.~\cite{Abreu:2014cla}, integrals are assigned to these configurations, and some configurations are discarded, according to the following rules:
\begin{enumerate}
    \item \label{cut1}
    Vertices with final value $c(v)=0$ are computed according
    to the usual Feynman rules; vertices with final value
    $c(v)=1$ are computed according to complex-conjugated
    Feynman rules.

    \item
    A propagator joining vertices $u$ and $v$ is uncut if
    $c_i(u)=c_i(v)$ for every $i$. It is then the usual
    Feynman propagator if the common final value of its
    endpoints is $0$, and the complex-conjugated propagator
    if that value is $1$.

    \item \label{cut3}
    A propagator joining vertices $u$ and $v$ is cut if
    $c_i(u)\ne c_i(v)$ for at least one $i$. It is put on
    shell, with a theta function imposing energy flow from
    $0$ to $1$ for each cut $i$ that separates its endpoints.
    If different cuts impose conflicting energy flows,
    the product of theta functions vanishes and the diagram
    gives no contribution.

    \item \label{crossedcutsconditions}
\emph{Crossed cuts} are excluded. These are  configurations
    in which, for some pair of cuts $i,j$, all four
    vertex-label pairs
    $(c_i(v),c_j(v))=(0,0),(0,1),(1,0),(1,1)$
    occur in the graph. This exclusion is additional to
    the energy-flow conditions above: crossed cuts are
    excluded from the prescription relating cuts to
    sequential discontinuities even when their energy-flow
    requirements do not conflict~\cite{Abreu:2014cla}.
    The associated time-ordering obstruction is explained
    below and illustrated in Fig.~\ref{fig:TOPT}.
\end{enumerate}
We call a set of cuts \emph{compatible} if no pair is crossed
and their on-shell and energy-flow conditions can be satisfied
simultaneously in the kinematic region under consideration. Note that any pair of cuts in partially overlapping momentum channels (as appear in the Steinmann relations~\cite{Steinmann1960a,Steinmann1960b,Cahill:1973qp}) is necessarily crossed; correspondingly, we do not consider such pairs of channels.

\begin{figure}
    \centering
    \scalebox{0.42}{
        \begin{picture}(451,252) (187,-210)
            \SetWidth{1.0}
            \SetColor{Black}
            \Line(424.545,-26.731)(424.545,-129.722)
            \SetWidth{3.0}
            \Line[arrow,arrowpos=0.5,arrowlength=12.5,arrowwidth=5,arrowinset=0.2](188.687,-77.833)(259.444,-77.833)
            \SetWidth{1.0}
            \Line(566.06,16.51)(566.06,-172.177)
            \SetWidth{3.0}
            \Line[arrow,arrowpos=0.5,arrowlength=12.5,arrowwidth=5,arrowinset=0.2,flip](566.06,-172.177)(636.817,-195.762)
            \Line[arrow,arrowpos=0.5,arrowlength=12.5,arrowwidth=5,arrowinset=0.2](566.06,16.51)(636.817,40.096)
            \SetWidth{1.0}
            \Line(566.06,-172.177)(259.444,-77.833)
            \Line(259.444,-77.833)(566.06,16.51)
            \SetWidth{3.0}
            \SetColor{ForestGreen}
            \Line[dash,dashsize=11.793](306.616,-7.076)(307.402,-167.459)
            \Line[arrow,arrowpos=1,arrowlength=12.5,arrowwidth=5,arrowinset=0.2](283.03,-101.419)(323.126,-115.571)
            \Line[arrow,arrowpos=1,arrowlength=12.5,arrowwidth=5,arrowinset=0.2](283.03,-54.247)(327.057,-42.454)
            \SetColor{Red}
            \Bezier[dash,dsize=11.793](360.077,-7.076)(355.36,-86.481)(394.67,-108.495)(620.307,-109.281)
            \Line[arrow,arrowpos=1,arrowlength=12.5,arrowwidth=5,arrowinset=0.2](345.926,-38.524)(383.663,-28.303)
            \Line[arrow,arrowpos=1,arrowlength=12.5,arrowwidth=5,arrowinset=0.2](436.338,-112.426)(436.338,-75.475)
            \Line[arrow,arrowpos=1,arrowlength=12.5,arrowwidth=5,arrowinset=0.2](577.853,-125.791)(577.853,-88.054)
            \Text(215,-110)[lb]{\Huge{\Black{$p_1$}}}
            \Text(595,0)[lb]{\Huge{\Black{$p_2$}}}
            \Text(595,-170)[lb]{\Huge{\Black{$p_3$}}}
        \end{picture}
    }
    \hspace{10pt}
    \scalebox{0.42}{
        \begin{picture}(513,276) (196,-207)
            \SetWidth{1.0}
            \SetColor{Black}
            \Line(270,-144)(270,6)
            \SetWidth{3.0}
            \Line[arrow,arrowpos=0.5,arrowlength=12.5,arrowwidth=5,arrowinset=0.2](630,-144)(690,-204)
            \Line[arrow,arrowpos=0.5,arrowlength=12.5,arrowwidth=5,arrowinset=0.2](450,-144)(450,-204)
            \Line[arrow,arrowpos=0.5,arrowlength=12.5,arrowwidth=5,arrowinset=0.2](270,-144)(210,-204)
            \Line[arrow,arrowpos=0.5,arrowlength=12.5,arrowwidth=5,arrowinset=0.2](210,66)(270,6)
            \SetWidth{1.0}
            \Line(270,6)(630,6)
            \Line(630,6)(630,-144)
            \Line(630,-144)(270,-144)
            \SetWidth{3.0}
            \Line[arrow,arrowpos=0.5,arrowlength=12.5,arrowwidth=5,arrowinset=0.2,flip](630,6)(690,66)
            \SetWidth{1.0}
            \Line(450,6)(450,-144)
            \SetWidth{3.0}
            \Line[arrow,arrowpos=0.5,arrowlength=12.5,arrowwidth=5,arrowinset=0.2,flip](450,6)(450,66)
            \SetColor{ForestGreen}
            \Bezier[dash,dsize=15](360,66)(360,-54)(510,-84)(690,-84)
            \Line[arrow,arrowpos=1,arrowlength=12.5,arrowwidth=5,arrowinset=0.2](388,17)(352,17)
            \Line[arrow,arrowpos=1,arrowlength=12.5,arrowwidth=5,arrowinset=0.2](440,-36)(440,-65)
            \Line[arrow,arrowpos=1,arrowlength=12.5,arrowwidth=5,arrowinset=0.2](640,-71)(640,-93)
            \SetColor{Red}
            \Bezier[dash,dsize=15](210,-24)(300,-84)(420,-114)(690,-114)
            \Line[arrow,arrowpos=1,arrowlength=12.5,arrowwidth=5,arrowinset=0.2](258,-40)(258,-74)
            \Line[arrow,arrowpos=1,arrowlength=12.5,arrowwidth=5,arrowinset=0.2](440,-92)(440,-122)
            \Line[arrow,arrowpos=1,arrowlength=12.5,arrowwidth=5,arrowinset=0.2](640,-105)(640,-128)
            \Text(210,25)[lb]{\Huge{\Black{$p_1$}}}
            \Text(460,25)[lb]{\Huge{\Black{$p_2$}}}
            \Text(680, 25)[lb]{\Huge{\Black{$p_3$}}}
            \Text(680,-180)[lb]{\Huge{\Black{$p_4$}}}
            \Text(460,-180)[lb]{\Huge{\Black{$p_5$}}}
            \Text(210,-175)[lb]{\Huge{\Black{$p_6$}}}
        \end{picture}
    }
    \caption{Examples of cuts in multiple channels that contribute to sequential discontinuities. The dashed lines cross the cut propagators, and the arrows specify the energy flow across them. The cut diagram on the left contributes to $\text{Disc}_{p_1^2, p_2^2}$ of the two-loop triangle, while the cut diagram on the right contributes to $\text{Disc}_{s_{23}, s_{123}}$ of the two-loop six-point diagram.}
    \label{fig: cuts multiple}
\end{figure}
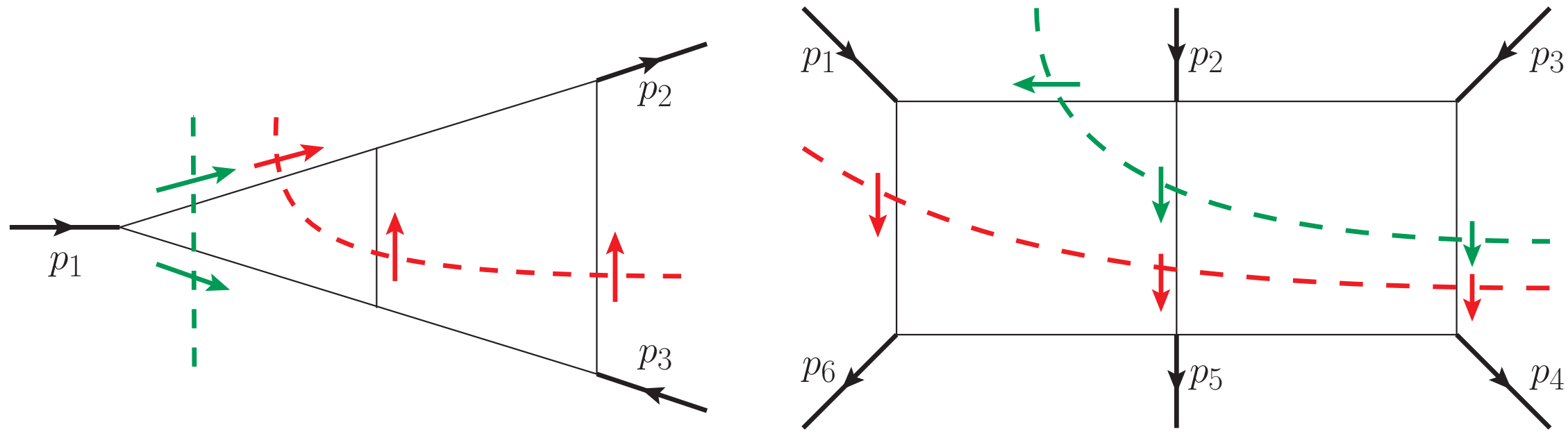

With these definitions, the relation between iterated
discontinuities and unitarity cuts takes the
form~\cite{Abreu:2014cla}
\begin{equation}
\label{eq: it_cuts=it_discs}
    \text{Disc}_{s_1,s_2,\dots,s_k}\mathcal{T}
    \Big|_{R_{s_1,s_2,\dots,s_k}}
    =
    (-1)^k
    \sum_{\mathcal{C}_{s_1,s_2,\dots,s_k}}
    \text{Cut}_{s_1,s_2,\dots,s_k}\mathcal{T}\,,
\end{equation}
where the sum runs over compatible configurations of cuts in
the specified channels. 
Both sides are evaluated in the region
$R_{s_1,s_2,\dots,s_k}$, where all cut invariants
$\{s_1,\dots,s_k\}$ are above their respective thresholds,
while all other Mandelstam invariants are below theirs.  The sequential discontinuities on the
left-hand side are defined recursively by
\begin{equation}
\label{eq: iterated_discs}
    \text{Disc}_{s_1,s_2,\dots,s_k}\mathcal{T}
    =
    \text{Disc}_{s_k}
    \left(
        \text{Disc}_{s_1,s_2,\dots,s_{k-1}}\mathcal{T}
    \right).
\end{equation}
The cutting prescription is independent of the order of the
channels. Accordingly, Eq.~\eqref{eq: it_cuts=it_discs} requires
the corresponding sequential discontinuities to be independent
of their order in the region considered
~\cite{Abreu:2014cla,Bourjaily:2020wvq}.
Examples of compatible cuts in multiple channels are shown
in Fig.~\ref{fig: cuts multiple}.

An immediate consequence of
Eq.~\eqref{eq: it_cuts=it_discs} is:
\begin{lemma}
\label{lemma}
Under the conditions of
Eq.~\eqref{eq: it_cuts=it_discs}, if no compatible set of cuts
exists in the specified channels, the corresponding iterated
discontinuity vanishes.
\end{lemma}

\noindent This lemma follows simply from the fact that, if no compatible cuts can be found, the sum in Eq.~\eqref{eq: it_cuts=it_discs} is empty. In Sec.~\ref{sec: new_constr}, we will see that this lemma has powerful consequences for the three-loop soft anomalous dimension.

The restriction of this prescription to cases without crossed cuts becomes natural when framed in terms of TOPT, where vertices are ordered in time and cuts impose energy conservation on intermediate states at specific times. 
Indeed, a pair of crossed cuts at times $\tau_a<\tau_b$
would require vertices at times $\tau_{v_1}$ and $\tau_{v_2}$ to satisfy $\tau_a<\tau_{v_1}<\tau_b$ and $\tau_b<\tau_{v_2}<\tau_a$, respectively. Since these conditions cannot be satisfied simultaneously, crossed cuts cannot arise in TOPT. Examples of the mapping between the covariant and time-ordered descriptions are shown in Fig.~\ref{fig:TOPT}. 

Ref.~\cite{Bourjaily:2020wvq} also addresses the mixed
$i\epsilon$ prescriptions arising when sequential
discontinuities are taken in TOPT. Using distributional
identities, the authors replace propagators carrying
$-i\epsilon$ by propagators carrying $+i\epsilon$ and
on-shell delta-function terms. The resulting representation
has $+i\epsilon$ prescriptions for all uncut propagators,
at the cost of including additional cuts in the specified
channels. Every term nevertheless requires at least one
intermediate-state cut in each specified channel within
a common time ordering. If no such time ordering exists,
introducing further cuts cannot resolve this obstruction.
Thus, this representation supports the same topological
vanishing criterion used here.

\subsection{Sequential discontinuities and the coaction}
\label{subsec:coaction}

In this subsection, we denote the function under consideration
by $F$ and assume that it is a combination of GPLs of weight $n$, with $d\log$ arguments drawn from
the alphabet $\mathcal{A}$~\cite{Goncharov:1998kja,goncharov2001multiple}.
For such functions, discontinuities act on the first factor
of the coaction~\cite{Duhr:2012fh}.
Following Ref.~\cite{Abreu:2014cla}, we review how this
property relates sequential discontinuities to specific
components of the iterated coaction. We begin by writing
\begin{align}
    \Delta_{\underbrace{1,1,\dots,1}_{k \, \text{times}},n-k}F=\sum_{(x_{i_1}, \dots, x_{i_k})\in\mathcal{A}^k}c_{i_1,...,i_k}\log x_{i_1} \otimes\dots\otimes\log x_{i_k}\otimes g_{x_{i_1}, \dots, x_{i_k}},
\end{align}
where the left-hand side denotes the $(1,1,\dots,1,n-k)$ coaction of $F$, and $g_{x_{i_1}, \dots, x_{i_k}}$ is a function of weight $n-k$. 
The sum runs over all ordered $k$-tuples of letters in
$\mathcal A$, or equivalently over their indices
$i_1,\ldots,i_k$.
Note that some of the coefficients $c_{i_1,...,i_k}$ may vanish. 

We now define an operator $\delta_{x_{j_1}, \dots, x_{j_k}}$ that extracts precisely the weight $n-k$ function that follows this sequence of logarithmic entries with arguments $\{x_{j_1}, \dots, x_{j_k}\}$:
\begin{align}
\label{coaction discontinuity}
    \delta_{x_{j_1}, \dots, x_{j_k}}F=\sum_{(x_{i_1}, \dots, x_{i_k})\in\mathcal{A}^k}\delta_{i_1\, j_1}\delta_{i_2\, j_2}\dots\delta_{i_k\, j_k}c_{i_1,...,i_k}g_{x_{i_1}, \dots, x_{i_k}}=c_{j_1,...,j_k}g_{x_{j_1}, \dots, x_{j_k}}.
\end{align}
Using this operator, we can isolate specific terms in the coaction, each of which can be interpreted as a sequential  discontinuity. 
Ref.~\cite{Abreu:2014cla} conjectured the following general
relation between sequential discontinuities and components
of the coaction, and verified it in several one- and
two-loop examples:
\begin{align}
\label{eq: it_discs=coaction}
    \text{Disc}_{s_1, \dots, s_k}F\Big|_{R_{s_1, \dots, s_k}}
    = \sum_{(x_1, \dots, x_k)\in\mathcal{A}^k}
    \left(\prod_{i=1}^k a_i(s_i, x_i)\right)
    \delta_{x_1, \dots, x_k}F
    + \mathcal{O}\left((i\pi)^{k+1} \right)\,.
\end{align}
This equality is expected to hold in the kinematic region $R_{s_1, \dots, s_k}$, modulo factors proportional to $(i \pi)^{k+1}$, and the factors $a_i(s_i, x_i)$ are given by $a_i(s_i, x_i) = \text{Disc}_{s_i} \left( \log x_i\right)$ after analytically continuing to the appropriate kinematic region. We refer the reader to Ref.~\cite{Abreu:2014cla} for a more detailed discussion.

At times, it is convenient to use the symbol \cite{Goncharov:2010jf}, which can be computed by maximally applying the coaction until only logarithms remain in each entry:
    \begin{equation}
  S(F)= \Delta_{\underbrace{1,1,\dots,1}_{n \, \text{times}}}F=\sum_{(x_{i_1}, \dots, x_{i_n})\in\mathcal{A}^n}c_{i_1,...,i_n}x_{i_1} \otimes\dots\otimes x_{i_n}. \end{equation}
Once in this form, it is conventional to just write the argument of each logarithm in each entry. The operator defined in \eqref{coaction discontinuity} can then be applied to the symbol as follows:
\begin{align}
\begin{split}
\delta_{x_{j_1},\ldots,x_{j_k}} S(F)
&= S\bigl(\delta_{x_{j_1},\ldots,x_{j_k}}F\bigr)
\\
&= \sum_{(x_{i_{k+1}},\ldots,x_{i_n})
          \in\mathcal A^{n-k}}
c_{j_1,\ldots,j_k,i_{k+1},\ldots,i_n}\,
x_{i_{k+1}}\otimes\cdots\otimes x_{i_n}.
     \end{split}
\end{align}
We will make extensive use of the symbol in later sections, when we bootstrap the massless and one-mass soft anomalous dimensions.

\section{The structure of the soft anomalous dimension}
\label{sec:struct}
In this section, we briefly review the structure of the soft
anomalous dimension and its description in terms of Wilson-line
correlators, preparing for the derivation of new constraints
on the three-loop quadrupole functions in
Section~\ref{sec: new_constr}. For a comprehensive review,
see Ref.~\cite{Agarwal:2021ais}.
We also derive a minimal colour basis\footnote{We retain the Jacobi redundancy to keep
permutation symmetry manifest. The Jacobi-free
combinations used in the bootstrap are defined in
Eq.~\eqref{eq:F123Defs}. } for the soft anomalous
dimension at three loops for amplitudes involving arbitrary
numbers of massive and massless external partons, extending
previous treatments of the massless and one-mass
cases~\cite{Almelid:2015jia,Liu:2022elt,Gardi:2025lws}
to configurations with several massive partons.
Uppercase indices denote massive external partons,
while lowercase indices denote massless external partons.

\subsection{The soft anomalous dimension at three loops}

QCD scattering amplitudes contain IR divergences originating
from soft and collinear regions of loop
momenta~\cite{Sen:1982bt,Catani:1998bh,Sterman:2002qn}.
In dimensional regularisation, all poles in $\epsilon$
of a UV-renormalised amplitude are of IR origin.
For an amplitude $\mathcal{M}$ involving $N$ massless and $M$ massive
external partons, these divergences
factorise as~\cite{Becher:2009cu,Becher:2009qa,Becher:2009kw}
\begin{equation}
    \mathcal{M}\Big(\hspace{-1pt}\{p_i, p_I\}, \mu, \alpha_s(\mu^2), \epsilon\hspace{-1pt}\Big)=\mathbf{Z}\Big(\hspace{-1pt}\{p_i, p_I\}, \mu_F, \alpha_s(\mu_F^2), \epsilon\hspace{-1pt}\Big)\mathcal{H}\Big(\hspace{-1pt}\{p_i, p_I\},  \mu, \mu_F, \alpha_s(\mu^2), \epsilon\hspace{-1pt}\Big),
    \label{eq:ampl_renorm}
\end{equation}
where $\mu_F$ is the factorisation scale introduced to separate the IR physics from the rest of the process, $\mu$ is the renormalisation scale, and $\epsilon$ is the dimensional regularisation parameter. 

The matrix-valued function $\mathbf{Z}$ captures all IR
singularities in $\mathcal{M}$, rendering the hard function
$\mathcal{H}$ free of IR poles in $\epsilon$~\cite{Aybat:2006mz,Dixon_2008}.
For all-order analyses of soft and collinear factorisation in
massless fixed-angle amplitudes, see
Refs.~\cite{Feige_2014,Ma:2019hjq}.
Factorisation implies that $\mathbf{Z}$ obeys a renormalisation
group (RG) equation of the form
\begin{equation}
    \frac{d}{d\log\mu_F}\mathbf{Z}\Big(\hspace{-1pt}\{p_i, p_I\}, \mu_F, \alpha_s(\mu_F^2), \epsilon\hspace{-1pt}\Big)=-\mathbf{Z}\Big(\hspace{-1pt}\{p_i, p_I\}, \mu_F, \alpha_s(\mu_F^2), \epsilon\hspace{-1pt}\Big) \, \mathbf{\Gamma}\Big(\hspace{-1pt}\{p_i, p_I\}, \mu_F, \alpha_s(\mu_F^2)\hspace{-1pt}\Big).
    \label{eq:RGforZ}
\end{equation}
The colour matrix $\mathbf{Z}$ retains all poles in $\epsilon$,\footnote{In a minimal subtraction scheme it consists exclusively of poles, while in a non-minimal scheme it may also include some finite and higher-order terms in $\epsilon$.} making $\mathcal{H}$ a finite function in four spacetime dimensions (while it still contains higher-order terms in the dimensional regularisation parameter). The finite function $\mathbf{\Gamma}$, known as the \emph{soft anomalous dimension}, governs all IR divergences; the poles in $\mathbf{Z}$ are generated by integrating to zero energy. Its analytic structure is the focus of this paper. 

Eq.~\eqref{eq:RGforZ} admits a formal solution as a path-ordered exponential
\begin{equation}
    \mathbf{Z}\Big(\hspace{-1pt}\{p_i, p_I\}, \mu_F, \alpha_s(\mu_F^2), \epsilon\hspace{-1pt}\Big)={\text P}\,\exp\left[-\int_0^{\mu_F}\frac{d\tau}{\tau}\mathbf{\Gamma}\Big(\{p_i, p_I\}, \tau, \alpha_s(\tau^2, \epsilon)\Big)\right],
    \label{eq:soft_factor_Z}
\end{equation}
where the path-ordering operator ${\text{P}}$ ensures that colour matrices in $\mathbf{Z}$ multiply each other in line with the ordering in the scale $\tau$, in such a way that matrices involving the hardest scale $\tau=\mu_F$ are closest to the hard interaction $\mathcal{H}$ in Eq.~(\ref{eq:ampl_renorm}). 
The soft anomalous dimension $\mathbf{\Gamma}$ in
Eq.~\eqref{eq:RGforZ} depends on the IR factorisation scale
$\mu_F$ through the $D=(4-2\epsilon)$ dimensional running coupling and, when massless
external partons are present, explicitly through a term
linear in $\log\mu_F$. 
The coefficient of this logarithm is colour diagonal and
determined by the lightlike cusp anomalous dimensions of
the massless external
partons~\cite{Korchemsky:1987wg,Korchemsky:1988hd}.
This explicit scale dependence reflects the overlap
of soft and collinear
singularities~\cite{Sterman:2002qn,Aybat:2006mz,Gardi:2009zv,Becher:2009kw}.
Through three loops, these anomalous dimensions take the
form $C_i\,\gamma_{\mathrm{cusp}}(\alpha_s)$, where
$C_i$ is the quadratic Casimir of the colour representation
of parton $i$. 
The $D$-dimensional running coupling satisfies
\begin{equation}
\label{eq:running-coupling}
    \frac{d}{d\log\mu_F^2}
    \frac{\alpha_s(\mu_F^2, \epsilon)}{4\pi}
    =
    -\epsilon\,\frac{\alpha_s(\mu_F^2, \epsilon)}{4\pi}
    -b_0\left(\frac{\alpha_s(\mu_F^2, \epsilon)}{4\pi}\right)^2
    -b_1\left(\frac{\alpha_s(\mu_F^2, \epsilon)}{4\pi}\right)^3
    +\mathcal{O}(\alpha_s^4),
\end{equation}
where $b_0$ and $b_1$ are the one- and two-loop
beta-function coefficients.

Rescaling symmetry severely restricts the allowed kinematic
dependence of the soft anomalous dimension. Correlations between
massive Wilson lines are invariant under independent rescalings
of their velocities.
For massless lines, soft--collinear singularities induce
an anomalous rescaling dependence, controlled by the
lightlike cusp anomalous dimension. 
After separating the dipole terms carrying this dependence, the
remaining multiparton correlations depend only on
rescaling-invariant combinations~\cite{Gardi:2009qi,Becher:2009qa,Becher:2009kw}.
For two massive lines, a convenient invariant is
\begin{equation}
\label{gamma_IJ}\gamma_{IJ}\equiv\frac{2\beta_I\cdot\beta_J}{\sqrt{\beta_I^2}\sqrt{\beta_J^2}}
\end{equation}
associated with the cusp angle between two parton momentum  directions (or Wilson lines),
where $\beta_K\sim p_K$ is the four-velocity of a massive parton $K$. It is also convenient to introduce the variable $\alpha_{AB}$ as the solution to the equation
\begin{equation}
\label{eq: alpha}
    -\,\gamma_{AB}=\alpha_{AB}+\frac{1}{\alpha_{AB}}, 
\end{equation}
which admits an inversion symmetry $\alpha_{AB}\rightarrow\alpha^{-1}_{AB}$ \cite{Gardi:2013saa}. This symmetry separates the $\alpha$ plane into two regions: $|\alpha|<1$, which is where we usually choose to work, and $|\alpha|>1$. Further details about the $\alpha_{AB}$ variables and their connection to the Mandelstam invariants $s_{AB}$ are provided in Appendix~\ref{sec:alpha_details}.

If some external lines become massless, the corresponding
cusp invariants in Eq.~\eqref{gamma_IJ} diverge, but suitable
ratios have finite, rescaling-invariant limits.
Examples include
\begin{equation}
    y_{IJk}\equiv\frac{\beta_I\cdot\beta_k\sqrt{\beta_J^2}}{\beta_J\cdot\beta_k\sqrt{\beta_I^2}}=\lim_{\beta_K^2\rightarrow0}\frac{\gamma_{IK}}{\gamma_{JK}}
\end{equation}
for correlators involving two massive partons and one massless parton~\hbox{\cite{Ferroglia:2009ep,Gardi:2025ule}}, and
\begin{equation}
    \label{eq: r-ratios}r_{jkI}=\frac{\beta^2_I\beta_j\cdot\beta_k}{2\beta_I\cdot\beta_{j}\beta_I\cdot\beta_k}=\lim_{\beta^2_J, \beta^2_K\rightarrow0}\frac{\gamma_{JK}}{\gamma_{IJ}\gamma_{IK}}
\end{equation}
for correlators involving two massless partons and one massive parton~\cite{Liu:2022elt,Gardi:2025lws}. 

For correlations involving four massless lines, the
nontrivial rescaling-invariant dependence can be expressed
in terms of \emph{conformally invariant cross-ratios}
(CICRs)~\cite{Becher:2009cu,Gardi:2009qi,Dixon:2009ur,Almelid:2015jia},
\begin{equation}
\label{eq: CICR}
    \rho_{ijkl}\equiv\frac{(\beta_i\cdot\beta_j)(\beta_k\cdot\beta_l)}{(\beta_i\cdot\beta_k)(\beta_j\cdot\beta_l)}=\lim_{\beta_I^2,\beta_J^2,\beta_K^2,\beta_L^2\rightarrow0}\frac{\gamma_{IJ}\gamma_{KL}}{\gamma_{IK}\gamma_{JL}},
\end{equation}
which are invariant under 
independent rescalings of each velocity. 
Their interpretation as cross-ratios of distances on the
celestial sphere is discussed in Ref.~\cite{Almelid:2017qju}.
These cross-ratios satisfy several permutation symmetry relations, for example
\begin{equation}
    \rho_{ijkl}=\rho_{jilk}=\rho_{klij}=\rho_{lkji}.
\end{equation}
For four distinct labels, the 24 permutations therefore
give six generically distinct cross-ratios. Writing
$u=\rho_{ijkl}$ and $v=\rho_{ilkj}$, these are
\begin{equation}
    u,\qquad v,\qquad \frac{1}{u},\qquad
    \frac{1}{v},\qquad \frac{u}{v},\qquad \frac{v}{u}.
\end{equation}
Thus, two independent cross-ratios suffice to parametrise
the kinematic dependence~\cite{Dixon:2009ur,Almelid:2015jia,Almelid}.

Note that rescaling invariance not only restricts the possible variables the soft anomalous dimension can depend on, but also -- when combined with other considerations, such as Bose symmetry -- may exclude entire colour structures. For example, the fully antisymmetric colour structure involving three partons cannot appear unless at least two of the partons it involves are massive~\cite{Ferroglia:2009ep, Ferroglia:2009ii,Gardi:2009qi, Becher:2009cu}.

Another key feature of the soft anomalous dimension is the fact that it consists exclusively of connected colour factors. 
This follows from the non-abelian exponentiation theorem~\cite{Gardi:2013ita}, which generalises earlier non-abelian diagrammatic exponentiation theorems to the case of  multileg scattering~\cite{Gatheral:1983cz, Frenkel:1984pz, sterman_talk, Mitov:2010rp, Gardi:2010rn, Gardi:2011wa}. 
The connected structures relevant through three loops are summarised in Table~\ref{tab:col_struct}.
\begin{table}
    \centering
    \begin{tabular}{ccc}
    \toprule
        In Eqs.~\eqref{eq:generalGamma} and \eqref{eq:Gamma}& Explicit form & Kinematic dependence \\
        \hline
        $\mathbf{D}_{(ij)}$ & $\mathbf{T}_i\cdot\mathbf{T}_j$ & Explicit in Eq.~\eqref{eq:generalGamma}\\
        $\mathbf{D}_{(Ij)}$ & $\mathbf{T}_I\cdot\mathbf{T}_j$ & Explicit in Eq.~\eqref{eq:generalGamma}\\
        $\mathbf{D}_{(IJ)}$ & $\mathbf{T}_I\cdot\mathbf{T}_J$ & $\{\gamma_{IJ}\}$\\
        $\mathbf{T}_{(IJk)}$ & $if_{abc}\mathbf{T}^a_I\mathbf{T}^b_J\mathbf{T}^c_k$ & $\{\gamma_{IJ}, y_{IJk}\}$\\
        $\mathbf{T}_{(IJK)}$ & $if_{abc}\mathbf{T}^a_I\mathbf{T}^b_J\mathbf{T}^c_K$ & $\{\gamma_{IJ}, \gamma_{JK}, \gamma_{IK}\}$\\
        $\mathbf{Q}_{(ij;ik)}$ & $\frac{1}{2}f_{abe}f_{cde}\left\{\mathbf{T}^a_i,\mathbf{T}^c_i\right\}\mathbf{T}^b_j\mathbf{T}^d_k$ & Constant\\
        $\mathbf{Q}_{(Ij;Ik)}$& $\frac{1}{2}f_{abe}f_{cde}\left\{\mathbf{T}^a_I,\mathbf{T}^c_I\right\}\mathbf{T}^b_j\mathbf{T}^d_k$  & $\{r_{jkI}\}$\\
        $\mathbf{Q}_{(IJ;Ik)}$ & $\frac{1}{2}f_{abe}f_{cde}\left\{\mathbf{T}^a_I,\mathbf{T}^c_I\right\}\mathbf{T}^b_J\mathbf{T}^d_k$ & $\{\gamma_{IJ}, y_{IJk}\}$\\
        $\mathbf{Q}_{(IJ;IK)}$ & $\frac{1}{2}f_{abe}f_{cde}\left\{\mathbf{T}^a_I,\mathbf{T}^c_I\right\}\mathbf{T}^b_J\mathbf{T}^d_K$ & $\{\gamma_{IJ}, \gamma_{JK}, \gamma_{IK}\}$\\
        $\mathbf{Q}_{(ij;kl)}$ & $f_{abe}f_{cde}\mathbf{T}^a_i\mathbf{T}^b_j\mathbf{T}_k^c\mathbf{T}_l^d$ & $\{\rho_{ijkl}, \rho_{ilkj}\}$\\
        $\mathbf{Q}_{(Ij;kl)}$ & $f_{abe}f_{cde}\mathbf{T}^a_I\mathbf{T}^b_j\mathbf{T}_k^c\mathbf{T}_l^d$ & $\{r_{jkI}, r_{klI}, r_{jlI}\}$\\
        $\mathbf{Q}_{(IJ;kl)}$ & $f_{abe}f_{cde}\mathbf{T}^a_I\mathbf{T}^b_J\mathbf{T}_k^c\mathbf{T}_l^d$ & $\{\gamma_{IJ}, y_{IJk}, y_{IJl}, r_{klI}\}$\\
        $\mathbf{Q}_{(Ij;Kl)}$ & $f_{abe}f_{cde}\mathbf{T}^a_I\mathbf{T}^b_j\mathbf{T}_K^c\mathbf{T}_l^d$ & $\{\gamma_{IK}, y_{IKj}, y_{IKl}, r_{jlI}\}$\\
        $\mathbf{Q}_{(IJ;Kl)}$ & $f_{abe}f_{cde}\mathbf{T}^a_I\mathbf{T}^b_J\mathbf{T}_K^c\mathbf{T}_l^d$ & $\{\gamma_{IJ}, \gamma_{IK}, \gamma_{JK}, y_{IJl}, y_{IKl}\}$\\
        $\mathbf{Q}_{(IJ;KL)}$ & $f_{abe}f_{cde}\mathbf{T}^a_I\mathbf{T}^b_J\mathbf{T}_K^c\mathbf{T}_L^d$ & $\{\gamma_{IJ}, \gamma_{IK}, \gamma_{IL}, \gamma_{JK}, \gamma_{JL}, \gamma_{KL}\}$\\
        \bottomrule
    \end{tabular}
    \caption{
The connected colour structures used to parametrise the
soft anomalous dimension through three loops for general
multileg amplitudes involving massive and massless external
partons. Colour conservation has been used to eliminate
some redundant structures; further reductions are possible
for amplitudes with fewer external legs.
We use the colour insertion operator notation introduced
in Refs.~\cite{Catani:1996vz,Bassetto:1983mvz}.
Appendix~\ref{app:color} shows a graphical representation
of each colour structure.}
    \label{tab:col_struct}
\end{table}

The soft anomalous dimension $\mathbf{\Gamma}$ is known through
two loops for any number of partons and any mass
configuration~\cite{Ferroglia:2009ep,Ferroglia:2009ii}.
At three loops, complete results are available for purely
massless amplitudes~\cite{Almelid:2015jia} and for amplitudes
with one massive and any number of massless
partons~\cite{Liu:2022elt,Gardi:2025lws}.
Through three loops, the soft anomalous dimension for a process
involving $N$ massless and $M$ massive partons can be written as
\begin{align}
\begin{split}
\label{eq:generalGamma}
    \mathbf{\Gamma}\Big(\{p_i, p_I\}, \mu_F, \alpha_s(\mu_F^2)\Big)&\hspace{-1pt}=\hspace{-1pt}\sum_{i=1}^N\mathbf{1}\gamma_i\hspace{-1pt}(\alpha_s)+\hspace{-1pt}\sum_{I=1}^M\mathbf{1}\Omega_I\hspace{-1pt}(\alpha_s)+\hspace{-1pt}\frac{1}{2}\sum_{(i,j)}\mathbf{D}_{(ij)}\gamma_{\text{cusp}}(\alpha_s)\log\frac{\mu_F^2}{-2p_i\cdot p_j}\\
    &+\sum_{i,J}\mathbf{D}_{(iJ)}\gamma_{\text{cusp}}(\alpha_s)\log\frac{\mu_F\sqrt{p_J^2}}{-2p_i\cdot p_J}+\frac{1}{2}\sum_{(I,J)}\mathbf{D}_{(IJ)}\mathcal{D}_{(IJ)}\\
    &+\frac{1}{2}\sum_{(I,J), \,k}\mathbf{T}_{(IJk)}\mathcal{T}_{(IJk)}+\frac{1}{6}\sum_{(I,J,K)}\mathbf{T}_{(IJK)}\mathcal{T}_{(IJK)}+\mathbf{\Gamma}_{\ge4}(\alpha_s).
\end{split}
\end{align}
Here, each boldface character (such as $\mathbf{1}, \mathbf{D}_{(ij)}, \dots$) represents a colour structure whose subscripts specify the number and nature of the partons participating in the interaction, while calligraphic letters (such as $\mathcal{D}_{(IJ)}, \mathcal{T}_{(IJk)}, \dots$) represent kinematic functions. Note that while we suppress kinematic dependence in the interest of readability, we keep track of the symmetries of the kinematic functions (and which partons they depend on) by assigning them the same subscripts as the colour matrices they multiply. The notation $(i, j, \dots)$ under the summation signs refers to ordered tuples of distinct parton indices, with permutations acting simultaneously on the colour factors and the arguments of the associated kinematic functions; the explicit symmetry factors compensate for repeated contributions. As an example, the tripole sums can equivalently be expressed as
\begin{subequations}
    \begin{align}
        \frac{1}{6}\sum_{(I,J,K)}\mathbf T_{(IJK)}\mathcal T_{(IJK)}&=\sum_{I<J<K}\mathbf T_{(IJK)}\mathcal T_{(IJK)},\\
        \frac{1}{2}\sum_{(I, J), k}\mathbf T_{(IJk)}\mathcal T_{(IJk)}&=\sum_{I<J}\sum_k\mathbf T_{(IJk)}\mathcal T_{(IJk)}.
    \end{align}
\end{subequations}
Uppercase indices label massive partons and run from \(1\) to \(M\), while lowercase indices label massless partons and run from \(1\) to \(N\). 

The colour-diagonal terms~$\gamma_i$ and $\Omega_I$ in
Eq.~\eqref{eq:generalGamma} have no kinematic dependence.
The~$\gamma_i$ are the collinear anomalous dimensions associated
with massless external partons~\cite{Aybat:2006mz,Gardi:2009qi,Becher:2009qa,Falcioni:2019nxk}
and are known in QCD through four loops from massless form
factors~\cite{Moch:2005id,Moch:2005tm,Baikov:2009bg,vonManteuffel:2020vjv,Agarwal:2021zft}.
The $\Omega_I$ describe single-parton soft contributions associated
with massive external lines. Their three-loop coefficients can
be extracted from the massive cusp anomalous
dimension~\cite{Korchemsky:1991zp,Grozin:2014hna,Grozin:2015kna}; explicit
heavy-quark expressions are also given in
Ref.~\cite{Bruser:2019yjk}.

The colour operators $\mathbf D$ involve two generators
and are referred to as \emph{dipoles}; they are symmetric under
exchange of the parton labels. For dipoles involving at least
one massless parton, soft--collinear factorisation fixes the
logarithmic kinematic dependence through three loops in terms
of the lightlike cusp anomalous dimension
$\gamma_{\mathrm{cusp}}$ introduced above, with quadratic colour
factors carried by the dipole
operators~\cite{Gardi:2009qi,Becher:2009cu,Becher:2009qa,Becher:2009kw}.
For two massive partons, $\mathcal D_{(IJ)}$ contains the
angle-dependent massive cusp contribution, with the single-parton
terms $\Omega_I$ kept separate~\cite{Korchemsky:1987wg,Korchemsky:1991zp}.
Massive dipoles are present already at one loop, with the expansion
\begin{equation}
    \mathcal{D}_{(IJ)}=\sum_{n=1}\left(\frac{\alpha_s}{4\pi}\right)^n\mathcal{D}^{(n)}_{(IJ)}.
\end{equation}
The full angular dependence of the massive cusp anomalous
dimension $\mathcal{D}^{(n)}_{(IJ)}$ was computed at two
loops~\cite{Korchemsky:1987wg,Kidonakis:2009ev} and at three
loops~\cite{Grozin:2014hna,Grozin:2015kna}.

The operators $\mathbf T$ involve three generators
and are referred to as \emph{tripoles}; they are fully
antisymmetric under permutations of the parton labels.
The functions $\mathcal T_{(IJk)}$ and $\mathcal T_{(IJK)}$
describe three-parton correlations involving at least two
massive partons. They first appear at two loops, where both
functions are known analytically~\cite{Ferroglia:2009ep,Ferroglia:2009ii}.
Their kinematic arguments and colour operators are listed in
Table~\ref{tab:col_struct}, and their perturbative expansions read
\begin{subequations}
    \begin{align}
        \mathcal{T}_{(IJK)}=\sum_{n=2}\left(\frac{\alpha_s}{4\pi}\right)^n\mathcal{T}^{\,(n)}_{(IJK)},\\
        \mathcal{T}_{(IJk)}=\sum_{n=2}\left(\frac{\alpha_s}{4\pi}\right)^n\mathcal{T}^{\,(n)}_{(IJk)},
    \end{align}
\end{subequations}

Finally, the remaining contribution
$\mathbf{\Gamma}_{\ge4}$ starts at three loops, where it comprises
connected colour structures involving four generators.
At this order, its decomposition reads
\begin{equation}
\label{eq:Gamma}
\begin{split}
    \mathbf{\Gamma}_{\ge4}&\Big( \alpha_s(\mu_F^2)\Big)=\frac{1}{2}\sum_{(i,j,k)}\mathbf{Q}_{(ij;ik)}\mathcal{F}_{(ij;ik)}+\frac{1}{8}\sum_{(i,j,k,l)}\mathbf{Q}_{(ij;kl)}\mathcal{F}_{(ij;kl)}+\frac{1}{2}\sum_{I, (j,k)}\mathbf{Q}_{(Ij;Ik)}\mathcal{F}_{(Ij;Ik)}\\
    &+\frac{1}{2}\sum_{I,(j,k,l)}\mathbf{Q}_{(Ij;kl)}\mathcal{F}_{(Ij;kl)}+ \sum_{(I,J),k}\mathbf{Q}_{(IJ;Ik)}\mathcal{F}_{(IJ;Ik)}\\
    &+\sum_{(I,J),\,(k,l)}\left(\frac{1}{4}\mathbf{Q}_{(IJ;kl)}\mathcal{F}^{A}_{(IJ;kl)}+\frac{1}{2}\mathbf{Q}_{(Ik;Jl)}\mathcal{F}^{S}_{(Ik;Jl)}\right)+\frac{1}{2}\sum_{(I,J,K),l}\mathbf{Q}_{(IJ;Kl)}\mathcal{F}_{(IJ;Kl)}\\
    &+\frac{1}{2}\sum_{(I,J,K)}\mathbf{Q}_{(IJ;IK)}\mathcal{F}_{(IJ;IK)}+\frac{1}{8}\sum_{(I,J,K,L)}\mathbf{Q}_{(IJ;KL)}\mathcal{F}_{(IJ;KL)}+\mathcal{O}(\alpha_s^4).
\end{split}
\end{equation}
The colour structure $\mathbf{Q}$ involves four generators, and is called a \emph{quadrupole}. Quadrupoles, whose explicit realisations are given in Table~\ref{tab:col_struct}, are antisymmetric under the exchange of the indices on either side of the semicolon, but symmetric under the exchange of the two pairs, so
\begin{align}
  \mathbf{Q}_{(kl;ij)}=  \mathbf{Q}_{(ij;kl)}=-\mathbf{Q}_{(ji;kl)}=\mathbf{Q}_{(ji;lk)}\,.
\end{align}
The Jacobi identity also relates different pairings of
the same four partons:
\begin{equation}
\label{eq:Jacobi}
    \mathbf{Q}_{(IJ;KL)}
    -\mathbf{Q}_{(IK;JL)}
    +\mathbf{Q}_{(IL;JK)}=0.
\end{equation}
This identity holds for any assignment of parton masses.
We retain all three pairings in Eq.~\eqref{eq:Gamma}
to keep permutation symmetry manifest. The corresponding
freedom in the quadrupole functions is discussed in
Section~\ref{sec:SADlevel_constr}.
The functions multiplying these colour structures -- the so-called quadrupole functions $\mathcal{F}$ -- will be the central objects of study in Section~\ref{sec: new_constr}; they admit an expansion in $\alpha_s$ starting at three loops
\begin{equation}
    \mathcal{F}_{(IJ;KL)}=\sum_{n=3}\left(\frac{\alpha_s}{4\pi}\right)^n\mathcal{F}^{(n)}_{(IJ;KL)}.
\end{equation}
Where relevant, we include an additional superscript indicating whether the kinematic function is symmetric or antisymmetric under the exchange of the two massive/massless lines. We summarise the kinematic dependence associated to each term in Table~\ref{tab:col_struct}, alongside the definition of the corresponding colour structure. 

Eq.~\eqref{eq:Gamma} was derived by considering all possible assignments of massive and massless legs in the quadrupoles and then using colour conservation, which states that
\begin{equation}
    \left(\sum_{i=1}^N\mathbf{T}_i+\sum_{I=1}^{M}\mathbf{T}_I\right)\mathcal{M}=0\, ,
\end{equation}
to obtain relations among the various $\mathbf{Q}$ structures. This allows us to pick a basis of independent colour structures, similarly to what was done in Ref.~\cite{Liu:2022elt} for the case of one massive leg. The terms in Eq.~\eqref{eq:Gamma} that do not involve any massive line were computed in Ref.~\cite{Almelid:2015jia}, while those containing one massive leg (thus, one uppercase index) have been computed in Refs.~\cite{Liu:2022elt, Gardi:2025lws}. No other terms have yet been computed.

\subsection{Infrared singularities from correlators of timelike Wilson lines}

The standard strategy for determining $\mathbf{\Gamma}$ entails computing correlators of semi-infinite Wilson lines $\langle\Phi_{\beta_I}\Phi_{\beta_J}\dots\rangle$, where each Wilson line
\begin{equation}
    \Phi_{\beta_I}=\text{P}\exp\left[ig_s\int_0^\infty ds\,\beta_I\cdot A(s\beta_I)\right]
\end{equation}
represents the radiation emitted from a coloured particle moving
along a classical trajectory with four-velocity $\beta_I$.
Here $\text{P}$ denotes path ordering of the colour matrices
along the trajectory $x^\mu=s\beta_I^\mu$.
Cusps and intersections of Wilson lines generate UV
divergences requiring renormalisation~\cite{Polyakov:1980ca,Arefeva:1980zd,Dotsenko:1979wb,Brandt:1981kf}.
The associated anomalous dimensions encode the soft IR
singularities of scattering amplitudes~\cite{Sen:1982bt,Korchemsky:1985xj,Ivanov:1985np}.
Their relation to Sudakov behaviour and heavy-quark limits was
developed for example in Refs.~\cite{Korchemsky:1988hd,Korchemsky:1991zp,Korchemsky:1992xv,Gardi:2005yi}.

The formulation used here to define the soft anomalous dimension of multileg amplitudes starts from timelike Wilson lines,
which describe the classical trajectories of massive partons, all meeting at a single hard-scattering effective vertex. 
Massless partons are then incorporated by taking lightlike limits. 
Two strategies have been employed to implement these limits. 
First, one can extract the relevant $1/\epsilon$ coefficients
from the timelike Wilson-line correlator and then expand their
integral representations near the lightcone,
$\beta_I^2\to 0$, as was done in the calculation of the fully
massless soft anomalous dimension in Ref.~\cite{Almelid:2015jia}.
Alternatively, one can perform an asymptotic expansion of the correlator integrand using the method of regions \cite{Beneke:1997zp, Smirnov:1994tg, Pak:2010pt, Jantzen:2011nz, Jantzen:2012mw,Semenova:2018cwy, Gardi:2022khw, Ma:2023hrt}, as was done in the computation of the soft anomalous dimension for amplitudes with one massive particle~\cite{Gardi:2025lws}. A detailed description of the second approach is given in Ref.~\cite{Gardi:2025ule}. 
In this paper we derive constraints on the analytic structure
of the correlator integrals that apply to any mass configuration.
Their derivation does not require separate treatments of
massive and massless partons.

The soft IR singularities of the scattering amplitude are
encoded in the UV renormalisation of the common Wilson-line
vertex. The correlator itself also contains long-distance
IR divergences. Without an additional regulator, its
perturbative corrections are scaleless and vanish in
dimensional regularisation, with UV and IR poles cancelling.
To isolate the UV divergences that determine the soft
anomalous dimension, we introduce an IR regulator 
by damping long-distance interactions~\cite{Gardi:2011yz}:\footnote{Here we follow the standard approach of regularising all Wilson lines, but notice that partially regularised correlators have been used in \cite{Gardi:2021gzz, Henn:2023pqn}.}
\begin{equation}
\label{Regularised_Wilson_line}
    \Phi^{(m)}_{\beta_I}=\text{P}\,\text{exp}\left[ig_s\int_0^\infty ds\,e^{-ims\sqrt{\beta_I^2-i\epsilon}}\,\beta_I\cdot A(s\beta_I)\right]\,,
\end{equation}
where $m$ is an auxiliary IR scale.
For timelike Wilson lines, this prescription preserves
rescaling invariance and makes the correlator IR  finite.
Its remaining poles are therefore UV divergences, which
can be removed by renormalisation.

Following the standard renormalisation procedure for UV-divergent correlators, we write
\begin{equation}
    \Big\langle\Phi^{(m)}_{\beta_I}\Phi^{(m)}_{\beta_J}\dots\Big\rangle_{\text{ren.}}=\Big\langle\Phi^{(m)}_{\beta_I}\Phi^{(m)}_{\beta_J}\dots\Big\rangle \mathbf{Z}\Big(\{\gamma_{IJ}\}, \alpha_s(\mu_F^2), \epsilon_{\text{UV}}\Big).
\end{equation}
For several Wilson lines, renormalisation mixes colour
structures and is consequently matrix valued~\cite{Brandt:1981kf,Korchemskaya:1994qp,Kidonakis:1997gm,Kidonakis:1998nf}.
The corresponding renormalisation group equation is
\begin{equation}
    \frac{d}{d\log\mu_F}\mathbf{Z}\Big(\{\gamma_{IJ}\}, \alpha_s(\mu_F^2), \epsilon\Big)=-\mathbf{Z}\Big(\{\gamma_{IJ}\}, \alpha_s(\mu_F^2), \epsilon\Big)\mathbf{\Gamma}\Big(\{\gamma_{IJ}\}, \alpha_s(\mu_F^2)\Big),
    \label{eq:renorm_reg_wilson_corr}
\end{equation}
which is solved, in terms of the expansion coefficients of the soft anomalous dimension $\mathbf{\Gamma}$
\begin{equation}
    \mathbf{\Gamma}=\sum_n\left(\frac{\alpha_s}{4\pi}\right)^n\mathbf{\Gamma}^{(n)},
\end{equation}
through three loops in minimal subtraction by~\cite{Gardi:2011yz}
\begin{equation}
    \begin{split}
        \mathbf{Z}=&\exp\bigg[\frac{\alpha_s}{4\pi}\frac{1}{2\epsilon}\mathbf{\Gamma}^{ (1)}+\left(\frac{\alpha_s}{4\pi}\right)^2\left(\frac{1}{4\epsilon}\mathbf{\Gamma}^{ (2)}-\frac{b_0}{4\epsilon^2}\mathbf{\Gamma}^{(1)}\right)+\left(\frac{\alpha_s}{4\pi}\right)^3\bigg(\frac{1}{6\epsilon}\mathbf{\Gamma}^{(3)}+\\
        &+\frac{1}{48\epsilon^2}\left[\mathbf{\Gamma}^{(1)},\mathbf{\Gamma}^{(2)}\right]-\frac{1}{6\epsilon^2}\left(b_0\mathbf{\Gamma}^{(2)}+b_1\mathbf{\Gamma}^{(1)}\right)+\frac{b_0^2}{6\epsilon^3}\mathbf{\Gamma}^{(1)}\bigg)+\mathcal{O}(\alpha_s^4)\bigg].
    \end{split}
    \label{eq:Z_UV_reg}
\end{equation}
The exponentiation dictated by the renormalisation group,
expressed above in terms of $\mathbf{\Gamma}$, can be matched
to the diagrammatic exponentiation of the Wilson-line
correlator~\cite{Gardi:2011yz,Mitov:2010rp,Gardi:2013ita,
Frenkel:1984pz,Gatheral:1983cz}:
 \begin{equation}  \Big\langle\Phi^{(m)}_{\beta_I}\Phi^{(m)}_{\beta_J}\dots\Big\rangle=\exp\left[\sum_n\left(\frac{\alpha_s}{4\pi}\right)^nw^{(n)}\right]=\exp\left[\sum_{n,l}\left(\frac{\alpha_s}{4\pi}\right)^n\epsilon^lw^{(n,l)}\right],
\end{equation}
 where the exponent is expressed in terms of \emph{webs}, namely sets of diagrams differing only by the order of attachment of gluons to a given Wilson line~\cite{Gardi:2011yz,Gardi:2010rn, Gardi:2011wa}. 
Here $\alpha_s$ is the renormalised QCD coupling, and
the coefficients $w^{(n)}$ include the usual QCD
counterterms. The multi-eikonal operator itself has
not yet been renormalised: the matrix $\mathbf{Z}$
removes its remaining UV divergences~\cite{Gardi:2010rn,Gardi:2011yz}.

The organisation into webs explains how colour-disconnected
diagrams contribute to the fully colour-connected soft
anomalous dimension: they enter in specific linear
combinations that form connected non-Abelian colour
structures. An efficient algorithm for determining these contributions has been derived using the replica trick in  Refs.~\cite{Gardi:2010rn, Gardi:2013ita}. The diagrams entering $\mathbf{\Gamma}$ at three loops were fully classified in Ref.~\cite{Almelid}; we briefly describe them below.

Using Eqs.~\eqref{eq:renorm_reg_wilson_corr} and \eqref{eq:Z_UV_reg} and the properties of the exponential of matrices, one finds expressions for the coefficients of the soft anomalous dimension in terms of webs \cite{Gardi:2011yz}:
\begin{subequations}
\label{eq:subtracted webs}
    \begin{align}
    \mathbf{\Gamma}^{(1)}=&-2w^{(1,-1)},\\
    \mathbf{\Gamma}^{(2)}=&-4w^{(2,-1)}-2\left[w^{(1,-1)},w^{(1,0)}\right],\\
    \begin{split}
        \mathbf{\Gamma}^{(3)}=&-6w^{(3,-1)}+\frac{3}{2}b_0\left[w^{(1,-1)},w^{(1,1)}\right]+3\left[w^{(1,0)},w^{(2,-1)}\right]+3\left[w^{(2,0)},w^{(1,-1)}\right]\\
        &+\left[w^{(1,0)},\left[w^{(1,-1)},w^{(1,0)}\right]\right]-\left[w^{(1,-1)},\left[w^{(1,-1)},w^{(1,1)}\right]\right]
    \end{split},
\end{align}
\end{subequations}
and so on, with increasingly complex expressions involving nested commutators at higher orders. The equations above provide a practical means of computing the soft anomalous dimension at the desired order through the computation of the appropriate terms in the diagrammatic expansion of the regularised correlators of Wilson lines. The method we present in this paper does not involve any direct calculation of the aforementioned diagrams, but relies crucially on the fact that each coefficient in the expansion of the soft anomalous dimension $\mathbf{\Gamma}$ can be obtained as a linear combination of webs at the desired order and lower orders.

At three loops, in $w^{(3)}$, we find webs that connect four distinct Wilson lines for the first time, giving rise to quadrupole colour structures $\mathbf{Q}$. Since these webs will lie at the heart of our analysis in the next section, we briefly introduce them here. Four of these webs are fully connected, namely the web involving a four-gluon vertex $w_{4g}$ and the three webs that involve a pair of three-gluon vertices (often referred to as $w_{(3g)^2}$ for brevity, when the colour channel is irrelevant or is clear from the context). These webs are depicted in Fig.~\ref{fig:FCwebs4lines}. (More specifically, the second web in this figure corresponds to $w_{(3g)^2(IK;JL)}$, which contributes to the colour structure $\mathbf{Q}_{(IK;JL)}$; the other $w_{(3g)^2}$ webs are permutations of this one.) All the other webs connecting four lines and contributing to the same colour structure $\mathbf{Q}_{(IK;JL)}$ through the non-abelian exponentiation theorem \cite{Gardi:2013ita} are shown in Figs.~\ref{fig: NFC4lines} and~\ref{fig:MGEWs}. Finally, webs connecting two or three Wilson lines also contribute to the quadrupole function through colour conservation, as described in Refs.~\cite{Almelid:2015jia, Almelid}; a representative web that connects three of the four lines is shown in Fig.~\ref{fig:wless4legs}.
\begin{figure}
    \centering
    \scalebox{0.45}{
        \begin{picture}(338,324) (147,-69)
            \SetWidth{2.8}
            \SetColor{Black}
            \Line[double,sep=5.5](150,234)(450,-66)
            \Line[double,sep=5.5](150,-66)(450,234)
            \SetWidth{1}
            \Line[double,sep=3.5](300,84)(300,24)
            \SetWidth{1.0}
            \GluonArc[clock](270,144)(67.082,153.435,63.435){7.5}{6}
            \GluonArc[clock](330,144)(67.082,116.565,26.565){7.5}{6}
            \GluonArc(360,54)(161.555,111.801,201.801){7.5}{15}
            \GluonArc[clock](205,39)(190.394,60.068,-13.671){7.5}{15}
            \GOval(300,84)(20,20)(0){0.882}
            \Vertex(300,204){8}
            \Vertex(211,-6){8}
            \Vertex(211,175){8}
            \Vertex(390,175){8}
            \Vertex(390,-6){8}
            \Text(166,230)[lb]{\LARGE{\Black{$I$}}}
            \Text(425,230)[lb]{\LARGE{\Black{$J$}}}
            \Text(175,-66)[lb]{\LARGE{\Black{$K$}}}
            \Text(415,-66)[lb]{\LARGE{\Black{$L$}}}
        \end{picture}
    \hspace{150pt}
        \begin{picture}(308,324) (207,-69)
            \SetWidth{2.8}
            \SetColor{Black}
            \Line[double,sep=5.5](210,234)(510,-66)
            \Line[double,sep=5.5](210,-66)(510,234)
            \SetWidth{1}
            \Line[double,sep=3.5](360,84)(360,24)
            \SetWidth{1.0}
            \GOval(360,84)(20,20)(0){0.882}
            \GluonArc(478,84)(240,155.3757,204.6243){7.5}{13}
            \GluonArc(240,84)(240,-24.6243,24.6243){7.5}{13}
            \GluonArc[clock](360,25.5)(133.5,154.011,25.989){7.5}{20}
            \Text(226,230)[lb]{\LARGE{\Black{$I$}}}
            \Text(485,230)[lb]{\LARGE{\Black{$J$}}}
            \Text(230,-66)[lb]{\LARGE{\Black{$K$}}}
            \Text(475,-66)[lb]{\LARGE{\Black{$L$}}}
            \Vertex(240,84){8}
            \Vertex(480,84){8}
            \Vertex(260,-16){8}
            \Vertex(260,184){8}
            \Vertex(460,184){8}
            \Vertex(460,-16){8}
        \end{picture}
    }
    \caption{Fully connected webs contributing to the quadrupole colour structure. The web on the left is $w_{4g}$, while the web $w_{(3g)^2(IK;JL)}$ that multiplies the colour structure~$\mathbf{Q}_{(IK;JL)}=f_{ace}f_{bde}\mathbf{T}_{I}^{a}\mathbf{T}_{J}^{b}\mathbf{T}_{K}^{c}\mathbf{T}_{L}^{d}$ is on the right. When the colour structure is irrelevant, or clear from context, we will refer to the diagram on the right (or an equivalent permutation of its external lines) as $w_{(3g)^2}$. These webs depend on all six cusp angles $(\gamma_{IJ}, \gamma_{IK}, \gamma_{IL}, \gamma_{JK}, \gamma_{JL}, \gamma_{KL})$.}
    \label{fig:FCwebs4lines}
\end{figure}
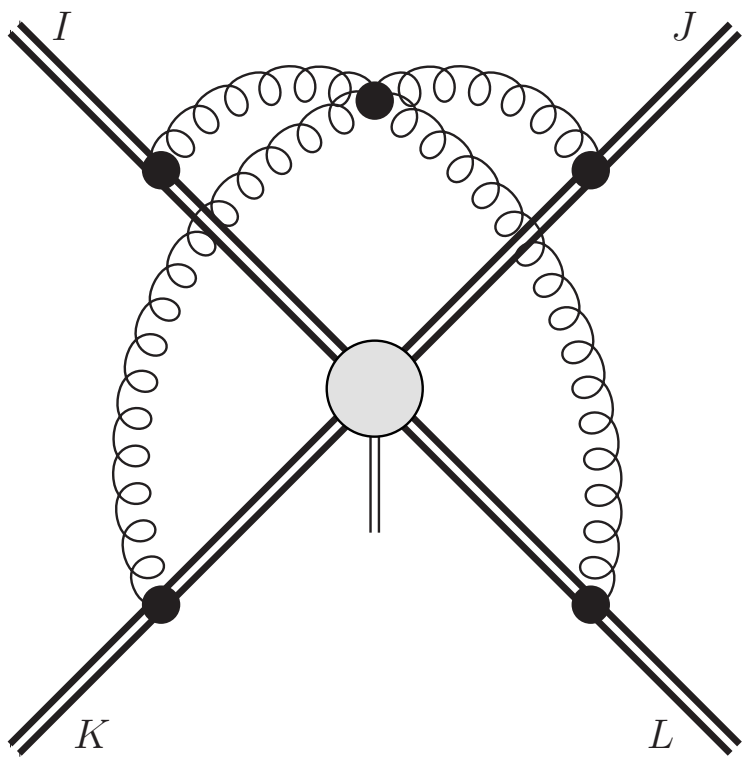
\begin{figure}
    \centering
    \scalebox{0.43}{
        \begin{picture}(315,306) (174,-87)
            \SetWidth{2.8}
            \SetColor{Black}
            \Line[double,sep=5.5](180,216)(480,-84)
            \Line[double,sep=5.5](180,-84)(480,216)
            \SetWidth{1}
            \Line[double,sep=3.5](330,66)(330,6)
            \SetWidth{1.0}
            \GluonArc[clock](225,66)(255,28.072,-28.072){7.5}{15}
            \GluonArc(463.167,66)(280.167,154.639,205.361){7.5}{15}
            \GluonArc(375,221)(249.098,-141.52,-79.592){7.5}{15}
            \Vertex(180,66){8}
            \Vertex(208,186){8}
            \Vertex(208,-55){8}
            \Vertex(450,186){8}
            \Vertex(422,-27){8}
            \Vertex(450,-55){8}
            \GOval(330,66)(20,20)(0){0.882}
            \Text(198,209)[lb]{\LARGE{\Black{$I$}}}
            \Text(450,209)[lb]{\LARGE{\Black{$J$}}}
            \Text(198,-85)[lb]{\LARGE{\Black{$K$}}}
            \Text(450,-90)[lb]{\LARGE{\Black{$L$}}}
        \end{picture}}
    \caption{A diagram which is part of the non-fully-connected web $w_{1112}(s_{IK}, s_{IL}, s_{KL}, s_{JL})$ that connects four lines at three loops. Permuting the Wilson lines creates equivalent webs that contribute to different colour structures.}
    \label{fig: NFC4lines}
\end{figure}
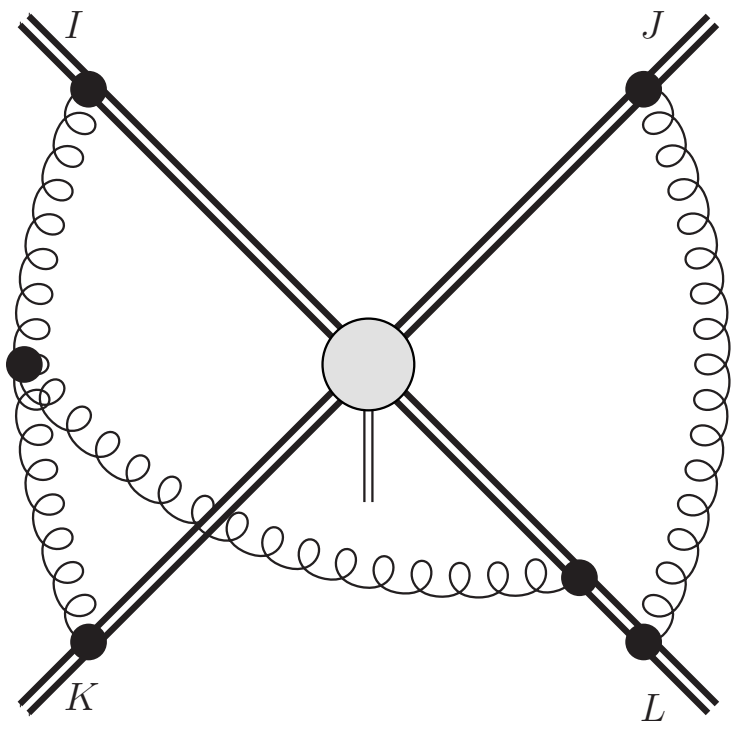
\begin{figure}
    \centering
    \scalebox{0.45}{
        \begin{picture}(306,311) (177,-72)
            \SetWidth{2.8}
            \SetColor{Black}
            \Line[double,sep=5.5](180,231)(480,-69)
            \Line[double,sep=5.5](180,-69)(480,231)
            \SetWidth{1}
            \Line[double,sep=3.5](330,81)(330,21)
            \SetWidth{1.0}
            \GOval(330,81)(20,20)(0){0.882}
            \GluonArc[clock](331.5,52.526)(177.785,128.841,51.159){7.5}{13}
            \GluonArc(332.392,58.031)(115.543,53.574,216.071){7.5}{17}
            \GluonArc[clock](300.512,60.899)(161.804,42.398,-32.484){7.5}{11}
            \Text(200,220)[lb]{\LARGE{\Black{$I$}}}
            \Text(450,220)[lb]{\LARGE{\Black{$J$}}}
            \Text(200,-66)[lb]{\LARGE{\Black{$K$}}}
            \Text(440,-66)[lb]{\LARGE{\Black{$L$}}}
            \Vertex(219,193){8}
            \Vertex(237,-13){8}
            \Vertex(442,194){8}
            \Vertex(422,172){8}
            \Vertex(402,152){8}
            \Vertex(438,-27){8}
        \end{picture}
    \hspace{150pt}
        \begin{picture}(306,306) (197,-77)
            \SetWidth{2.8}
            \SetColor{Black}
            \Line[double,sep=5.5](500,-74)(200,226)
            \Line[double,sep=5.5](200,-74)(500,226)
            \SetWidth{1.0}
            \Line[double,sep=3.5](350,76)(350,16)
            \GluonArc(340.526,113.895)(123.712,144.349,240.709){7.5}{11}
            \GluonArc(261.519,294.165)(328.87,-93.752,-61.191){7.5}{10}
            \GluonArc(158.696,73.739)(321.538,-19.388,20.435){7.5}{11}
            \GOval(350,77)(20,20)(0){0.882}
            \Text(220,220)[lb]{\LARGE{\Black{$I$}}}
            \Text(470,220)[lb]{\LARGE{\Black{$J$}}}
            \Text(226,-66)[lb]{\LARGE{\Black{$K$}}}
            \Text(460,-70)[lb]{\LARGE{\Black{$L$}}}
            \Vertex(240,187){8}
            \Vertex(275,2){8}
            \Vertex(240,-33){8}
            \Vertex(458,185){8}
            \Vertex(459,-32){8}
            \Vertex(422,5){8}
        \end{picture}
    }
    \caption{Representative diagrams contributing to two multiple gluon exchange webs that connect four external legs at three loops~\cite{falcioni_multiple_2014}. The one on the left is part of $w_{1113}(s_{IJ}, s_{JK}, s_{JL})$, while that on the right is part of $w_{1221}(s_{IK}, s_{JL}, s_{KL})$. Permuting the Wilson lines creates equivalent webs that contribute to different colour structures.}
    \label{fig:MGEWs}
\end{figure}
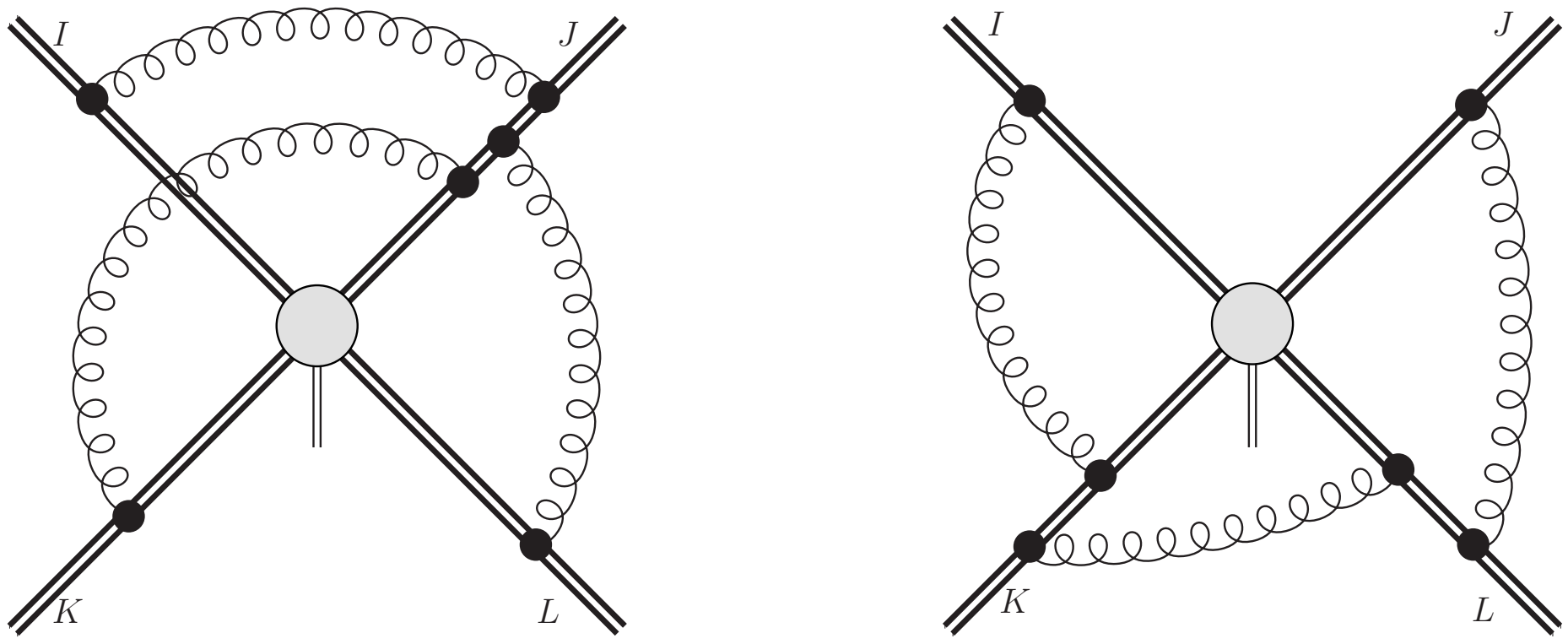
\begin{figure}
    \centering
    \scalebox{0.45}{
        \begin{picture}(363,246) (288,-267)
            \SetWidth{2.8}
            \SetColor{Black}
            \Line[double,sep=5.5](300,-144)(480,-144)
            \Line[double,sep=5.5](480,-144)(630,-24)
            \Line[double,sep=5.5](480,-144)(630,-264)
            \SetWidth{1.0}
            \Line[double,sep=3](480,-144)(480,-204)
            \Vertex(480,-84){8}
            \Vertex(450,-204){8}
            \Vertex(354,-144){8}
            \Vertex(327,-144){8}
            \Vertex(585,-58){8}
            \Vertex(570,-215){8}
            \GluonArc[clock](536.007,-94.705)(57.021,169.179,36.235){7.5}{7}
            \GluonArc(485,-149)(65.192,94.399,237.529){7.5}{9}
            \GluonArc(516.933,-145.134)(89.136,-138.669,-52.658){7.5}{7}
            \GluonArc(420,-151.5)(90.312,48.366,175.236){7.5}{11}
            \GluonArc[clock](420,-151.5)(60.467,-60.255,-187.125){7.5}{8}
            \GOval(480,-144)(20,20)(0){0.882}
            \Text(295,-170)[lb]{\LARGE{\Black{$I$}}}
            \Text(616,-60)[lb]{\LARGE{\Black{$J$}}}
            \Text(620,-240)[lb]{\LARGE{\Black{$L$}}}
        \end{picture}
    }
    \caption{A representative  diagram that is part of the fully-connected $w_{211}(s_{IL}, s_{IJ}, s_{JL})$ web. It connects just three Wilson lines, but that contributes to quadrupole functions through colour conservation. 
    }
    \label{fig:wless4legs}
\end{figure}
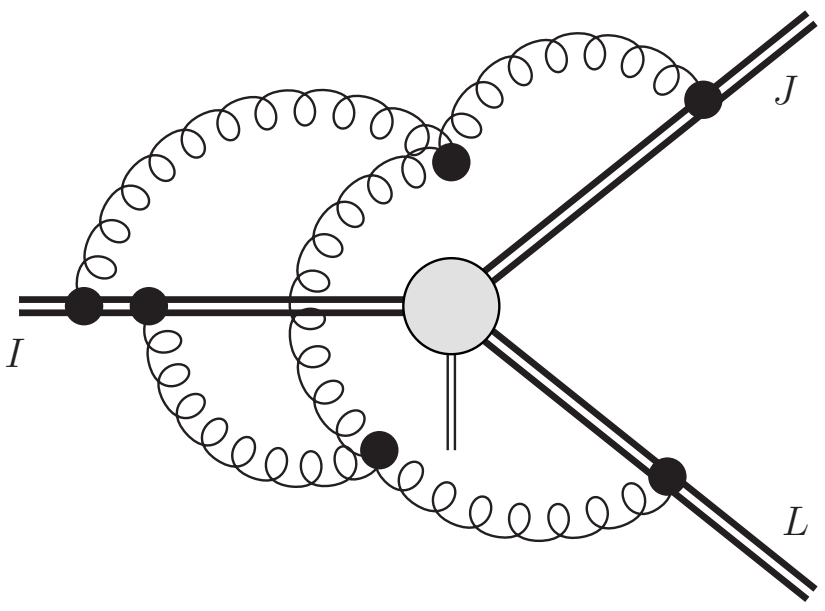

\subsection{Cutting Wilson lines}
In Section~\ref{sec: cutsanddiscsW}, we discussed the cuts and discontinuities of Feynman integrals, without specialising to correlators of Wilson lines (or webs). Cuts and discontinuities generalise straightforwardly to integrals involving Wilson lines, as discussed in Refs.~\cite{andries,Laenen:2015jia}.  There is, however, one important difference -- since Wilson lines are represented by eikonal propagators, which are linear rather than quadratic in the loop momenta, the definition of positive and negative energy components differs from that for standard quadratic propagators. 
For the cut analysis in this subsection, we take $m$
to be real and positive, with the long-distance damping prescription
inherited from Eq.~\eqref{Regularised_Wilson_line}.

For an incoming Wilson line that has emitted $n$ soft gluons with momenta $\{q_j\}$, the configuration-space propagator is given by 
\begin{equation}
    E^\mu(x)=\theta(x^0)\int\frac{d^4k}{(2\pi)^4}e^{-ik\cdot x}E^{\mu,(+)}(k)+\theta(-x^0)\int\frac{d^4k}{(2\pi)^4}e^{ik\cdot x}E^{\mu,(-)}(k),
\end{equation}
where the positive- and negative-energy components in momentum space are given by~\cite{andries, Gardi:2025ule} 
\begin{equation}
    E^{\mu, (\pm)}(k)=i\beta^\mu\,\theta(\pm\beta^0)\,(2\pi i)\delta\bigg(-\beta\cdot\sum_{j=1}^nq_j-nm\bigg), \qquad\quad \beta^\mu=\frac{k^\mu}{\sqrt{k^2}}\,.
\end{equation} 
We see that an eikonal propagator has only one energy component -- either positive or negative -- depending on the sign of the energy of the on-shell particle it represents. This follows directly from the linear dependence of the eikonal propagator on the virtual momentum, which implies that it has only a single pole in the complex \(k_0\)-plane, and hence that at most one residue can be picked up when deforming the integration contour. Consequently, although the standard prescriptions for defining cut integrals still apply, additional cuts can be expected to vanish when the Heaviside functions arising from the cuts are incompatible with the fixed energy flow through the eikonal lines~\cite{andries}. While the constraints derived in this paper do not rely on any special property of webs beyond their topology, the single energy component of eikonal propagators makes the soft anomalous dimension a natural object to investigate and compute using unitarity cuts~\cite{andries}.

In fact, the restriction on energy flow through eikonal lines implies that cutting an eikonal line (which we assume to be incoming, without loss of generality), along with all the
gluons already emitted or absorbed by it, always gives a vanishing
contribution, as shown in \cite{andries} and briefly reviewed here. To derive this result,
we need only keep track of the delta functions and Heaviside
functions that arise from cutting the parton and soft gluon lines,
\begin{subequations}
    \begin{align}
        \frac{1}{k^2+i\varepsilon}
        &\mapsto \delta(k^2)\theta(k^{(0)}),
        \label{eq: k-cut-contraints}\\
        \frac{i\beta^\mu}{-\beta\cdot k-m+i\varepsilon}
        &\mapsto \delta(-\beta\cdot k-m).
        \label{eq: wilson-cut-contraints}
    \end{align}
\end{subequations}
(Since the energy flow through the Wilson line is already fixed
by the on-shell particle it represents, we do not need to assign an
additional Heaviside function to the cut eikonal propagator.)
We now introduce the parametrisation
\begin{subequations}
    \begin{align}
        k&=(k^{(0)},k^{(1)}\mathbf{v}),
        \qquad \mathbf{v}^2=1,\\
        \beta_i&=(\cosh\phi,\sinh\phi\,\mathbf{v}'),
        \qquad \mathbf{v}'^2=1,
    \end{align}
    \vspace{-1.5\baselineskip}
    \begin{equation}
        \mathbf{v}\cdot\mathbf{v}'=\cos\sigma.
    \end{equation}
\end{subequations}
The on-shell condition and positive-energy constraint in
Eq.~\eqref{eq: k-cut-contraints} enforce
\begin{equation}
    k^{(0)}=k^{(1)}>0.
\end{equation}
Substituting this into the argument of the delta function
in Eq.~\eqref{eq: wilson-cut-contraints}, we find
\begin{equation}
    -\beta_i\cdot k-m
    =-k^{(0)}(\cosh\phi-\sinh\phi\cos\sigma)-m<0,
\end{equation}
where the inequality follows from the positivity of the
factor in parentheses, assuming $m\geq0$. Thus, the
delta function associated with the cut eikonal propagator
has no support, since its argument can never vanish.
The constraints imposed by the cut gluon and eikonal
propagators are therefore incompatible, and the
corresponding cut vanishes. 

In the next section, we apply this 
unitarity cut 
framework
to pairs of non-overlapping two-particle channels of
the webs. We use the web topologies to identify pairs
of channels that admit no compatible cuts, thereby
deriving constraints on the sequential discontinuities
of the quadrupole functions.


\section{New constraints on the three-loop soft anomalous dimension}
\label{sec: new_constr}
In this section we derive new analytic constraints on the quadrupole contributions to the three-loop soft anomalous dimension. Focusing on pairs of non-overlapping two-particle channels, we use Lemma~\ref{lemma} to establish that the sequential discontinuity with respect to these channels vanishes for all contributing webs (depicted in Figs.~\ref{fig:FCwebs4lines}, \ref{fig: NFC4lines}, \ref{fig:MGEWs}, and \ref{fig:wless4legs}), and thus for the complete quadrupole functions in Eq.~\eqref{eq:Gamma}. As these constraints are expected to hold for any assignment of external parton masses, we modify our notation in this section by allowing uppercase indices to denote either massive or massless partons. Thus, $\mathcal{F}^{(3)}_{(IK;JL)}$ should be understood below as representing the generic quadrupole function associated with the colour structure $\mathbf{Q}_{(IK;JL)}$ for any assignment of external partons (massive or massless) at three loops.

Note that, in the case of webs, the effective central vertex is a generic hard subdiagram that may involve additional external particles. This means that, in the analysis below, we do not impose either momentum conservation or colour conservation among the four coloured Wilson lines. Thus, the two-particle invariants formed by non-overlapping pairs of the four connected Wilson lines are truly independent. We include a fifth Wilson line at the central vertex of all web diagrams to remind us of this fact.

\subsection{Cuts of fully connected webs}
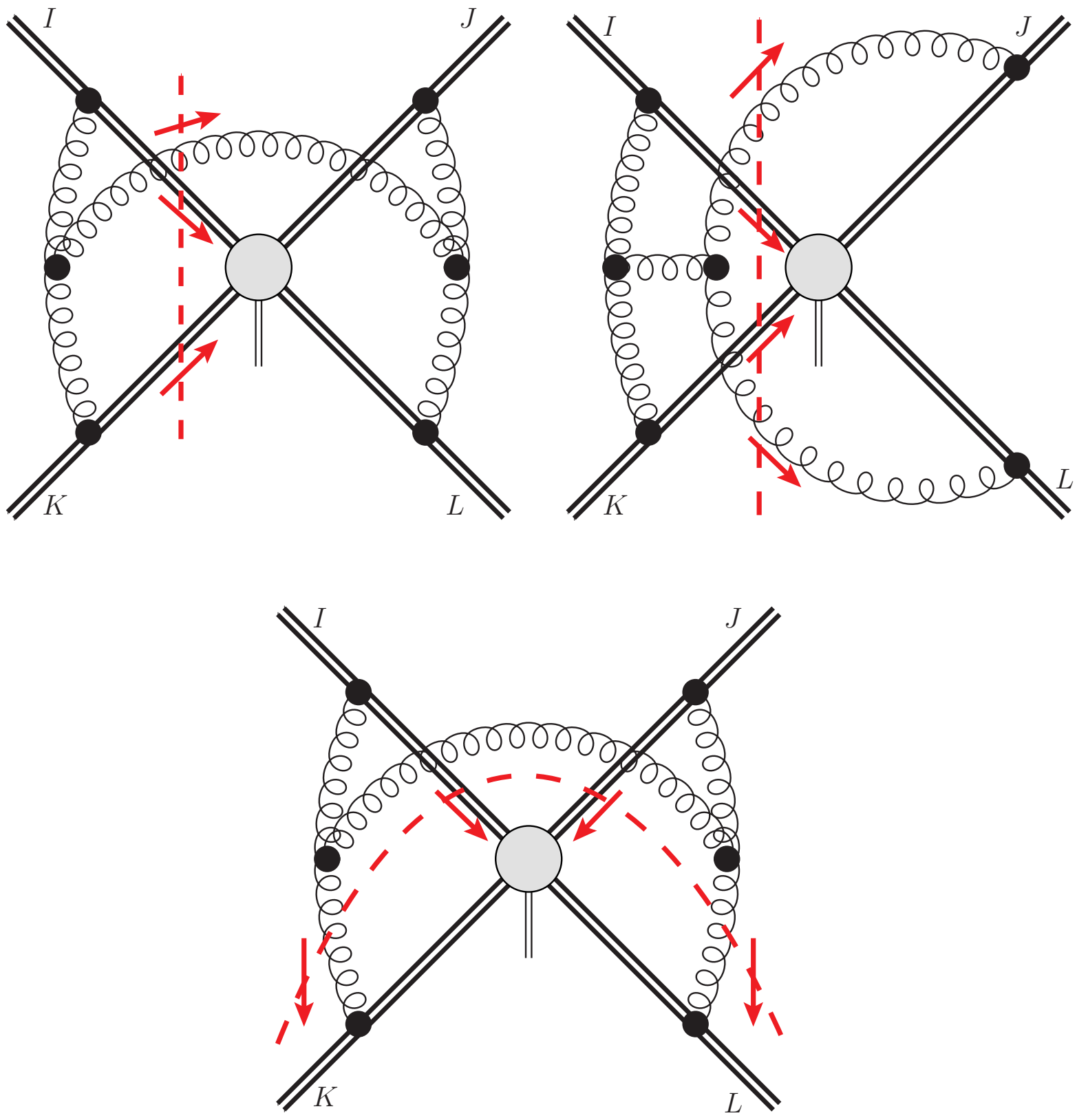
\begin{figure}
    \centering
    \scalebox{0.45}{
        \begin{picture}(308,324) (207,-69)
            \SetWidth{2.8}
            \SetColor{Black}
            \Line[double,sep=5.5](210,234)(510,-66)
            \Line[double,sep=5.5](210,-66)(510,234)
            \SetWidth{1.0}
            \Line[double,sep=3.5](360,84)(360,24)
            \GluonArc(478,84)(240,155.3757,204.6243){7.5}{13}
            \GluonArc(240,84)(240,-24.6243,24.6243){7.5}{13}
            \GluonArc[clock](360,25.5)(133.5,154.011,25.989){7.5}{20}
            \Text(230,230)[lb]{\LARGE{\Black{$I$}}}
            \Text(483,230)[lb]{\LARGE{\Black{$J$}}}
            \Text(230,-66)[lb]{\LARGE{\Black{$K$}}}
            \Text(475,-66)[lb]{\LARGE{\Black{$L$}}}
            \Vertex(238,84){8}
            \Vertex(480,84){8}
            \Vertex(257,185){8}
            \Vertex(461,-16){8}
            \Vertex(461,185){8}
            \Vertex(257,-16){8}
            \SetWidth{3.0}
            \SetColor{Red}
            \Line[dash,dashsize=12](313,200)(313,-20)
            \SetWidth{1.0}
            \GOval(360,84)(20,20)(0){0.882}
            \SetWidth{3.0}
            \SetColor{Red}
            \Line[arrow,arrowpos=1,arrowlength=12.5,arrowwidth=5,arrowinset=0.2](297,165)(330,175)
            \Line[arrow,arrowpos=1,arrowlength=12.5,arrowwidth=5,arrowinset=0.2](301,7)(330,35)
            \Line[arrow,arrowpos=1,arrowlength=12.5,arrowwidth=5,arrowinset=0.2](300,127)(327,103)
        \end{picture}
    \hspace{25pt}
        \begin{picture}(312,324) (207,-69)
            \SetWidth{2.8}
            \SetColor{Black}
            \Line[double,sep=5.5](210,234)(510,-66)
            \Line[double,sep=5.5](210,-66)(510,234)
            \Text(230,225)[lb]{\LARGE{\Black{$I$}}}
            \Text(480,225)[lb]{\LARGE{\Black{$J$}}}
            \Text(230,-66)[lb]{\LARGE{\Black{$K$}}}
            \Text(505,-50)[lb]{\LARGE{\Black{$L$}}}
            \SetWidth{1.0}
            \Vertex(237,84){8}
            \Vertex(298,84){8}
            \Vertex(257,185){8}
            \Vertex(257,-16){8}
            \Vertex(480,205){8}
            \Vertex(480,-36){8}
            \SetWidth{1}
            \Line[double,sep=3.5](360,84)(360,24)
            \SetWidth{1.0}
            \GOval(360,84)(20,20)(0){0.882}
            \SetWidth{3.0}
            \SetColor{Red}
            \Line[arrow,arrowpos=1,arrowlength=12.5,arrowwidth=5,arrowinset=0.2](317,27)(340,50)
            \Line[arrow,arrowpos=1,arrowlength=12.5,arrowwidth=5,arrowinset=0.2](312,119)(334,98)
            \SetWidth{1.0}
            \SetColor{Black}
            \Gluon(240,84)(300,84){7.5}{3}
            \GluonArc(478,84)(240,155.3757,204.6243){7.5}{15}
            \GluonArc[clock](420,98.409)(121.447,-175.083,-299.607){7.5}{15}
            \GluonArc(420,69)(120.934,172.875,299.745){7.5}{13}
            \SetWidth{3.0}
            \SetColor{Red}
            \Line[dash,dashsize=15](324,234)(324,-66)
            \Line[arrow,arrowpos=1,arrowlength=12.5,arrowwidth=5,arrowinset=0.2](307,187)(334,215)
            \Line[arrow,arrowpos=1,arrowlength=12.5,arrowwidth=5,arrowinset=0.2](318,-19)(344,-44)
        \end{picture}
    }\\
    \vspace{15pt}
    \scalebox{0.45}{
        \begin{picture}(324,324) (207,-69)
            \SetWidth{2.8}
            \SetColor{Black}
            \Line[double,sep=5.5](210,234)(510,-66)
            \Line[double,sep=5.5](210,-66)(510,234)
            \SetWidth{1.0}
            \Line[double,sep=3.5](360,84)(360,24)
            \GluonArc(478,84)(240,155.3757,204.6243){7.5}{13}
            \GluonArc(240,84)(240,-24.6243,24.6243){7.5}{13}
            \GluonArc[clock](360,25.5)(133.5,154.011,25.989){7.5}{20}
            \Text(230,225)[lb]{\LARGE{\Black{$I$}}}
            \Text(480,225)[lb]{\LARGE{\Black{$J$}}}
            \Text(230,-66)[lb]{\LARGE{\Black{$K$}}}
            \Text(480,-70)[lb]{\LARGE{\Black{$L$}}}
            \Vertex(238,84){8}
            \Vertex(480,84){8}
            \Vertex(257,185){8}
            \Vertex(461,-16){8}
            \Vertex(461,185){8}
            \Vertex(257,-16){8}
            \GOval(360,84)(20,20)(0){0.882}
            \SetWidth{3.0}
            \SetColor{Red}
            \Line[arrow,arrowpos=1,arrowlength=12.5,arrowwidth=5,arrowinset=0.2](416,125)(392,100)
            \Line[arrow,arrowpos=1,arrowlength=12.5,arrowwidth=5,arrowinset=0.2](304,125)(330,100)
            \Line[arrow,arrowpos=1,arrowlength=12.5,arrowwidth=5,arrowinset=0.2](496,36)(496,-10)
            \Bezier[dash,dsize=15](208,-28)(304,212)(432,164)(515,-28)
            \Line[arrow,arrowpos=1,arrowlength=12.5,arrowwidth=5,arrowinset=0.2](224,36)(224,-10)
        \end{picture}
    }
    \caption{Representative cuts for $w_{(3g)^2(IK;JL)}$ web: the two cuts at the top both correspond to the $s_{IK}$ channel, while the one at the bottom is the only cut on the $s_{IJ}$ channel. In the top-right diagram, we have chosen to draw the web $w_{(3g)^2(IK;JL)}$ in a different way, to make the connection with the TOPT picture more evident.}
    \label{fig: cut on the web}
\end{figure}

We start our analysis by considering the cuts of the $w_{(3g)^2}$ webs. For this diagram topology there are two types of two-particle channels -- those formed by legs that are connected to the same three-gluon vertex, and those formed by legs that are connected to opposite three-particle vertices. For concreteness, we consider the web $w_{(3g)^2(IK;JL)}$. In Fig.~\ref{fig: cut on the web}, we enumerate the complete set of cuts that exist for this web in the $s_{IK}$ and $s_{IJ}$ channels, recalling from Sec.~\ref{sec:unitarity_cuts} that the subgraphs on each side of the cut must remain connected in order for the cut to be valid. Thus, we see that a pair of cuts exist in the $s_{IK}$ channel, while just a single cut exists in the $s_{IJ}$ channel. By symmetry, there also exist two cuts in the $s_{JL}$ channel, and one cut in each of the $s_{IL}$, $s_{JK}$, and $s_{KL}$ channels. 

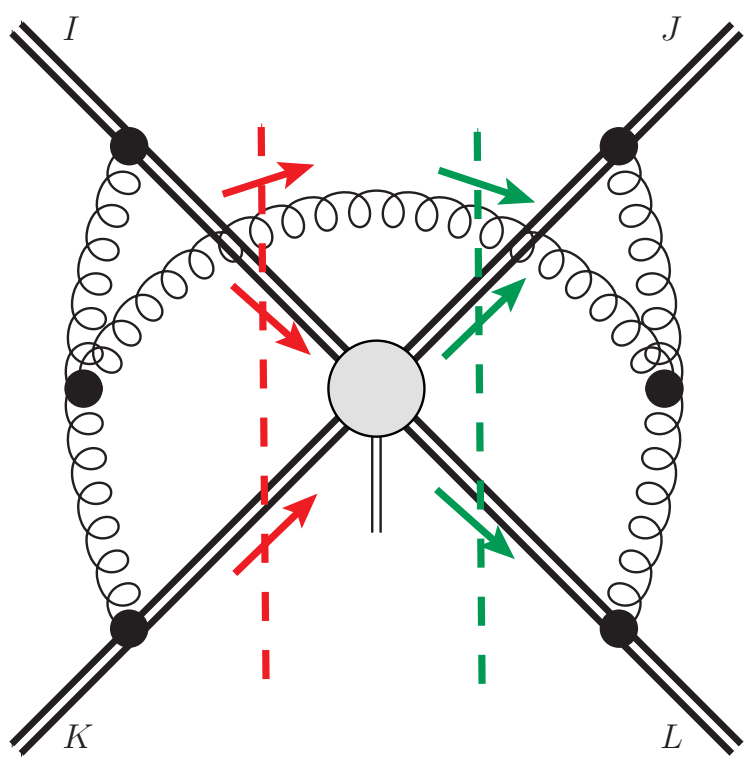
\begin{figure}
    \centering
    \scalebox{0.45}{
        \begin{picture}(308,324) (207,-69)
            \SetWidth{2.8}
            \SetColor{Black}
            \Line[double,sep=5.5](210,234)(510,-66)
            \Line[double,sep=5.5](210,-66)(510,234)
            \SetWidth{1.0}
            \Line[double,sep=3.5](360,84)(360,24)
            \GluonArc(478,84)(240,155.3757,204.6243){7.5}{13}
            \GluonArc(240,84)(240,-24.6243,24.6243){7.5}{13}
            \GluonArc[clock](360,25.5)(133.5,154.011,25.989){7.5}{20}
            \Text(230,230)[lb]{\Large{\Black{$I$}}}
            \Text(480,230)[lb]{\Large{\Black{$J$}}}
            \Text(230,-66)[lb]{\Large{\Black{$K$}}}
            \Text(480,-66)[lb]{\Large{\Black{$L$}}}
            \Vertex(238,84){8}
            \Vertex(480,84){8}
            \Vertex(257,185){8}
            \Vertex(461,-16){8}
            \Vertex(461,185){8}
            \Vertex(257,-16){8}
            \SetWidth{3.0}
            \SetColor{Red}
            \Line[dash,dashsize=12](312,193)(314,-37)
            \SetWidth{1.0}
            \GOval(360,84)(20,20)(0){0.882}
            \SetWidth{3.0}
            \SetColor{Red}
            \Line[arrow,arrowpos=1,arrowlength=12.5,arrowwidth=5,arrowinset=0.2](301,7)(330,35)
            \Line[arrow,arrowpos=1,arrowlength=12.5,arrowwidth=5,arrowinset=0.2](300,127)(327,103)
            \SetColor{ForestGreen}
            \Line[dash,dashsize=12](402,191)(404,-39)
            \Line[arrow,arrowpos=1,arrowlength=12.5,arrowwidth=5,arrowinset=0.2](388,97)(417,125)
            \Line[arrow,arrowpos=1,arrowlength=12.5,arrowwidth=5,arrowinset=0.2](385,42)(412,18)
            \SetColor{Red}
            \Line[arrow,arrowpos=1,arrowlength=12.5,arrowwidth=5,arrowinset=0.2](296,165)(327,175)
            \SetColor{ForestGreen}
            \Line[arrow,arrowpos=1,arrowlength=12.5,arrowwidth=5,arrowinset=0.2](386,175)(419,164)
        \end{picture}
    }
    \caption{Compatible double cut in the channels $s_{IK},s_{JL}$ for the web $w_{(3g)^2(IK;JL)}$. 
    The red and green dashed lines isolate the \(s_{IK}\) and \(s_{JL}\) channels, respectively.
     The fifth Wilson line represents the generic momentum (and colour) flow of the effective central vertex, which makes $s_{IK}$ and $s_{JL}$ are independent channels.}
    \label{fig: non-zero_double-cut}
\end{figure}

\begin{figure}
    \centering
    \scalebox{0.45}{
        \begin{picture}(397,412) (168,-82)
            \SetWidth{3.0}
            \SetColor{ForestGreen}
            \Bezier[dash,dsize=12](208,268)(320,60)(400,60)(512,268)
            \SetWidth{2.8}
            \SetColor{Black}
            \Line[double,sep=5.5](210,322)(510,22)
            \Line[double,sep=5.5](210,22)(510,322)
            \SetWidth{1.0}
            \Line[double,sep=3.5](360,172)(360,120)
            \GluonArc(478,172)(240,155.3757,204.6243){7.5}{13}
            \GluonArc(240,172)(240,-24.6243,24.6243){7.5}{13}
            \Text(225,320)[lb]{\LARGE{\Black{$I$}}}
            \Text(485,320)[lb]{\LARGE{\Black{$J$}}}
            \Text(225,14)[lb]{\LARGE{\Black{$K$}}}
            \Text(485,15)[lb]{\LARGE{\Black{$L$}}}
            \Vertex(238,172){8}
            \Vertex(480,172){8}
            \Vertex(257,273){8}
            \Vertex(461,72){8}
            \Vertex(461,273){8}
            \Vertex(257,72){8}
            \GOval(360,172)(20,20)(0){0.882}
            \SetWidth{3.0}
            \SetColor{Red}
            \Bezier[dash,dsize=12](203,83)(314,241)(387,281)(518,78)
            \Line[arrow,arrowpos=1,arrowlength=12.5,arrowwidth=5,arrowinset=0.2](225,90)(225,120)
            \Line[arrow,arrowpos=1,arrowlength=12.5,arrowwidth=5,arrowinset=0.2](495,90)(495,120)
            \Line[arrow,arrowpos=1,arrowlength=12.5,arrowwidth=5,arrowinset=0.2](372,200)(397,225)
            \Line[arrow,arrowpos=1,arrowlength=12.5,arrowwidth=5,arrowinset=0.2](347,200)(324,225)
            \Text(245,40)[lb]{\Large{\Black{$\{0,0\}$}}}
            \Text(440,40)[lb]{\Large{\Black{$\{0,0\}$}}}
            \Text(195,170)[lb]{\Large{\Black{$\{1,0\}$}}}
            \Text(495,170)[lb]{\Large{\Black{$\{1,0\}$}}}
            \Text(385,170)[lb]{\Large{\Black{$\{0,1\}$}}}
            \Text(245,295)[lb]{\Large{\Black{$\{1,1\}$}}}
            \Text(440,295)[lb]{\Large{\Black{$\{1,1\}$}}}
            \SetWidth{1.0}
            \SetColor{Black}
             \Bezier(480,172)(482.622,174.681)(488.854,173.763)(492.464,170.165) \Bezier(492.464,170.165)(499.684,162.968)(488.897,152.544)(484.214,157.256) \Bezier(484.214,157.256)(479.53,161.968)(490.019,172.691)(499.305,163.072) \Bezier(499.305,163.072)(508.592,153.452)(497.497,143.358)(492.957,148.199) \Bezier(492.957,148.199)(488.418,153.04)(499.205,163.463)(508.006,153.778) \Bezier(508.006,153.778)(516.807,144.093)(505.392,134.362)(501.011,139.336) \Bezier(501.011,139.336)(496.631,144.31)(507.726,154.404)(516.032,144.629) \Bezier(516.032,144.629)(524.339,134.854)(512.561,125.564)(508.372,130.672) \Bezier(508.372,130.672)(504.183,135.78)(515.598,145.511)(524.038,134.771) \Bezier(524.038,134.771)(532.479,124.03)(520.295,115.281)(516.338,120.537) \Bezier(516.338,120.537)(512.381,125.793)(524.158,135.082)(531.917,124.254) \Bezier(531.917,124.254)(539.675,113.426)(527.077,105.284)(523.38,110.708) \Bezier(523.38,110.708)(519.683,116.132)(531.867,124.882)(538.901,113.938) \Bezier(538.901,113.938)(545.935,102.995)(532.886,95.597)(529.503,101.182) \Bezier(529.503,101.182)(526.119,106.767)(538.717,114.908)(545.403,103.038) \Bezier(545.403,103.038)(552.089,91.167)(538.564,84.68)(535.559,90.432) \Bezier(535.559,90.432)(532.554,96.183)(545.603,103.581)(551.307,91.584) \Bezier(551.307,91.584)(557.011,79.588)(543.007,74.215)(540.458,80.13) \Bezier(540.458,80.13)(537.909,86.046)(551.434,92.533)(556.314,79.667) \Bezier(556.314,79.667)(561.194,66.802)(546.743,62.782)(544.741,68.846) \Bezier(544.741,68.846)(542.739,74.911)(556.744,80.284)(560.286,67.334) \Bezier(560.286,67.334)(563.827,54.384)(549.021,51.984)(547.665,58.162) \Bezier(547.665,58.162)(546.31,64.34)(560.761,68.36)(562.889,54.67) \Bezier(562.889,54.67)(565.018,40.981)(550.023,40.581)(549.44,46.786) \Bezier(549.44,46.786)(548.857,52.991)(563.664,55.391)(563.984,41.153) \Bezier(563.984,41.153)(564.303,26.915)(549.417,28.754)(549.713,34.918) \Bezier(549.713,34.918)(550.01,41.081)(565.005,41.482)(563.277,27.603) \Bezier(563.277,27.603)(561.549,13.724)(547.116,17.81)(548.336,23.845) \Bezier(548.336,23.845)(549.556,29.88)(564.443,28.042)(560.691,14.755) \Bezier(560.691,14.755)(556.94,1.468)(543.282,7.67)(545.404,13.461) \Bezier(545.404,13.461)(547.525,19.253)(561.958,15.166)(556.315,2.667) \Bezier(556.315,2.667)(550.673,-9.832)(538.034,-1.754)(540.992,3.694) \Bezier(540.992,3.694)(543.951,9.142)(557.609,2.94)(550.271,-8.625) \Bezier(550.271,-8.625)(542.933,-20.19)(531.455,-10.533)(535.155,-5.501) \Bezier(535.155,-5.501)(538.855,-0.47)(551.494,-8.547)(542.684,-19.096) \Bezier(542.684,-19.096)(533.875,-29.644)(523.61,-18.706)(527.941,-14.134) \Bezier(527.941,-14.134)(532.271,-9.562)(543.749,-19.22)(533.688,-28.728) \Bezier(533.688,-28.728)(523.628,-38.237)(514.562,-26.287)(519.412,-22.191) \Bezier(519.412,-22.191)(524.262,-18.094)(534.526,-29.032)(523.414,-37.519) \Bezier(523.414,-37.519)(512.301,-46.006)(504.379,-33.269)(509.647,-29.644) \Bezier(509.647,-29.644)(514.916,-26.018)(523.982,-37.968)(512.007,-45.479) \Bezier(512.007,-45.479)(500.031,-52.99)(493.128,-39.673)(498.736,-36.482) \Bezier(498.736,-36.482)(504.343,-33.292)(512.265,-46.029)(500.223,-52.318) \Bezier(500.223,-52.318)(488.18,-58.607)(482.18,-44.86)(488.044,-42.064) \Bezier(488.044,-42.064)(493.908,-39.269)(500.811,-52.586)(488.259,-58.101) \Bezier(488.259,-58.101)(475.707,-63.615)(470.505,-49.546)(476.565,-47.106) \Bezier(476.565,-47.106)(482.624,-44.666)(488.624,-58.414)(476.357,-62.978) \Bezier(476.357,-62.978)(464.09,-67.542)(459.595,-53.231)(465.804,-51.11) \Bezier(465.804,-51.11)(472.013,-48.989)(477.214,-63.058)(464.637,-67.021) \Bezier(464.637,-67.021)(452.06,-70.983)(448.235,-56.479)(454.551,-54.656) \Bezier(454.551,-54.656)(460.867,-52.832)(465.361,-67.143)(452.514,-70.541) \Bezier(452.514,-70.541)(439.667,-73.939)(436.479,-59.282)(442.881,-57.742) \Bezier(442.881,-57.742)(449.283,-56.203)(453.108,-70.707)(440.035,-73.558) \Bezier(440.035,-73.558)(426.962,-76.41)(424.385,-61.633)(430.855,-60.366) \Bezier(430.855,-60.366)(437.325,-59.098)(440.513,-73.755)(427.254,-76.074) \Bezier(427.254,-76.074)(413.996,-78.393)(412.007,-63.526)(418.53,-62.521) \Bezier(418.53,-62.521)(425.053,-61.516)(427.63,-76.293)(414.226,-78.091) \Bezier(414.226,-78.091)(400.821,-79.889)(399.404,-64.956)(405.966,-64.206) \Bezier(405.966,-64.206)(412.528,-63.456)(414.517,-78.324)(401.003,-79.609) \Bezier(401.003,-79.609)(387.489,-80.894)(386.634,-65.918)(393.223,-65.418) \Bezier(393.223,-65.418)(399.812,-64.917)(401.228,-79.85)(387.64,-80.628) \Bezier(387.64,-80.628)(374.053,-81.405)(373.752,-66.408)(380.358,-66.154) \Bezier(380.358,-66.154)(386.963,-65.899)(387.818,-80.875)(374.192,-81.148) \Bezier(374.192,-81.148)(360.566,-81.421)(360.819,-66.423)(367.43,-66.413) \Bezier(367.43,-66.413)(374.04,-66.403)(374.34,-81.4)(360.711,-81.169) \Bezier(360.711,-81.169)(347.083,-80.937)(347.893,-65.959)(354.498,-66.194) \Bezier(354.498,-66.194)(361.104,-66.428)(360.85,-81.426)(347.253,-80.688) \Bezier(347.253,-80.688)(333.657,-79.951)(335.031,-65.014)(341.621,-65.495) \Bezier(341.621,-65.495)(348.211,-65.977)(347.401,-80.955)(333.871,-79.706) \Bezier(333.871,-79.706)(320.342,-78.457)(322.295,-63.584)(328.858,-64.317) \Bezier(328.858,-64.317)(335.42,-65.05)(334.045,-79.987)(320.62,-78.219) \Bezier(320.62,-78.219)(307.194,-76.451)(309.743,-61.669)(316.266,-62.659) \Bezier(316.266,-62.659)(322.789,-63.649)(320.837,-78.522)(307.552,-76.225) \Bezier(307.552,-76.225)(294.268,-73.928)(297.436,-59.266)(303.906,-60.522) \Bezier(303.906,-60.522)(310.376,-61.778)(307.827,-76.56)(294.723,-73.72) \Bezier(294.723,-73.72)(281.62,-70.881)(285.434,-56.374)(291.835,-57.907) \Bezier(291.835,-57.907)(298.237,-59.439)(295.069,-74.101)(282.187,-70.703) \Bezier(282.187,-70.703)(269.305,-67.304)(273.8,-52.994)(280.114,-54.815) \Bezier(280.114,-54.815)(286.428,-56.635)(282.613,-71.142)(269.997,-67.166) \Bezier(269.997,-67.166)(257.381,-63.19)(262.596,-49.126)(268.801,-51.249) \Bezier(268.801,-51.249)(275.005,-53.373)(270.51,-67.683)(258.209,-63.106) \Bezier(258.209,-63.106)(245.908,-58.53)(251.885,-44.772)(257.955,-47.214) \Bezier(257.955,-47.214)(264.024,-49.655)(258.809,-63.72)(246.87,-58.499) \Bezier(246.87,-58.499)(234.931,-53.278)(241.769,-39.927)(247.653,-42.708) \Bezier(247.653,-42.708)(253.537,-45.49)(247.56,-59.248)(235.428,-52.991) \Bezier(235.428,-52.991)(223.296,-46.735)(231.102,-33.926)(236.746,-37.085) \Bezier(236.746,-37.085)(242.39,-40.245)(235.552,-53.596)(224.011,-46.506) \Bezier(224.011,-46.506)(212.469,-39.416)(221.359,-27.334)(226.688,-30.909) \Bezier(226.688,-30.909)(232.018,-34.484)(224.211,-47.293)(212.877,-38.881) \Bezier(212.877,-38.881)(201.542,-30.47)(211.622,-19.361)(216.542,-23.386) \Bezier(216.542,-23.386)(221.463,-27.41)(212.573,-39.492)(202.234,-30.05) \Bezier(202.234,-30.05)(191.894,-20.608)(203.178,-10.724)(207.6,-15.224) \Bezier(207.6,-15.224)(212.022,-19.725)(201.943,-30.834)(192.81,-20.338) \Bezier(192.81,-20.338)(183.677,-9.842)(196.118,-1.462)(199.937,-6.424) \Bezier(199.937,-6.424)(203.757,-11.386)(192.474,-21.27)(184.756,-9.739) \Bezier(184.756,-9.739)(177.038,1.791)(190.508,8.392)(193.621,3.007) \Bezier(193.621,3.007)(196.735,-2.378)(184.294,-10.758)(178.202,1.736) \Bezier(178.202,1.736)(172.111,14.23)(186.391,18.823)(188.707,13.082) \Bezier(188.707,13.082)(191.023,7.34)(177.554,0.739)(173.276,14.068) \Bezier(173.276,14.068)(168.999,27.396)(183.797,29.849)(185.253,23.842) \Bezier(185.253,23.842)(186.709,17.835)(172.429,13.241)(170.099,27.229) \Bezier(170.099,27.229)(167.769,41.218)(182.766,41.52)(183.336,35.354) \Bezier(183.336,35.354)(183.907,29.188)(169.109,26.735)(168.763,41.169) \Bezier(168.763,41.169)(168.417,55.604)(183.327,53.962)(183.047,47.719) \Bezier(183.047,47.719)(182.767,41.476)(167.77,41.174)(169.223,55.128) \Bezier(169.223,55.128)(170.677,69.082)(185.323,65.844)(184.291,59.593) \Bezier(184.291,59.593)(183.258,53.342)(168.348,54.984)(171.227,68.254) \Bezier(171.227,68.254)(174.106,81.524)(188.386,76.934)(186.716,70.764) \Bezier(186.716,70.764)(185.046,64.593)(170.4,67.832)(174.609,81.089) \Bezier(174.609,81.089)(178.818,94.347)(192.686,88.632)(190.471,82.58) \Bezier(190.471,82.58)(188.255,76.528)(173.975,81.117)(179.044,93.532) \Bezier(179.044,93.532)(184.113,105.947)(197.561,99.303)(194.885,93.389) \Bezier(194.885,93.389)(192.21,87.475)(178.341,93.19)(184.417,105.536) \Bezier(184.417,105.536)(190.493,117.882)(203.503,110.416)(200.437,104.668) \Bezier(200.437,104.668)(197.372,98.921)(183.924,105.565)(190.94,117.859) \Bezier(190.94,117.859)(197.956,130.153)(210.555,122.013)(207.147,116.419) \Bezier(207.147,116.419)(203.738,110.826)(190.728,118.292)(198.023,129.635) \Bezier(198.023,129.635)(205.319,140.979)(217.542,132.285)(213.848,126.838) \Bezier(213.848,126.838)(210.154,121.392)(197.555,129.532)(205.549,140.801) \Bezier(205.549,140.801)(213.543,152.071)(225.413,142.9)(221.481,137.598) \Bezier(221.481,137.598)(217.55,132.296)(205.327,140.99)(213.605,151.739) \Bezier(213.605,151.739)(221.882,162.487)(233.438,152.923)(229.301,147.751) \Bezier(229.301,147.751)(225.164,142.578)(213.294,151.749)(221.729,161.96) \Bezier(221.729,161.96)(230.164,172.171)(241.428,162.266)(237.121,157.217) \Bezier(237.121,157.217)(232.813,152.168)(221.257,161.732)(227.812,169.342) \Bezier(227.812,169.342)(231.09,173.147)(237.184,174.476)(240,172)
            \SetWidth{3.0}
            \SetColor{ForestGreen}
            \Line[arrow,arrowpos=1,arrowlength=12.5,arrowwidth=5,arrowinset=0.2](225,220)(225,250)
            \Line[arrow,arrowpos=1,arrowlength=12.5,arrowwidth=5,arrowinset=0.2](495,220)(495,250)
            \SetColor{ForestGreen}
            \Line[arrow,arrowpos=1,arrowlength=12.5,arrowwidth=5,arrowinset=0.2](310,105)(335,130)
            \Line[arrow,arrowpos=1,arrowlength=12.5,arrowwidth=5,arrowinset=0.2](405,105)(382,130)
        \end{picture}
    }
    \caption{The crossed cut in the channels $s_{IJ}$ and $s_{KL}$ on $w_{(3g)^2(IK;JL)}$. We have drawn the central gluon around the rest of the diagram to stress that it is not cut. For the channels $s_{IL}$ and $s_{JK}$, the situation is analogous.}
    \label{fig: crossed cut}
\end{figure}
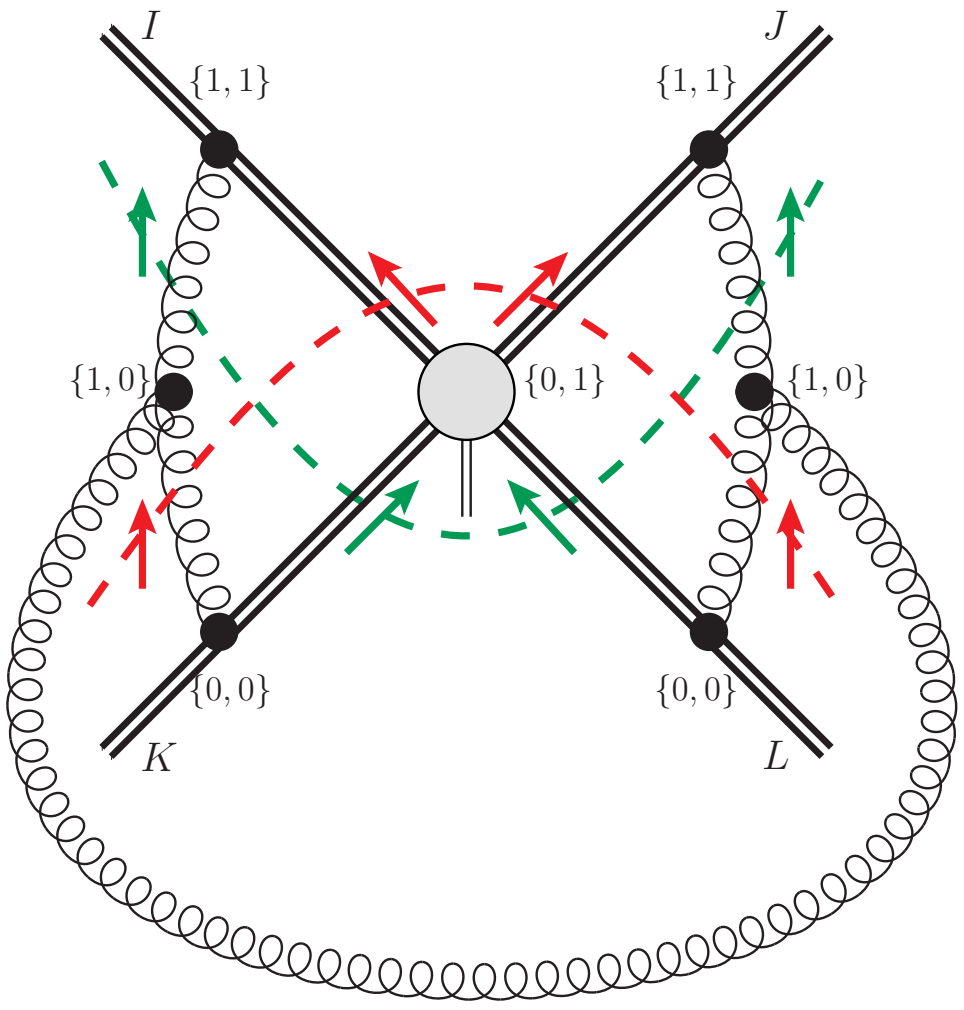

Now let us consider computing discontinuities with respect to non-overlapping pairs of these channels, beginning with $\text{Disc}_{s_{IK}, s_{JL}}w_{(3g)^2(IK;JL)}$. Recalling the rules \ref{cut1}--\ref{crossedcutsconditions} in Sec.~\ref{sec:iterateddiscsandcuts} for constructing the cuts associated with this sequential discontinuity, we find the compatible cut shown in Fig.~\ref{fig: non-zero_double-cut}. As such, we conclude that this double discontinuity can be nonzero. In contrast, for $\text{Disc}_{s_{IJ},s_{KL} }w_{(3g)^2(IK;JL)}$, we find that there is no compatible cut; overlaying the single cut associated with each channel gives rise to a crossed cut, as depicted in Fig.~\ref{fig: crossed cut}. By Lemma~\ref{lemma}, this double discontinuity must vanish. The same conclusion can be reached for $\text{Disc}_{s_{IL},s_{JK}}w_{(3g)^2(IK;JL)}$:
\begin{subequations}
\label{eq: discsw13}
    \begin{align}
        \text{Disc}_{s_{IJ},s_{KL}}w_{(3g)^2(IK;JL)}=0, \label{eq: discsw13a} \\
        \text{Disc}_{s_{IL},s_{JK}}w_{(3g)^2(IK;JL)}=0. \label{eq: discsw13b} 
    \end{align}
\end{subequations}
We emphasise that each channel entering these relations
separately admits a unitarity cut. The double discontinuities
vanish because the cuts in the two channels cannot be
combined into a compatible pair. 
By symmetry, one can deduce a similar pair of constraints on $w_{(3g)^2(IJ;KL)}$ and $w_{(3g)^2(IL;JK)}$:
\begin{subequations}
\label{eq: discsw12}
    \begin{align}
        \text{Disc}_{s_{IK},s_{JL}}w_{(3g)^2(IJ;KL)}=0,\\
        \text{Disc}_{s_{IL},s_{JK}}w_{(3g)^2(IJ;KL)}=0,
    \end{align}
\end{subequations}
and
\begin{subequations}
\label{eq: discsw14}
    \begin{align}    
        \text{Disc}_{s_{IJ},s_{KL}}w_{(3g)^2(IL;JK)}=0, \\ \text{Disc}_{s_{IK},s_{JL}}w_{(3g)^2(IL;JK)}=0.
    \end{align}
\end{subequations}
Thus, leveraging the connection between cuts in multiple channels and sequential discontinuities, we have been able to find two constraints on the sequential discontinuities of each of the three $w_{(3g)^2}$ webs.

Now consider the four-gluon-vertex web $w_{4g}$ shown on the left side of Fig.~\ref{fig:FCwebs4lines}. This is precisely the diagram topology that one gets by contracting the central gluon propagator in $w_{(3g)^2}$. Referring back to Fig~\ref{fig: cut on the web}, it is easy to convince oneself that only one of the cuts associated with each channel will still exist after this contraction, and consequently each two-particle channels is associated with a single cut. Again following the prescription in Sec.~\ref{sec:iterateddiscsandcuts}, we then find there are no compatible cuts in any pair of non-overlapping channels, allowing us to write
\begin{subequations}
\label{eq:disc4g}
    \begin{align}
        \text{Disc}_{s_{IJ},s_{KL}}w_{4g}=0, \label{eq:disc4ga}\\
        \text{Disc}_{s_{IK},s_{JL}}w_{4g}=0,\\
        \text{Disc}_{s_{IL},s_{JK}}w_{4g}=0, \label{eq:disc4gc}
    \end{align}
\end{subequations}
an even stronger result than for $w_{(3g)^2}$.

\subsection{Cuts of additional webs}

The remaining classes of webs that contribute to the quadrupole terms in the soft anomalous dimension are the non-fully-connected webs that correlate four partons shown in Fig.~\ref{fig: NFC4lines}, the multiple gluon exchange webs shown in Fig.~\ref{fig:MGEWs}, and the webs that only correlate two or three lines shown in Fig.~\ref{fig:wless4legs}. We now consider the cuts of each of these classes of webs in turn.

Let's start with the simplest class of webs, namely those that connect only two or three Wilson lines.
These webs cannot simultaneously depend on a pair of
two-particle Mandelstam invariants associated with
non-overlapping channels, since such dependence would
require the web's soft-gluon attachments to involve
all four corresponding Wilson lines.
This automatically allows us to conclude that
\begin{subequations}
    \begin{align}
        \text{Disc}_{s_{IJ},s_{KL}}w_{<4}=0,\\
        \text{Disc}_{s_{IK},s_{JL}}w_{<4}=0,\\
        \text{Disc}_{s_{IL},s_{JK}}w_{<4}=0,
    \end{align}
\end{subequations}
where $w_{<4}$ represents any web connecting fewer than 4 Wilson lines.

Next consider $w_{1112}(s_{IK}, s_{IL}, s_{KL}, s_{JL})$, shown in Fig.~\ref{fig: NFC4lines}, which contributes to $\mathbf{Q}_{(IK;JL)}$. In this diagram, we see that partons $I$ and $J$ are not connected by soft gluons, and similarly for partons $K$ and $J$. From this we can infer that the diagram does not depend on either $s_{IJ}$ or $s_{JK}$ \cite{Almelid,harley_multiparton_nodate}. Thus, similar to the case of webs connecting fewer than four lines, we can write 
\begin{subequations}
\begin{align}
    &\text{Disc}_{s_{IJ}}w_{1112}(s_{IK}, s_{IL}, s_{KL}, s_{JL})\,=0,\\
    &\text{Disc}_{s_{JK}}w_{1112}(s_{IK}, s_{IL}, s_{KL}, s_{JL})=0,
\end{align}   
\end{subequations}
where in these equations $w_{1112}(s_{IK}, s_{IL}, s_{KL}, s_{JL})$ should be understood as referring to the web that contributes to $\mathbf{Q}_{(IK;JL)}$. Thus, the same double discontinuities that vanished for the connected webs (in~\eqref{eq: discsw13a} and~\eqref{eq: discsw13b} as well as in~\eqref{eq:disc4ga}~and~\eqref{eq:disc4gc}) must also vanish for $w_{1112}(s_{IK}, s_{IL}, s_{KL}, s_{JL})$:
\begin{subequations}
\begin{align}
    \text{Disc}_{s_{IJ},s_{KL}}w_{1112}(s_{IK}, s_{IL}, s_{KL}, s_{JL})=0,\\
    \text{Disc}_{s_{IL},s_{JK}}w_{1112}(s_{IK}, s_{IL}, s_{KL}, s_{JL})=0.
\end{align}  
\end{subequations}
By symmetry, analogous relations hold for the other permutations of $w_{1112}$, which contribute to different quadrupole colour structures.

Finally, in the two webs in Fig.~\ref{fig:MGEWs}, we again notice that there are pairs of Wilson lines that are not connected by soft gluons. Repeating the same argument as above, $w_{1113}(s_{IJ}, s_{JK}, s_{JL})$ (which contributes to $\mathbf{Q}_{(IK;JL)}$ and $\mathbf{Q}_{(IJ;KL)}$) must have
\begin{subequations}
    \begin{align}
    &\text{Disc}_{s_{IK}}w_{1113}(s_{IJ}, s_{JK}, s_{JL})\,=0,\\
    &\text{Disc}_{s_{IL}}w_{1113}(s_{IJ}, s_{JK}, s_{JL})\,=0,\\
    &\text{Disc}_{s_{KL}}w_{1113}(s_{IJ}, s_{JK}, s_{JL})=0,
\end{align}
\end{subequations}
which similarly implies that all sequential discontinuities involving these channels vanish, namely
\begin{subequations}
    \begin{align}
    &\text{Disc}_{s_{IJ},s_{KL}}w_{1113}(s_{IJ}, s_{JK}, s_{JL})=0,\\
    &\text{Disc}_{s_{IK},s_{JL}}w_{1113}(s_{IJ}, s_{JK}, s_{JL})=0,\\
    &\text{Disc}_{s_{IL},s_{JK}}w_{1113}(s_{IJ}, s_{JK}, s_{JL})=0.
\end{align}
\end{subequations}
In the case of $w_{1221}(s_{IK}, s_{JL}, s_{KL})$ in Fig.~\ref{fig:MGEWs} (which contributes to $\mathbf{Q}_{(IK;JL)}$), we have that
\begin{subequations}
    \begin{align}
    &\text{Disc}_{s_{IL}}w_{1221}(s_{IK}, s_{JL}, s_{KL})=0,\\
    &\text{Disc}_{s_{IJ}}w_{1221}(s_{IK}, s_{JL}, s_{KL})\,=0,\\
    &\text{Disc}_{s_{JK}}w_{1221}(s_{IK}, s_{JL}, s_{KL})=0,
\end{align}
\end{subequations}
which readily implies that
\begin{subequations}
    \begin{align}
    &\text{Disc}_{s_{IJ},s_{KL}}w_{1221}(s_{IK}, s_{JL}, s_{KL})=0,\\
    &\text{Disc}_{s_{IL},s_{JK}}w_{1221}(s_{IK}, s_{JL}, s_{KL})=0.
\end{align}
\end{subequations}
As before, analogous relations hold for all permutations of $w_{1113}$ and $w_{1221}$. Together with the results of the previous subsection, this establishes that every three-loop web that contributes to $\mathbf{Q}_{(IK;JL)}$ has a vanishing double discontinuity with respect to $s_{IJ}$ and $s_{KL}$, and similarly with respect to $s_{IL}$ and $s_{JK}$. Similar results hold for the colour structures $\mathbf{Q}_{(IJ;KL)}$ and $\mathbf{Q}_{(IL;JK)}$.

\subsection{New constraints on the quadrupole functions at three loops}
\label{sec:SADlevel_constr}

While these web-level results provide useful  information about
the analytic structure of individual web functions, our
aim is to obtain constraints on the full soft anomalous
dimension that can be used in a bootstrap.
Indeed, many of the physical constraints and simplifications that have been observed or exploited in previous calculations hold only after summing over webs~\cite{Almelid:2015jia, Almelid:2017qju, Liu:2022elt, Gardi:2025lws}. Therefore, we now use the results that we have established for individual webs to formulate constraints on the full quadrupole functions that appear in the soft anomalous dimension. 
We stress that these constraints follow just from
considering diagram topologies, and therefore apply to any assignment of
massive or massless external partons.

In order to connect to the physical answer, we now briefly recall how the webs considered above combine to give the quadrupole functions. As already discussed, the three-loop soft anomalous dimension receives contributions from diagrams involving two, three, and four external partons~\cite{Almelid:2015jia, Almelid}. Webs that connect fewer than four partons can still contribute to colour structures that involve four different colour generator upon imposing colour conservation, as explained in Refs.~\cite{Almelid:2015jia,Almelid}. 
Let us schematically denote the sum of all contributions to the three-loop quadrupole colour structure $\mathbf{Q}_{(IJ;KL)}$ by
\begin{equation}
    \sum w^{(3)}\Big|_{(IJ;KL)}.
\end{equation}
For a given quadrupole colour structure, the sum of webs
need not coincide with the corresponding quadrupole function
introduced in Eq.~\eqref{eq:Gamma}; the difference can be
parametrised by a common function $\mathcal{J}$ as
\cite{Almelid,Almelid:2015jia,Maher:2023jqy,Gardi:2025lws}
\begin{subequations}
\label{eq:sumofwebs}
\begin{align}
\label{eq:sumofwebsIJ}
\sum w^{(3)}\Big|_{(IJ;KL)}
    &=\mathcal{F}^{(3)}_{(IJ;KL)}+\mathcal{J},\\
\label{eq:sumofwebsIK}
\sum w^{(3)}\Big|_{(IK;JL)}
    &=\mathcal{F}^{(3)}_{(IK;JL)}-\mathcal{J},\\
\label{eq:sumofwebsIL}
\sum w^{(3)}\Big|_{(IL;JK)}
    &=\mathcal{F}^{(3)}_{(IL;JK)}+\mathcal{J}.
\end{align}
\end{subequations}
To see why $\mathcal{J}$ leaves the soft anomalous dimension
unchanged, we rewrite one of the quadrupole terms in
Eq.~\eqref{eq:Gamma} as
\begin{equation}
\begin{split}
\frac{1}{8}\sum_{(I,J,K,L)}
\mathbf{Q}_{(IJ;KL)}\mathcal{F}_{(IJ;KL)}
={}&\sum_{1\le I<J<K<L\le M}\Big[
\mathbf{Q}_{(IJ;KL)}\mathcal{F}_{(IJ;KL)}
\\
&+\mathbf{Q}_{(IK;JL)}\mathcal{F}_{(IK;JL)}
+\mathbf{Q}_{(IL;JK)}\mathcal{F}_{(IL;JK)}
\Big],
\end{split}
\end{equation}
where the restriction $I<J<K<L$ on the right-hand side ensures that each set of four distinct partons appears once in the sum, while 
the three terms in the square brackets display the three possible pairings
of these partons.
The shifts $+\mathcal{J}$, $-\mathcal{J}$ and
$+\mathcal{J}$ in the three coefficient functions cancel
by the Jacobi identity, Eq.~\eqref{eq:Jacobi}.
Thus, different choices of $\mathcal{J}$ correspond
to different decompositions of the same gauge-invariant
soft anomalous dimension.

By inspecting Eqs.~\eqref{eq:sumofwebs}, we find that we can build combinations of webs that contribute to different colour structures and that are free of $\mathcal{J}$ \cite{Almelid:2015jia, Almelid:2017qju, Maher:2023jqy, Duhr:2025cye}:
\begin{subequations}
\label{eq:F123Defs}
    \begin{align}
        \Fboot_1&\equiv+\mathcal{F}^{(3)}_{(IJ;KL)}+\mathcal{F}^{(3)}_{(IK;JL)}=+\sum w^{(3)}\Big|_{(IJ;KL)}+\sum w^{(3)}\Big|_{(IK;JL)},\\
        \Fboot_2&\equiv+\mathcal{F}^{(3)}_{(IL;JK)}-\mathcal{F}^{(3)}_{(IJ;KL)}=+\sum w^{(3)}\Big|_{(IL;JK)}-\sum w^{(3)}\Big|_{(IJ;KL)},\label{eq:jacobi-free-F2-definition}\\
        \Fboot_3&\equiv-\mathcal{F}^{(3)}_{(IL;JK)}-\mathcal{F}^{(3)}_{(IK;JL)}=-\sum w^{(3)}\Big|_{(IL;JK)}-\sum w^{(3)}\Big|_{(IK;JL)}.
\end{align}
\end{subequations}
By construction, these functions satisfy
\begin{equation}
\label{eq:Fboot-sum}
    \Fboot_1+\Fboot_2+\Fboot_3=0.
\end{equation}
This relation mirrors the colour Jacobi identity,
Eq.~\eqref{eq:Jacobi}, and suggests a parallel with the
kinematic Jacobi identities employed in
colour--kinematics duality~\cite{Bern:2008qj,Bern:2010ue}. These Jacobi-free combinations may also provide an efficient
organisation of direct calculations. When the common
$\mathcal{J}$ contribution can be identified at integrand
level, forming these combinations eliminates it before
integration. Such cancellations occur between the
four-gluon-vertex web and the webs containing two
three-gluon vertices, where the common contribution
is more complicated than the surviving combinations~\cite{Almelid:2015jia,Almelid}.

Using Eq.~\eqref{eq:Fboot-sum}, the quadrupole contribution
can be expressed in terms of any two of these functions,
for example~\cite{Almelid:2015jia, Almelid:2017qju, Maher:2023jqy, Duhr:2025cye}:
\begin{equation}
    \begin{split}
        &\mathbf{Q}_{(IJ;KL)}\mathcal{F}^{(3)}_{(IJ;KL)}+\mathbf{Q}_{(IK;JL)}\mathcal{F}^{(3)}_{(IK;JL)}+\mathbf{Q}_{(IL;JK)}\mathcal{F}^{(3)}_{(IL;JK)}=\\
        &\hspace{70pt}=\frac{1}{2}\left[\mathbf{Q}_{(IJ;KL)}\left(\Fboot_1-\Fboot_2\right)-\left(\mathbf{Q}_{(IK;JL)}+\mathbf{Q}_{(IL;JK)}\right)\Fboot_3\right]=\\
        &\hspace{240pt}=\mathbf{Q}_{(IK;JL)}\Fboot_1+\mathbf{Q}_{(IL;JK)}\Fboot_2.
    \end{split}
\end{equation}
More generally, each of the functions $\Fboot_1$, $\Fboot_2$, and $\Fboot_3$ combines the web contributions from a different pair of quadrupole colour structures. Importantly, referring back to the previous subsections, we notice that the webs entering each pair always share a vanishing double discontinuity. Leveraging the linearity of the discontinuity operator, we therefore arrive at the main result of this paper, the double-discontinuity constraints:
\begin{subequations}
\label{eq:final_constraints}
    \begin{align}
    \label{eq: final_constraint14}
    \text{Disc}_{s_{IL},s_{JK}}\left(\mathcal{F}^{(3)}_{(IJ;KL)}+\mathcal{F}^{(3)}_{(IK;JL)}\right)&={\text{Disc}_{s_{IL},s_{JK}}\Fboot_1}=0,\\
    \label{eq: final_constraint13}
    \text{Disc}_{s_{IK},s_{JL}}\left(\mathcal{F}^{(3)}_{(IL;JK)}-\mathcal{F}^{(3)}_{(IJ;KL)}\right)&={\text{Disc}_{s_{IK},s_{JL}}\Fboot_2}=0,\\
    \label{eq: final_constraint12}
    \text{Disc}_{s_{IJ},s_{KL}}\left(\mathcal{F}^{(3)}_{(IL;JK)}+\mathcal{F}^{(3)}_{(IK;JL)}\right)&={-\text{Disc}_{s_{IJ},s_{KL}}\Fboot_3}=0.
\end{align}
\end{subequations}
Thus, the new constraints act directly on the
gauge-invariant, Jacobi-free functions $\Fboot_1$,
$\Fboot_2$ and $\Fboot_3$ defined in
Eq.~\eqref{eq:F123Defs}.
These equations severely constrain the analytic structure of the quadrupole functions that appear in the soft anomalous dimension at three loops, and since they follow purely from the topology of the contributing webs (rather than the explicit form of the resulting kinematic functions), they are expected to hold for any mass configuration. 

We test the constraints in Eq.~\eqref{eq:final_constraints}
in the massless and one-mass bootstrap computations
presented in Secs.~\ref{sec: 4LLbootstrap} and
\ref{sec: 1m3ll-bootstrap}, comparing the resulting
quadrupole functions with the known
results~\cite{Almelid:2015jia,Gardi:2025lws}.
These computations also demonstrate how strongly the
new constraints restrict the bootstrap ansatz, motivating
their future use in determining the unknown quadrupole
functions involving two or more massive partons.
The next section develops the kinematic formulation
needed to implement these constraints.


\section{Implementation of the double-discontinuity constraint}
\label{sec: implement}
In Section~\ref{sec: new_constr}, we formulated the double-discontinuity constraints~\eqref{eq:final_constraints} in terms of two-particle Mandelstam invariants. However, to actually apply these constraints to the soft anomalous dimension for a given pattern of massive and massless partons, we must first translate them into a rescaling-invariant language -- namely, into the variables on which the quadrupole functions actually depend. As shown in Table~\ref{tab:col_struct}, the number of such variables changes with the mass configuration; in the fully massless case, only two cross-ratios appear, while an additional variable can be formed every time one of the partons is made massive. In this section, we show how to implement the constraints from Eq.~\eqref{eq:final_constraints} in practice, by expressing them directly in these kinematic variables. Where the symbol alphabet is known, we also present the explicit combinations of symbol letters that these constraints force to vanish.

\subsection{The fully massive case}
Let's start by considering the fully massive case. Unlike the massless and one-mass configurations, in which the quadrupole functions evaluate to pure linear combinations of GPLs~\cite{Almelid:2015jia, Gardi:2025lws}, nontrivial rational prefactors are expected to appear when two or more massive partons are present. Indeed, this behaviour has been seen in three-loop massive webs~\cite{Gardi:2013saa, falcioni_multiple_2014, harley_multiparton_nodate, Almelid, Henn:2023pqn}, the angle-dependent cusp anomalous dimension~\cite{Grozin:2014hna, Grozin:2015kna, Henn:2019swt}, and the two-loop fully massive soft anomalous dimension~\cite{Ferroglia:2009ep, Ferroglia:2009ii}. Given this, we write the expected form of the fully massive quadrupole $\mathcal{F}^{(3)}$ at three loops as
\begin{equation}
    \mathcal{F}^{(3)}_{(IJ;KL)}\left(\{\alpha\}\right)
    =\sum_r \mathcal{R}_r\left(\{\alpha\}\right)
    \mathcal{P}^{(3),r}_{(IJ;KL)}\left(\{\alpha\}\right).
\end{equation}
Here the $\mathcal{R}_r$ are rational prefactors and the
$\mathcal{P}^{(3),r}_{(IJ;KL)}$ are pure transcendental
functions. Both may depend on all six $\alpha$ variables
defined in Eq.~\eqref{eq: alpha}.

We now determine how the discontinuity in $s_{AB}$ acts on the full quadrupole function
$\mathcal{F}^{(3)}_{(IJ;KL)}$. This is done by tracking how the value of $\alpha_{AB}$ changes as $s_{AB}$ approaches the threshold branch cut $s_{AB} > (m_A+m_B)^2$ from either above or below, holding all other invariants $\{\bar s\}$ fixed between their threshold and pseudo-threshold.\footnote{Keeping the invariants $\{\bar s\}$ fixed between their threshold and pseudo-threshold means that the corresponding $\{\bar\alpha\}$ variables sit on the unit circle (cfr. Fig.~\ref{fig:alpha-map}). We stress, however, that this is not the only possible choice. On the first Riemann sheet, the pseudo-threshold is not a branch point of the function and choosing to fix all $\{\bar s\}$ below the pseudo-threshold is an equivalently valid choice. In that case, the fixed $\{\bar\alpha\}$ variables sit between $0$ and $1$ on the real axis.} Using the definition in Eq.~\eqref{eq:kinematic_region}, we denote this kinematic region by $R_{s_{AB}}$. As described in more detail in Appendix~\ref{app:alpha-boundaries}, one can deduce from  Eqs.~\eqref{gamma_IJ} and \eqref{eq: alpha} that a consistent choice of branches results in the values  
\begin{subequations}
\label{eq:boundary-alpha}
    \begin{align}
    \alpha_{AB}^+ \equiv \alpha_{AB}\big|_{s_{AB}+i\varepsilon} &= -\tau + i\varepsilon,\\
    \alpha_{AB}^- \equiv \alpha_{AB}\big|_{s_{AB}-i\varepsilon}&=-\tau^{-1}+i\varepsilon,
\end{align}
\end{subequations}
where $0<\tau<1$. Recalling Eqs.~\eqref{eq:def_disc} and suppressing the colour and loop indices, we then have that the discontinuity with respect to $s_{AB}$ is given by
\begin{align}
    \text{Disc}_{s_{AB}}\mathcal{F}\left(\{\alpha(s)\}\right)
    \Big|_{R_{s_{\scalebox{0.7}{$\scriptscriptstyle AB$}}}}
    =\mathcal{F}\left(\alpha_{AB}^+,\{\bar{\alpha}\}\right)
    -\mathcal{F}\left(\alpha_{AB}^-,\{\bar{\alpha}\}\right),
\end{align}
where $\{\bar{\alpha}\}$ denotes all the $\alpha$ variables other than $\alpha_{AB}$. Using the inversion symmetry $\alpha_{AB} \leftrightarrow \alpha_{AB}^{-1}$ implied by the definition in Eq.~\eqref{eq: alpha}, the second term can then be rewritten as
\begin{align}
\begin{split}
    \mathcal{F}\left(\alpha^-_{AB},\{\bar{\alpha}\}\right)&=\mathcal{F}\left((\alpha^-_{AB})^{-1},\{\bar{\alpha}\}\right) \\ 
    &=\mathcal{F}\left(-\tau-i\varepsilon, \{\bar{\alpha}\}\right)
    \\ 
    &=\mathcal{F}\left((\alpha_{AB}^+)^*), \{\bar{\alpha}\}\right).
    \end{split}
\end{align}
This allows us to recast the discontinuity across the threshold cut as
\begin{align}
    \label{eq: s-to-alpha-single-disc}\text{Disc}_{s_{AB}}\mathcal{F}\left(\{\alpha(s)\}\right)
    \Big|_{R_{s_{\scalebox{0.7}{$\scriptscriptstyle AB$}}}}=\text{Disc}_{\alpha_{AB}}\mathcal{F}\left(\alpha_{AB},\{\bar{\alpha}\}\right)\Big|_{R_{\alpha_{\scalebox{0.7}{$\scriptscriptstyle AB$}}}},
\end{align}
where $R_{\alpha_{AB}}$ is the region in which $\alpha_{AB}<0$ and all other $\alpha$ variables satisfy $|\alpha|=1$. Again, we refer the reader to Appendix~\ref{sec:alpha_details} for the details of the mapping between the Mandelstam and $\alpha$ variables. The key point in the derivation above is that the inversion symmetry $\alpha_{AB}\leftrightarrow \alpha_{AB}^{-1}$ allows us to relate the values of \(\mathcal{F}(\{\alpha\})\) outside the unit circle to those inside it, flipping also the small imaginary part of $\alpha_{AB}$~\cite{Gardi:2013saa}.

We now extend this argument to the double discontinuities of $\mathcal{F}\left(\{\alpha\}\right)$. This involves evaluating four boundary values of the function $\mathcal{F}\left(\{\alpha\}\right)$ at different kinematic points, which we denote by 
\begin{align}
    \mathcal{F}_{\sigma_1,\sigma_2}
    =\mathcal{F}\left(
    \alpha_{AB}^{\sigma_1},
    \alpha_{CD}^{\sigma_2},
    \{\bar{\alpha}\}\right),
    \qquad \sigma_1,\sigma_2\in\{+,-\}.
\end{align}
Using the definition of sequential discontinuities from
Eq.~\eqref{eq: iterated_discs}, we have that
\begin{align}
    \label{eq:doubleboundary}
    \text{Disc}_{s_{AB},s_{CD}}
    \mathcal{F}\left(\{\alpha(s)\}\right)
    \Big|_{R_{s_{\scalebox{0.7}{$\scriptscriptstyle AB$}}},R_{s_{\scalebox{0.7}{$\scriptscriptstyle CD$}}}}
    =\mathcal{F}_{+,+}
    -\mathcal{F}_{-,+}
    -\mathcal{F}_{+,-}
    +\mathcal{F}_{-,-},
\end{align}
where we compute first and second discontinuities in the regions $R_{s_{AB}}$ and $R_{s_{CD}}$, respectively (for more discussion on this point, see Ref.~\cite{Abreu:2014cla}). Applying the inversion symmetry as above to each variable,
we can express the double discontinuity on the left of~\eqref{eq:doubleboundary} in
terms of the $\alpha$ variables as
\begin{align}
    \label{eq: s-to-alpha}
    \text{Disc}_{s_{AB},s_{CD}}
    \mathcal{F}\left(\{\alpha(s)\}\right)
    \Big|_{R_{s_{\scalebox{0.7}{$\scriptscriptstyle AB$}}},R_{s_{\scalebox{0.7}{$\scriptscriptstyle CD$}}}}
    =
    \text{Disc}_{\alpha_{AB},\alpha_{CD}}
    \mathcal{F}\left(\{\alpha\}\right)
    \Big|_{R_{\alpha_{\scalebox{0.7}{$\scriptscriptstyle AB$}}},R_{\alpha_{\scalebox{0.7}{$\scriptscriptstyle CD$}}}},
\end{align}
where the discontinuities with respect to $\alpha_{AB}$ and $\alpha_{CD}$ are computed in the regions $R_{\alpha_{AB}}$ and $R_{\alpha_{CD}}$, respectively.

Eq.~\eqref{eq: s-to-alpha} allows us to express the general
constraints from Eq.~\eqref{eq:final_constraints}
directly in terms of the $\alpha$ variables. For the fully-massive quadrupole functions, we obtain
\begin{subequations}
    \label{eq:alpha_discs}
    \begin{align}
    \label{eq: alpha_ddisc1}
    \text{Disc}_{\alpha_{IL},\alpha_{JK}}
    \underbrace{\left(\mathcal{F}^{(3)}_{(IJ;KL)}
    +\mathcal{F}^{(3)}_{(IK;JL)}\right)}_{{\Fboot_1}}
    \Big|_{R_{\alpha_{\scalebox{0.7}{$\scriptscriptstyle IL$}}},R_{\alpha_{\scalebox{0.7}{$\scriptscriptstyle JK$}}}}
    &=0,\\
    \label{eq: alpha_ddisc2}
    \text{Disc}_{\alpha_{IK},\alpha_{JL}}
    \underbrace{\left(\mathcal{F}^{(3)}_{(IL;JK)}
    -\mathcal{F}^{(3)}_{(IJ;KL)}\right)}_{{\Fboot_2}}
    \Big|_{R_{\alpha_{\scalebox{0.7}{$\scriptscriptstyle IK$}}},R_{\alpha_{\scalebox{0.7}{$\scriptscriptstyle JL$}}}}
    &=0,\\
    \label{eq: alpha_ddisc3}
    \text{Disc}_{\alpha_{IJ},\alpha_{KL}}
    \underbrace{\left(\mathcal{F}^{(3)}_{(IL;JK)}
    +\mathcal{F}^{(3)}_{(IK;JL)}\right)}_{{-\Fboot_3}}
    \Big|_{R_{\alpha_{\scalebox{0.7}{$\scriptscriptstyle IJ$}}},R_{\alpha_{\scalebox{0.7}{$\scriptscriptstyle KL$}}}}
    &=0.
    \end{align}
\end{subequations}
We stress once more that taking a single discontinuity
in an individual two-particle channel generally gives
a non-zero result. It is only the specified double
discontinuities of the appropriate gauge-invariant
combinations of quadrupole functions that we expect
to vanish.

\subsection{Sequential massless limits}
\label{sec:function-massless-limits}
We now consider the massless limits of Eq.~\eqref{eq:alpha_discs}, in order to formulate the constraint from Eq.~\eqref{eq:final_constraints} in terms of the kinematic variables appropriate to configurations with both massive and massless external legs. In addition to making new predictions, this will allow us to verify these constraints in the one-mass and massless cases, where the answer is known~\cite{Almelid:2015jia, Gardi:2025lws}.

Without loss of generality, we fix the order in which we take the masses to zero as $\{m_I, m_J, m_K, m_L\}$ (namely, $m_I$ first, followed by $m_J$, and so on). Since the cusp angles involving a given leg become ill-defined when its mass vanishes, we work in a parametrisation that utilises finite ratios which remain finite in the relevant massless limits:
\begin{equation}
    (\alpha_{IJ},\alpha_{IK},\alpha_{IL},
   \alpha_{JK},\alpha_{JL},\alpha_{KL})\longrightarrow(\alpha_{IL},\alpha_{JL},\alpha_{KL},
   \eta_{IJKL},\rho_{IJKL},\rho_{ILKJ}),
 \label{eq:fv-chart}
\end{equation}
where 
\begin{subequations}
\label{eqs:fv}
    \begin{align}
        \eta_{IJKL}&\equiv\frac{\gamma_{IJ}\gamma_{JK}}{\gamma_{IK}(\gamma_{JL})^{2}}=\frac{r_{IJL}r_{JKL}}{r_{IKL}},\label{eq:fv-eta}\\
        \rho_{IJKL}&\equiv\frac{\gamma_{IJ}\gamma_{KL}}{\gamma_{IK}\gamma_{JL}}=\frac{r_{IJL}}{r_{IKL}},\label{eq:fv-rho}\\
        \rho_{ILKJ}&\equiv\frac{\gamma_{IL}\gamma_{JK}}{\gamma_{IK}\gamma_{JL}}=\frac{r_{JKL}}{r_{IKL}} \label{eq:fv-rhot}.
    \end{align}
\end{subequations}
In the last equality of each line, we have highlighted that each of these new variables is related to the \(r\) variables defined in Eq.~\eqref{eq: r-ratios}, which provide a parametrisation of the one-mass kinematics used in Ref.~\cite{Gardi:2025lws}. In the fully massless limit, $\rho_{IJKL}$ and $\rho_{ILKJ}$ reduce to the cross ratios seen in Eq.~\eqref{eq: CICR}. As stated above, the advantage of this parametrisation is that each massless limit simply sets one additional variable to zero:
\begin{equation}
\begin{aligned}
 &(\alpha_{IL},\alpha_{JL},\alpha_{KL},\eta_{IJKL},\rho_{IJKL},\rho_{ILKJ})\\
 &\xrightarrow{m_I\to0}
 (0,\alpha_{JL},\alpha_{KL},\eta_{iJKL},\rho_{iJKL},\rho_{iLKJ})\\
 &\xrightarrow{m_J\to0}
 (0,0,\alpha_{KL},\eta_{ijKL},\rho_{ijKL},\rho_{iLKj})\\
 &\xrightarrow{m_K\to0}
 (0,0,0,\eta_{ijkL},\rho_{ijkL},\rho_{iLkj})\\
 &\xrightarrow{m_L\to0}
 (0,0,0,0,\rho_{ijkl},\rho_{ilkj}),
\end{aligned}
\label{eq:fv-sequence}
\end{equation}
where we change labels from uppercase to lowercase as the corresponding partons become massless.

\begin{figure}[tbp]
  \centering
  \resizebox{\textwidth}{!}{%
\begingroup%
\definecolor{ALPlus}{HTML}{2468A0}%
\definecolor{ALMinus}{HTML}{CF642F}%
\definecolor{ALAxis}{HTML}{7C858E}%
\definecolor{ALCircle}{HTML}{BAC3CC}%
\definecolor{ALDisk}{HTML}{F7F9FB}%
\def\ALSetRGB#1,#2,#3\relax{\SetColor{#1 #2 #3 setrgbcolor}}%
\def\ALSetColor#1{%
  \color{#1}%
  \convertcolorspec{named}{#1}{rgb}{\ALRGB}%
  \expandafter\ALSetRGB\ALRGB\relax}%
\def\ALDraw#1{\put(0,0){#1}\ignorespaces}%
\setlength{\unitlength}{1pt}%
\SetScale{0.99626400996}%
\SetOffset(0,0)\unskip%
\SetScaledOffset(0,0)\unskip%
\SetArrowStroke{0}%
\fontsize{15}{18}\selectfont%
\begin{picture}(760,410)%
%
\ALSetColor{black}%
\ALDraw{\Text(170,389)[c]{\fontsize{21}{24}\selectfont $\alpha_{AB}$ plane}}%
\ALDraw{\Text(590,389)[c]{\fontsize{21}{24}\selectfont $\chi_{AB}$ plane}}%
%
\ALSetColor{ALDisk}%
\ALDraw{\Vertex(170,225){112}}%
\ALSetColor{ALCircle}%
\SetWidth{1}%
\ALDraw{\Arc[dash,dashsize=3](170,225)(112,0,360)}%
\ALDraw{\Text(80,330)[c]{$|\alpha_{AB}|=1$}}%
%
\ALSetColor{ALAxis}%
\SetWidth{0.7}%
\ALDraw{\LongArrow[arrowwidth=4,arrowlength=6](35,225)(305,225)}%
\ALDraw{\LongArrow[arrowwidth=4,arrowlength=6](170,94)(170,354)}%
\ALDraw{\Text(330,240)[r]{$\operatorname{Re}\alpha_{AB}$}}%
\ALDraw{\Text(180,352)[l]{$\operatorname{Im}\alpha_{AB}$}}%
\ALDraw{\LongArrow[arrowwidth=4,arrowlength=6](455,225)(725,225)}%
\ALDraw{\LongArrow[arrowwidth=4,arrowlength=6](590,94)(590,354)}%
\ALDraw{\Text(750,240)[r]{$\operatorname{Re}\chi_{AB}$}}%
\ALDraw{\Text(600,352)[l]{$\operatorname{Im}\chi_{AB}$}}%
\ALDraw{\Line(58,222)(58,228)}%
\ALDraw{\Line(282,222)(282,228)}%
\ALDraw{\Text(49,210)[r]{$-1$}}%
\ALDraw{\Text(291,210)[l]{$1$}}%
\ALDraw{\Text(177,209)[l]{\fontsize{12}{14}\selectfont $0$}}%
\ALDraw{\Text(597,209)[l]{\fontsize{12}{14}\selectfont $0$}}%
%
\ALSetColor{ALPlus}%
\SetWidth{2.35}%
\ALDraw{\Arc[arrow,arrowpos=0.28,arrowwidth=6,arrowlength=8](170,225)(33.6,0,180)}%
\ALDraw{\Text(210,265)[c]{$\alpha_{AB}^+(t)$}}%
\ALSetColor{ALMinus}%
\SetWidth{2.35}%
\ALDraw{\Arc[clock,arrow,arrowpos=0.28,arrowwidth=6,arrowlength=8](170,225)(33.6,0,180)}%
\ALDraw{\Text(210,180)[c]{$\alpha_{AB}^-(t)$}}%
\ALSetColor{ALMinus}%
\SetWidth{2.35}%
\ALDraw{\Bezier(702,225)(702,249.79896)(690.20003,273.58222)(669.19596,291.11773)}%
\ALDraw{\Bezier(669.19596,291.11773)(648.19189,308.65324)(619.70425,318.50459)(590,318.50459)}%
\ALDraw{\Bezier(590,318.50459)(560.29575,318.50459)(531.80811,308.65324)(510.80404,291.11773)}%
\ALDraw{\Bezier(510.80404,291.11773)(489.79997,273.58222)(478,249.79896)(478,225)}%
\ALDraw{\ArrowLine[arrowpos=0.5,arrowwidth=6,arrowlength=8](666.38382,293.38493)(657.40328,299.67608)}%
\ALDraw{\Text(654,327)[c]{$\chi_{AB}^-(t)$}}%
\ALSetColor{ALPlus}%
\SetWidth{2.35}%
\ALDraw{\Bezier(702,225)(702,200.20104)(690.20003,176.41778)(669.19596,158.88227)}%
\ALDraw{\Bezier(669.19596,158.88227)(648.19189,141.34676)(619.70425,131.49541)(590,131.49541)}%
\ALDraw{\Bezier(590,131.49541)(560.29575,131.49541)(531.80811,141.34676)(510.80404,158.88227)}%
\ALDraw{\Bezier(510.80404,158.88227)(489.79997,176.41778)(478,200.20104)(478,225)}%
\ALDraw{\ArrowLine[arrowpos=0.5,arrowwidth=6,arrowlength=8](666.38382,156.61507)(657.40328,150.32392)}%
\ALDraw{\Text(654,123)[c]{$\chi_{AB}^+(t)$}}%
%
\ALSetColor{black}%
\ALDraw{\Vertex(203.6,225){2.8}}%
\ALDraw{\Vertex(136.4,225){2.8}}%
\ALDraw{\Vertex(702,225){2.8}}%
\ALDraw{\Vertex(478,225){2.8}}%
\ALDraw{\Text(214,209)[l]{$0.3$}}%
\ALDraw{\Text(126,209)[r]{$-0.3$}}%
\ALDraw{\Text(712,209)[l]{$1$}}%
\ALDraw{\Text(468,209)[r]{$-1$}}%
%
\ALSetColor{black}%
\ALDraw{\Text(170,61)[c]{\fontsize{17}{20}\selectfont
  $\alpha_{AB}^\pm(t)=0.3\,e^{\pm i\pi t}$}}%
\ALDraw{\Text(590,61)[c]{\fontsize{17}{20}\selectfont
  $\chi_{AB}^\pm(t)=\cos(\pi t)\mp i\,\frac{91}{109}\sin(\pi t)$}}%
\end{picture}%
\endgroup
  }
  \caption{The two boundary paths $\alpha_{AB}^{\pm}(t)$
  and their corresponding images $\chi_{AB}^{\pm}(t)$.
  The semicircular paths in the $\alpha_{AB}$ plane
  map to the two halves of an ellipse in the
  $\chi_{AB}$ plane. Here $\alpha_{AB}(0)$ is chosen to be $0.3$.}
  \label{fig:lambda-map}
\end{figure}
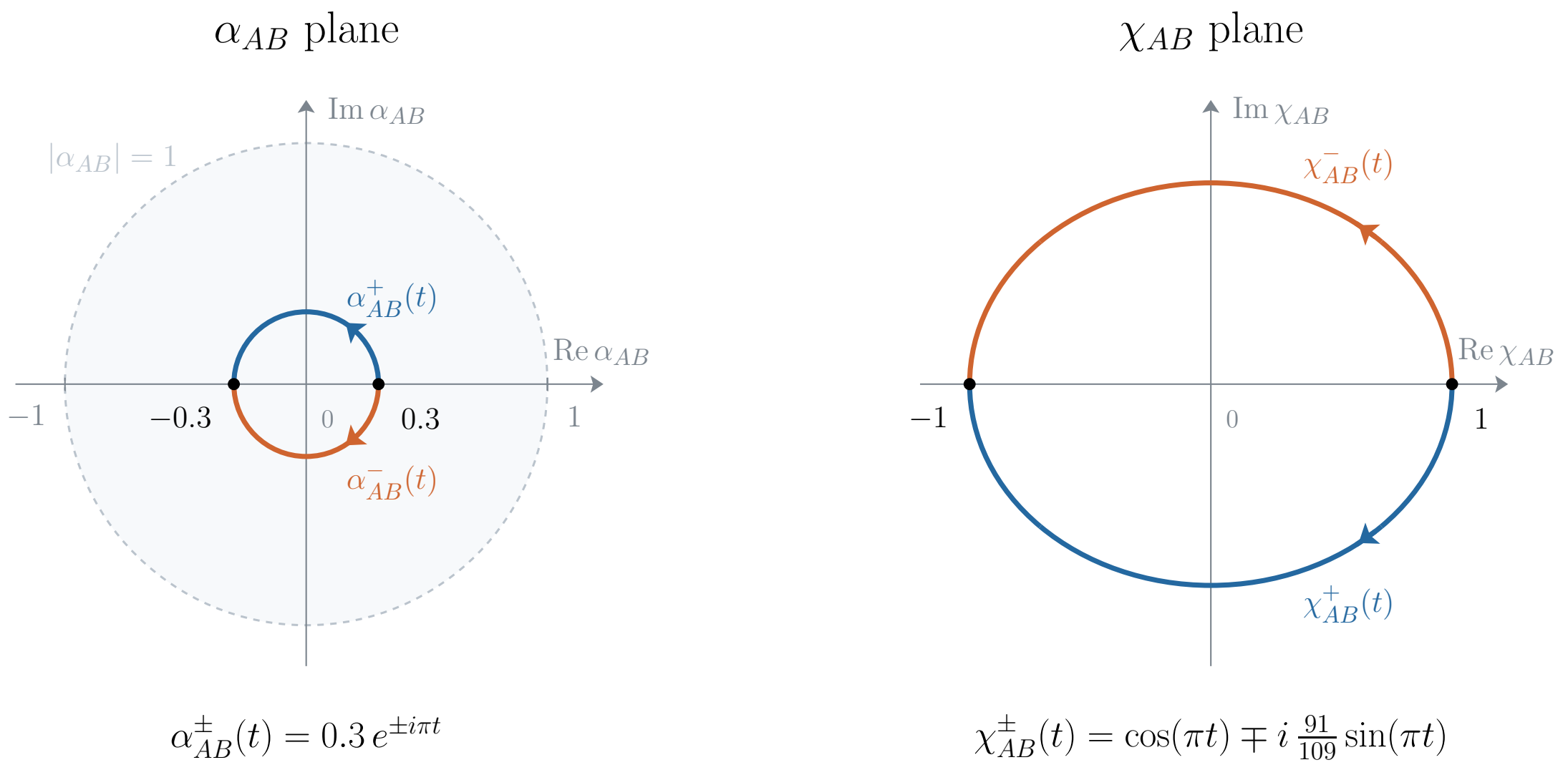

To express our discontinuity constraints in this new parametrisation, we must determine how the new variables behave under the appropriate analytic continuation. We do this by tracking how $\alpha_{AB}$ (and thus, similarly, \(\gamma_{AB}\)) changes as \(s_{AB}\) approaches the two sides of its threshold branch cut for \(s_{AB}>(m_A+m_B)^2\). Consider a pair of paths \(\alpha_{AB}^{\pm}(t)\) that start at \(t=0\) from a common reference point in the Euclidean region, and approach these respective boundary values as \(t\to1\), with all other cusp variables and masses held fixed.
Along each path, we can define the ratio
\begin{equation}
    \chi_{AB}^{\pm}(t)
    =\frac{\gamma_{AB}^{\pm}(t)}{\gamma_{AB}(0)}
    =\frac{\alpha_{AB}(0)}{\alpha_{AB}^{\pm}(t)}
    \frac{1+[\alpha_{AB}^{\pm}(t)]^2}
    {1+[\alpha_{AB}(0)]^2},
    \label{eq:fv-lambda}
\end{equation}
where the second equality follows directly from
Eq.~\eqref{eq: alpha}. Since the variables defined in Eq.~\eqref{eqs:fv} are
ratios of cusp variables, their induced
paths are uniquely fixed by the corresponding powers of
$\chi_{AB}^{\pm}(t)$. More concretely, we can take $\alpha_{AB}^{\pm}(t)$ to be the two
semicircular arcs that start at a point $0 < \alpha_{AB}^+(0) = \alpha_{AB}^-(0) < 1$ and rotate around $\alpha_{AB}=0$, namely $\alpha_{AB}^{\pm}(t)= \alpha_{AB}^{\pm}(0) e^{\pm i\pi t}$, while all other variables $\{\bar\alpha\}$ are on the unit circle.\footnote{Notice that here we choose to start with $0<\alpha^{\pm}_{AB}(0)<1$ -- meaning that $s^{\pm}_{AB}(0)<s^{\text{ps.}}_{AB}$ -- because this choice allows for a simpler analytic expression of the paths $\alpha_{AB}^{\pm}(t)$. As previously discussed, this is a completely equivalent choice to $|\alpha^{\pm}_{AB}(0)|=1$, because $\alpha^{\pm}_{AB}(t)$ does not cross any branch cut.} In this case,
\begin{equation}
    \label{eq:path-lambda}
    \chi_{AB}^{\pm}(t)
    =\frac{e^{\mp i\pi t}+[\alpha_{AB}(0)]^2e^{\pm i\pi t}}
    {1+[\alpha_{AB}(0)]^2}
    =\cos(\pi t)\mp i\frac{1-[\alpha_{AB}(0)]^2}{1+[\alpha_{AB}(0)]^2}
    \sin(\pi t).
\end{equation}
Thus, the two semicircular paths in the $\alpha_{AB}$ plane
map to the two halves of an ellipse in the
$\chi_{AB}$ plane, as illustrated in
Fig.~\ref{fig:lambda-map}.

To express a discontinuity in the physical channel $s_{AB}$
in the new variables, we must determine their paths under
the corresponding analytic continuation. Along the paths
defined in Eq.~\eqref{eq:path-lambda}, only the cusp
invariant $\gamma_{AB}$ varies, according to
\begin{equation}
    \gamma_{AB}\to\gamma_{AB}^{\pm}(t)
    =\gamma_{AB}(0)\chi_{AB}^{\pm}(t),
\end{equation}
while all other cusp invariants remain fixed.

Introducing the shorthand notation
$\eta\equiv\eta_{IJKL}$, $\rho\equiv\rho_{IJKL}$ and
$\tilde{\rho}\equiv\rho_{ILKJ}$, we denote the net powers
of $\gamma_{AB}$ in these three ratios by
$e_{AB}$, $u_{AB}$ and $v_{AB}$, respectively.
These powers can be read directly from
Eq.~\eqref{eqs:fv}: a factor of $\gamma_{AB}$ in a
numerator contributes $+1$, while a factor in a
denominator contributes $-1$, with repeated factors
counted according to their multiplicity.
The induced continuation paths therefore take the form
\begin{equation}
\bigl(\eta(t),\rho(t),\tilde{\rho}(t)\bigr)
=\bigl(
\eta\,\chi_{AB}(t)^{e_{AB}},
\rho\,\chi_{AB}(t)^{u_{AB}},
\tilde{\rho}\,\chi_{AB}(t)^{v_{AB}}
\bigr),
\label{eq:fv-path-map}
\end{equation}
where we have suppressed the $\pm$ labels distinguishing
the two continuation paths $\chi_{AB}(t)$; the same choice of sign is understood throughout.
The exponents for each channel are listed in
Table~\ref{tab:winding-numbers}.

For example, $\gamma_{JL}$ appears squared in the
denominator of $\eta$ and once in the denominators
of both $\rho$ and $\tilde{\rho}$. Continuation in the
$JL$ channel therefore gives
$(e_{JL},u_{JL},v_{JL})=(-2,-1,-1)$ and maps
$(\eta,\rho,\tilde{\rho})$ to
$(\eta/\chi_{JL}^{2},\rho/\chi_{JL},
\tilde{\rho}/\chi_{JL})$.
It also deforms the explicit coordinate $\alpha_{JL}$
if the massless limit $m_J\to0$ has not yet been taken.
The same transformation rules apply after each successive
massless limit, restricted to the variables that remain
nonzero.

\begin{table}
    \centering
    \begin{tabular}{c r r r}
    \toprule
    $AB$ &\(e_{AB}\)&\(u_{AB}\)&\(v_{AB}\)\\
    \midrule
    $IJ$& 1& 1& 0\\
    $IK$&-1&-1&-1\\
    $IL$& 0& 0& 1\\
    $JK$& 1& 0& 1\\
    $JL$&-2&-1&-1\\
    $KL$& 0& 1& 0\\
    \bottomrule
    \end{tabular}
    \caption{The net powers of the cusp invariant $\gamma_{AB}$
in the ratios $(\eta,\rho,\tilde{\rho})$. The exponents $\{e_{AB}, u_{AB}, v_{AB}\}$
determine the induced continuation paths for $(\eta,\rho,\Tilde{\rho})$ in
Eq.~\eqref{eq:fv-path-map} for each two-particle channel
$s_{AB}$.
    }
    \label{tab:winding-numbers}
\end{table}

\subsection{The one-mass and massless constraints}

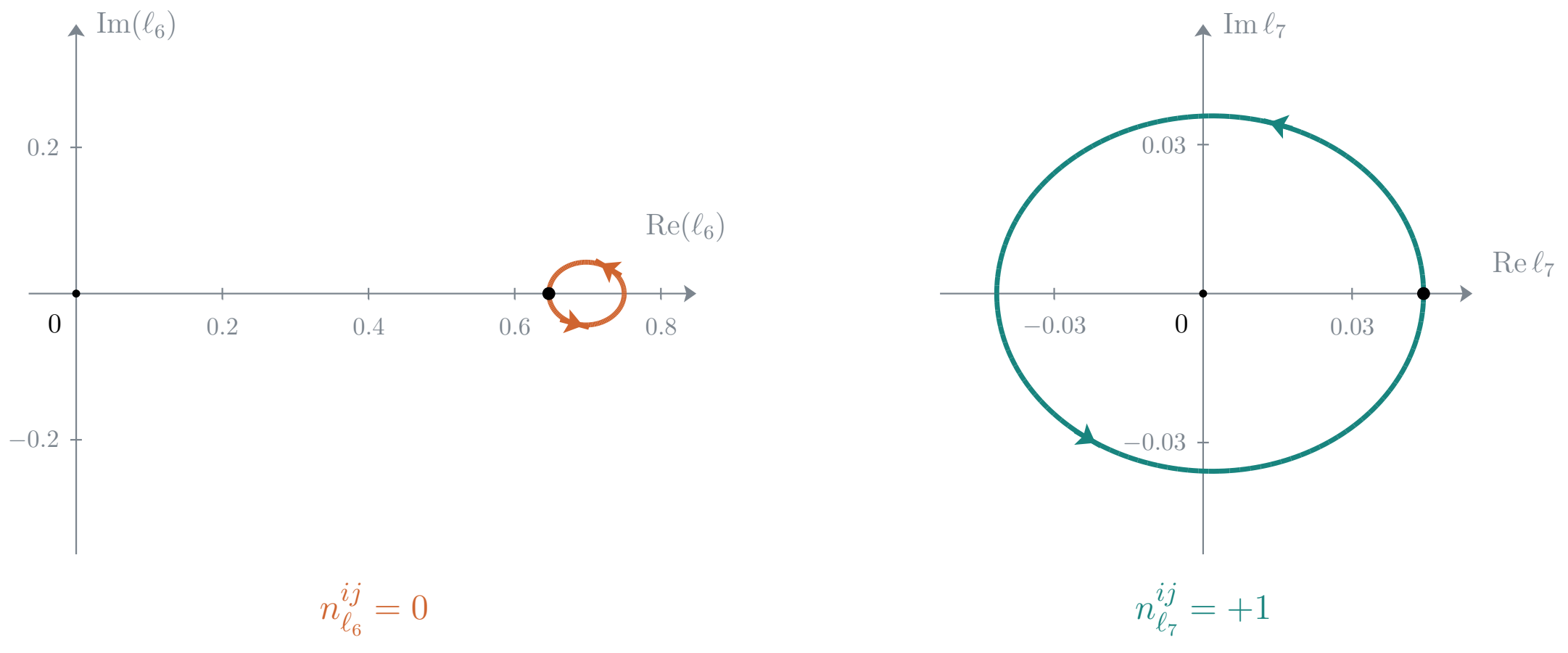
\begin{figure}[tbp]
    \centering
    \resizebox{\textwidth}{!}{%
\begingroup%
\definecolor{IJOne}{HTML}{18847E}%
\definecolor{IJOther}{HTML}{CF642F}%
\definecolor{IJAxis}{HTML}{7C858E}%
\def\IJSetRGB#1,#2,#3\relax{\SetColor{#1 #2 #3 setrgbcolor}}%
\def\IJSetColor#1{\color{#1}\convertcolorspec{named}{#1}{rgb}{\IJRGB}\expandafter\IJSetRGB\IJRGB\relax}%
\def\IJDraw#1{\put(0,0){#1}\ignorespaces}%
\setlength{\unitlength}{1pt}%
\SetScale{0.99626400996}%
\SetOffset(0,0)\unskip%
\SetScaledOffset(0,0)\unskip%
\SetArrowStroke{0}%
\begin{picture}(668,265)%
\IJSetColor{IJOne}%
\IJDraw{\Text(489,14)[c]{\fontsize{15}{18}\selectfont $n^{ij}_{\ell_7}=+1$}}%
\IJSetColor{IJAxis}%
\SetWidth{0.65}%
\IJDraw{\LongArrow[arrowwidth=3.5,arrowlength=5](383,141)(595,141)}%
\IJDraw{\LongArrow[arrowwidth=3.5,arrowlength=5](489,36)(489,247)}%
\IJDraw{\Text(631,153)[r]{\fontsize{12}{15}\selectfont $\operatorname{Re}\ell_7$}}%
\IJDraw{\Text(497,249)[l]{\fontsize{12}{15}\selectfont $\operatorname{Im}\ell_7$}}%
\IJDraw{\Line(429,138.7)(429,143.3)}%
\IJDraw{\Text(429,128)[c]{\fontsize{10}{13}\selectfont $-0.03$}}%
\IJDraw{\Line(549,138.7)(549,143.3)}%
\IJDraw{\Text(549,128)[c]{\fontsize{10}{13}\selectfont $0.03$}}%
\IJDraw{\Line(486.7,81)(491.3,81)}%
\IJDraw{\Text(482,81)[r]{\fontsize{10}{13}\selectfont $-0.03$}}%
\IJDraw{\Line(486.7,201)(491.3,201)}%
\IJDraw{\Text(482,201)[r]{\fontsize{10}{13}\selectfont $0.03$}}%
\IJSetColor{IJOne}%
\SetWidth{2}%
\IJDraw{\Bezier(577.81065,141)(577.81065,142.2601)(577.77145,143.5202)(577.69314,144.77837)}%
\IJDraw{\Bezier(577.69314,144.77837)(577.61482,146.03654)(577.49738,147.29279)(577.34103,148.5452)}%
\IJDraw{\Bezier(577.34103,148.5452)(577.18469,149.79762)(576.98943,151.04619)(576.75562,152.28903)}%
\IJDraw{\Bezier(576.75562,152.28903)(576.52182,153.53188)(576.24945,154.76899)(575.93903,155.99852)}%
\IJDraw{\Bezier(575.93903,155.99852)(575.62861,157.22804)(575.28013,158.44998)(574.89422,159.66251)}%
\IJDraw{\Bezier(574.89422,159.66251)(574.5083,160.87504)(574.08496,162.07816)(573.62493,163.27011)}%
\IJDraw{\Bezier(573.62493,163.27011)(573.16491,164.46206)(572.6682,165.64283)(572.1357,166.81072)}%
\IJDraw{\Bezier(572.1357,166.81072)(571.6032,167.97861)(571.03489,169.13362)(570.43178,170.27411)}%
\IJDraw{\Bezier(570.43178,170.27411)(569.82866,171.4146)(569.19074,172.54057)(568.51911,173.65045)}%
\IJDraw{\Bezier(568.51911,173.65045)(567.84748,174.76033)(567.14215,175.85412)(566.4043,176.93034)}%
\IJDraw{\Bezier(566.4043,176.93034)(565.66646,178.00657)(564.89611,179.06521)(564.09454,180.10488)}%
\IJDraw{\Bezier(564.09454,180.10488)(563.29298,181.14456)(562.4602,182.16526)(561.59758,183.16567)}%
\IJDraw{\Bezier(561.59758,183.16567)(560.73496,184.16608)(559.8425,185.1462)(558.92165,186.10482)}%
\IJDraw{\Bezier(558.92165,186.10482)(558.0008,187.06343)(557.05155,188.00054)(556.07544,188.91501)}%
\IJDraw{\Bezier(556.07544,188.91501)(555.09932,189.82948)(554.09633,190.72131)(553.06803,191.58948)}%
\IJDraw{\Bezier(553.06803,191.58948)(552.03973,192.45764)(550.98613,193.30214)(549.90885,194.12203)}%
\IJDraw{\Bezier(549.90885,194.12203)(548.83156,194.94192)(547.7306,195.7372)(546.60761,196.50703)}%
\IJDraw{\Bezier(546.60761,196.50703)(545.48462,197.27687)(544.33961,198.02125)(543.17428,198.73944)}%
\IJDraw{\Bezier(543.17428,198.73944)(542.00895,199.45763)(540.82329,200.14962)(539.61903,200.81475)}%
\IJDraw{\Bezier(539.61903,200.81475)(538.41477,201.47989)(537.19191,202.11818)(535.95218,202.72904)}%
\IJDraw{\Bezier(535.95218,202.72904)(534.71245,203.33991)(533.45587,203.92336)(532.18417,204.47892)}%
\IJDraw{\Bezier(532.18417,204.47892)(530.91247,205.03447)(529.62567,205.56214)(528.32552,206.06152)}%
\IJDraw{\Bezier(528.32552,206.06152)(527.02538,206.5609)(525.71188,207.032)(524.38681,207.47451)}%
\IJDraw{\Bezier(524.38681,207.47451)(523.06174,207.91702)(521.72508,208.33095)(520.37861,208.71606)}%
\IJDraw{\Bezier(520.37861,208.71606)(519.03213,209.10117)(517.67584,209.45747)(516.31148,209.78482)}%
\IJDraw{\Bezier(516.31148,209.78482)(514.94712,210.11216)(513.57469,210.41054)(512.19594,210.6799)}%
\IJDraw{\Bezier(512.19594,210.6799)(510.81719,210.94925)(509.43212,211.18958)(508.04244,211.40088)}%
\IJDraw{\Bezier(508.04244,211.40088)(506.65277,211.61217)(505.2585,211.79444)(503.86134,211.94775)}%
\IJDraw{\Bezier(503.86134,211.94775)(502.46419,212.10106)(501.06414,212.2254)(499.66289,212.32093)}%
\IJDraw{\Bezier(499.66289,212.32093)(498.26164,212.41645)(496.85918,212.48315)(495.45719,212.52123)}%
\IJDraw{\Bezier(495.45719,212.52123)(494.05519,212.5593)(492.65365,212.56875)(491.25421,212.54983)}%
\IJDraw{\Bezier(491.25421,212.54983)(489.85476,212.53092)(488.45741,212.48363)(487.06375,212.4083)}%
\IJDraw{\Bezier(487.06375,212.4083)(485.67009,212.33296)(484.28013,212.22958)(482.89543,212.09851)}%
\IJDraw{\Bezier(482.89543,212.09851)(481.51074,211.96745)(480.13131,211.8087)(478.75869,211.6227)}%
\IJDraw{\Bezier(478.75869,211.6227)(477.38607,211.4367)(476.02025,211.22344)(474.66274,210.9834)}%
\IJDraw{\Bezier(474.66274,210.9834)(473.30524,210.74336)(471.95604,210.47654)(470.61661,210.18345)}%
\IJDraw{\Bezier(470.61661,210.18345)(469.27719,209.89036)(467.94754,209.57101)(466.62909,209.22596)}%
\IJDraw{\Bezier(466.62909,209.22596)(465.31063,208.88091)(464.00338,208.51016)(462.70872,208.11432)}%
\IJDraw{\Bezier(462.70872,208.11432)(461.41405,207.71848)(460.13197,207.29755)(458.86382,206.85218)}%
\IJDraw{\Bezier(458.86382,206.85218)(457.59567,206.40681)(456.34144,205.937)(455.10245,205.44343)}%
\IJDraw{\Bezier(455.10245,205.44343)(453.86346,204.94987)(452.6397,204.43255)(451.43243,203.8922)}%
\IJDraw{\Bezier(451.43243,203.8922)(450.22516,203.35185)(449.03437,202.78847)(447.86128,202.20283)}%
\IJDraw{\Bezier(447.86128,202.20283)(446.6882,201.61718)(445.53281,201.00926)(444.39629,200.37987)}%
\IJDraw{\Bezier(444.39629,200.37987)(443.25976,199.75047)(442.1421,199.0996)(441.04442,198.42807)}%
\IJDraw{\Bezier(441.04442,198.42807)(439.94675,197.75655)(438.86905,197.06437)(437.81241,196.35239)}%
\IJDraw{\Bezier(437.81241,196.35239)(436.75576,195.64041)(435.72017,194.90863)(434.70665,194.15793)}%
\IJDraw{\Bezier(434.70665,194.15793)(433.69313,193.40723)(432.70168,192.63762)(431.73327,191.84999)}%
\IJDraw{\Bezier(431.73327,191.84999)(430.76486,191.06236)(429.8195,190.25672)(428.89809,189.434)}%
\IJDraw{\Bezier(428.89809,189.434)(427.97669,188.61129)(427.07924,187.77149)(426.20663,186.91558)}%
\IJDraw{\Bezier(426.20663,186.91558)(425.33401,186.05966)(424.48623,185.18762)(423.66408,184.30043)}%
\IJDraw{\Bezier(423.66408,184.30043)(422.84194,183.41325)(422.04544,182.51092)(421.27535,181.59445)}%
\IJDraw{\Bezier(421.27535,181.59445)(420.50525,180.67797)(419.76157,179.74735)(419.04499,178.8036)}%
\IJDraw{\Bezier(419.04499,178.8036)(418.32842,177.85985)(417.63896,176.90296)(416.97727,175.93398)}%
\IJDraw{\Bezier(416.97727,175.93398)(416.31557,174.96501)(415.68165,173.98393)(415.07609,172.99181)}%
\IJDraw{\Bezier(415.07609,172.99181)(414.47054,171.99969)(413.89335,170.99652)(413.34507,169.98337)}%
\IJDraw{\Bezier(413.34507,169.98337)(412.79679,168.97022)(412.27742,167.94708)(411.78745,166.91504)}%
\IJDraw{\Bezier(411.78745,166.91504)(411.29747,165.883)(410.8369,164.84206)(410.40616,163.79329)}%
\IJDraw{\Bezier(410.40616,163.79329)(409.97541,162.74453)(409.5745,161.68795)(409.20379,160.62465)}%
\IJDraw{\Bezier(409.20379,160.62465)(408.83307,159.56135)(408.49257,158.49133)(408.18258,157.41569)}%
\IJDraw{\Bezier(408.18258,157.41569)(407.87259,156.34006)(407.59313,155.25882)(407.34445,154.17307)}%
\IJDraw{\Bezier(407.34445,154.17307)(407.09576,153.08733)(406.87786,151.99708)(406.69094,150.90346)}%
\IJDraw{\Bezier(406.69094,150.90346)(406.50403,149.80984)(406.34809,148.71284)(406.22329,147.61358)}%
\IJDraw{\Bezier(406.22329,147.61358)(406.09848,146.51433)(406.0048,145.41281)(405.94234,144.31017)}%
\IJDraw{\Bezier(405.94234,144.31017)(405.87988,143.20754)(405.84864,142.10377)(405.84864,141)}%
\IJDraw{\Bezier(405.84864,141)(405.84864,139.89623)(405.87988,138.79246)(405.94234,137.68983)}%
\IJDraw{\Bezier(405.94234,137.68983)(406.0048,136.58719)(406.09848,135.48567)(406.22329,134.38642)}%
\IJDraw{\Bezier(406.22329,134.38642)(406.34809,133.28716)(406.50403,132.19016)(406.69094,131.09654)}%
\IJDraw{\Bezier(406.69094,131.09654)(406.87786,130.00292)(407.09576,128.91267)(407.34445,127.82693)}%
\IJDraw{\Bezier(407.34445,127.82693)(407.59313,126.74118)(407.87259,125.65994)(408.18258,124.58431)}%
\IJDraw{\Bezier(408.18258,124.58431)(408.49257,123.50867)(408.83307,122.43865)(409.20379,121.37535)}%
\IJDraw{\Bezier(409.20379,121.37535)(409.5745,120.31205)(409.97541,119.25547)(410.40616,118.20671)}%
\IJDraw{\Bezier(410.40616,118.20671)(410.8369,117.15794)(411.29747,116.117)(411.78745,115.08496)}%
\IJDraw{\Bezier(411.78745,115.08496)(412.27742,114.05292)(412.79679,113.02978)(413.34507,112.01663)}%
\IJDraw{\Bezier(413.34507,112.01663)(413.89335,111.00348)(414.47054,110.00031)(415.07609,109.00819)}%
\IJDraw{\Bezier(415.07609,109.00819)(415.68165,108.01607)(416.31557,107.03499)(416.97727,106.06602)}%
\IJDraw{\Bezier(416.97727,106.06602)(417.63896,105.09704)(418.32842,104.14015)(419.04499,103.1964)}%
\IJDraw{\Bezier(419.04499,103.1964)(419.76157,102.25265)(420.50525,101.32203)(421.27535,100.40555)}%
\IJDraw{\Bezier(421.27535,100.40555)(422.04544,99.48908)(422.84194,98.58675)(423.66408,97.69957)}%
\IJDraw{\Bezier(423.66408,97.69957)(424.48623,96.81238)(425.33401,95.94034)(426.20663,95.08442)}%
\IJDraw{\Bezier(426.20663,95.08442)(427.07924,94.22851)(427.97669,93.38871)(428.89809,92.566)}%
\IJDraw{\Bezier(428.89809,92.566)(429.8195,91.74328)(430.76486,90.93764)(431.73327,90.15001)}%
\IJDraw{\Bezier(431.73327,90.15001)(432.70168,89.36238)(433.69313,88.59277)(434.70665,87.84207)}%
\IJDraw{\Bezier(434.70665,87.84207)(435.72017,87.09137)(436.75576,86.35959)(437.81241,85.64761)}%
\IJDraw{\Bezier(437.81241,85.64761)(438.86905,84.93563)(439.94675,84.24345)(441.04442,83.57193)}%
\IJDraw{\Bezier(441.04442,83.57193)(442.1421,82.9004)(443.25976,82.24953)(444.39629,81.62013)}%
\IJDraw{\Bezier(444.39629,81.62013)(445.53281,80.99074)(446.6882,80.38282)(447.86128,79.79717)}%
\IJDraw{\Bezier(447.86128,79.79717)(449.03437,79.21153)(450.22516,78.64815)(451.43243,78.1078)}%
\IJDraw{\Bezier(451.43243,78.1078)(452.6397,77.56745)(453.86346,77.05013)(455.10245,76.55657)}%
\IJDraw{\Bezier(455.10245,76.55657)(456.34144,76.063)(457.59567,75.59319)(458.86382,75.14782)}%
\IJDraw{\Bezier(458.86382,75.14782)(460.13197,74.70245)(461.41405,74.28152)(462.70872,73.88568)}%
\IJDraw{\Bezier(462.70872,73.88568)(464.00338,73.48984)(465.31063,73.11909)(466.62909,72.77404)}%
\IJDraw{\Bezier(466.62909,72.77404)(467.94754,72.42899)(469.27719,72.10964)(470.61661,71.81655)}%
\IJDraw{\Bezier(470.61661,71.81655)(471.95604,71.52346)(473.30524,71.25664)(474.66274,71.0166)}%
\IJDraw{\Bezier(474.66274,71.0166)(476.02025,70.77656)(477.38607,70.5633)(478.75869,70.3773)}%
\IJDraw{\Bezier(478.75869,70.3773)(480.13131,70.1913)(481.51074,70.03255)(482.89543,69.90149)}%
\IJDraw{\Bezier(482.89543,69.90149)(484.28013,69.77042)(485.67009,69.66704)(487.06375,69.5917)}%
\IJDraw{\Bezier(487.06375,69.5917)(488.45741,69.51637)(489.85476,69.46908)(491.25421,69.45017)}%
\IJDraw{\Bezier(491.25421,69.45017)(492.65365,69.43125)(494.05519,69.4407)(495.45719,69.47877)}%
\IJDraw{\Bezier(495.45719,69.47877)(496.85918,69.51685)(498.26164,69.58355)(499.66289,69.67907)}%
\IJDraw{\Bezier(499.66289,69.67907)(501.06414,69.7746)(502.46419,69.89894)(503.86134,70.05225)}%
\IJDraw{\Bezier(503.86134,70.05225)(505.2585,70.20556)(506.65277,70.38783)(508.04244,70.59912)}%
\IJDraw{\Bezier(508.04244,70.59912)(509.43212,70.81042)(510.81719,71.05075)(512.19594,71.3201)}%
\IJDraw{\Bezier(512.19594,71.3201)(513.57469,71.58946)(514.94712,71.88784)(516.31148,72.21518)}%
\IJDraw{\Bezier(516.31148,72.21518)(517.67584,72.54253)(519.03213,72.89883)(520.37861,73.28394)}%
\IJDraw{\Bezier(520.37861,73.28394)(521.72508,73.66905)(523.06174,74.08298)(524.38681,74.52549)}%
\IJDraw{\Bezier(524.38681,74.52549)(525.71188,74.968)(527.02538,75.4391)(528.32552,75.93848)}%
\IJDraw{\Bezier(528.32552,75.93848)(529.62567,76.43786)(530.91247,76.96553)(532.18417,77.52108)}%
\IJDraw{\Bezier(532.18417,77.52108)(533.45587,78.07664)(534.71245,78.66009)(535.95218,79.27096)}%
\IJDraw{\Bezier(535.95218,79.27096)(537.19191,79.88182)(538.41477,80.52011)(539.61903,81.18525)}%
\IJDraw{\Bezier(539.61903,81.18525)(540.82329,81.85038)(542.00895,82.54237)(543.17428,83.26056)}%
\IJDraw{\Bezier(543.17428,83.26056)(544.33961,83.97875)(545.48462,84.72313)(546.60761,85.49297)}%
\IJDraw{\Bezier(546.60761,85.49297)(547.7306,86.2628)(548.83156,87.05808)(549.90885,87.87797)}%
\IJDraw{\Bezier(549.90885,87.87797)(550.98613,88.69786)(552.03973,89.54236)(553.06803,90.41052)}%
\IJDraw{\Bezier(553.06803,90.41052)(554.09633,91.27869)(555.09932,92.17052)(556.07544,93.08499)}%
\IJDraw{\Bezier(556.07544,93.08499)(557.05155,93.99946)(558.0008,94.93657)(558.92165,95.89518)}%
\IJDraw{\Bezier(558.92165,95.89518)(559.8425,96.8538)(560.73496,97.83392)(561.59758,98.83433)}%
\IJDraw{\Bezier(561.59758,98.83433)(562.4602,99.83474)(563.29298,100.85544)(564.09454,101.89512)}%
\IJDraw{\Bezier(564.09454,101.89512)(564.89611,102.93479)(565.66646,103.99343)(566.4043,105.06966)}%
\IJDraw{\Bezier(566.4043,105.06966)(567.14215,106.14588)(567.84748,107.23967)(568.51911,108.34955)}%
\IJDraw{\Bezier(568.51911,108.34955)(569.19074,109.45943)(569.82866,110.5854)(570.43178,111.72589)}%
\IJDraw{\Bezier(570.43178,111.72589)(571.03489,112.86638)(571.6032,114.02139)(572.1357,115.18928)}%
\IJDraw{\Bezier(572.1357,115.18928)(572.6682,116.35717)(573.16491,117.53794)(573.62493,118.72989)}%
\IJDraw{\Bezier(573.62493,118.72989)(574.08496,119.92184)(574.5083,121.12496)(574.89422,122.33749)}%
\IJDraw{\Bezier(574.89422,122.33749)(575.28013,123.55002)(575.62861,124.77196)(575.93903,126.00148)}%
\IJDraw{\Bezier(575.93903,126.00148)(576.24945,127.23101)(576.52182,128.46812)(576.75562,129.71097)}%
\IJDraw{\Bezier(576.75562,129.71097)(576.98943,130.95381)(577.18469,132.20238)(577.34103,133.4548)}%
\IJDraw{\Bezier(577.34103,133.4548)(577.49738,134.70721)(577.61482,135.96346)(577.69314,137.22163)}%
\IJDraw{\Bezier(577.69314,137.22163)(577.77145,138.4798)(577.81065,139.7399)(577.81065,141)}%
\IJDraw{\ArrowLine[arrowpos=0.5,arrowwidth=5,arrowlength=7](524.90395,207.49695)(513.32298,210.64037)}%
\IJDraw{\ArrowLine[arrowpos=0.5,arrowwidth=5,arrowlength=7](437.42021,85.66759)(447.78197,79.615)}%
\IJSetColor{black}%
\IJDraw{\Vertex(577.81065,141){2.6}}%
\IJDraw{\Vertex(489,141){1.6}}%
\IJDraw{\Text(483,129)[r]{\fontsize{11}{14}\selectfont $0$}}%
\IJSetColor{IJOther}%
\IJDraw{\Text(155,14)[c]{\fontsize{15}{18}\selectfont $n^{ij}_{\ell_6}=0$}}%
\IJSetColor{IJAxis}%
\SetWidth{0.65}%
\IJDraw{\LongArrow[arrowwidth=3.5,arrowlength=5](15.86111,141)(282.33333,141)}%
\IJDraw{\LongArrow[arrowwidth=3.5,arrowlength=5](35,36)(35,247)}%
\IJDraw{\Text(297,168)[r]{\fontsize{12}{15}\selectfont $\operatorname{Re}(\ell_6)$}}%
\IJDraw{\Text(43,249)[l]{\fontsize{12}{15}\selectfont $\operatorname{Im}(\ell_6)$}}%
\IJDraw{\Line(93.88889,138.7)(93.88889,143.3)}%
\IJDraw{\Text(93.88889,128)[c]{\fontsize{10}{13}\selectfont $0.2$}}%
\IJDraw{\Line(152.77778,138.7)(152.77778,143.3)}%
\IJDraw{\Text(152.77778,128)[c]{\fontsize{10}{13}\selectfont $0.4$}}%
\IJDraw{\Line(211.66667,138.7)(211.66667,143.3)}%
\IJDraw{\Text(211.66667,128)[c]{\fontsize{10}{13}\selectfont $0.6$}}%
\IJDraw{\Line(270.55556,138.7)(270.55556,143.3)}%
\IJDraw{\Text(270.55556,128)[c]{\fontsize{10}{13}\selectfont $0.8$}}%
\IJDraw{\Line(32.7,82.11111)(37.3,82.11111)}%
\IJDraw{\Text(28,82.11111)[r]{\fontsize{10}{13}\selectfont $-0.2$}}%
\IJDraw{\Line(32.7,199.88889)(37.3,199.88889)}%
\IJDraw{\Text(28,199.88889)[r]{\fontsize{10}{13}\selectfont $0.2$}}%
\IJSetColor{IJOther}%
\SetWidth{2}%
\IJDraw{\Bezier(225.40356,141)(225.40356,140.77695)(225.4105,140.5539)(225.42437,140.33119)}%
\IJDraw{\Bezier(225.42437,140.33119)(225.43824,140.10848)(225.45904,139.88612)(225.48674,139.66443)}%
\IJDraw{\Bezier(225.48674,139.66443)(225.51443,139.44274)(225.54901,139.22173)(225.59043,139.00174)}%
\IJDraw{\Bezier(225.59043,139.00174)(225.63184,138.78175)(225.68008,138.56277)(225.73507,138.34514)}%
\IJDraw{\Bezier(225.73507,138.34514)(225.79005,138.12751)(225.85178,137.91123)(225.92014,137.69661)}%
\IJDraw{\Bezier(225.92014,137.69661)(225.98849,137.48199)(226.06348,137.26904)(226.14497,137.05807)}%
\IJDraw{\Bezier(226.14497,137.05807)(226.22646,136.8471)(226.31444,136.63811)(226.40877,136.43141)}%
\IJDraw{\Bezier(226.40877,136.43141)(226.5031,136.22471)(226.60376,136.02029)(226.7106,135.81844)}%
\IJDraw{\Bezier(226.7106,135.81844)(226.81744,135.6166)(226.93044,135.41733)(227.04942,135.22092)}%
\IJDraw{\Bezier(227.04942,135.22092)(227.16839,135.0245)(227.29334,134.83094)(227.42404,134.6405)}%
\IJDraw{\Bezier(227.42404,134.6405)(227.55475,134.45006)(227.69121,134.26273)(227.83321,134.07878)}%
\IJDraw{\Bezier(227.83321,134.07878)(227.9752,133.89482)(228.12273,133.71423)(228.27554,133.53724)}%
\IJDraw{\Bezier(228.27554,133.53724)(228.42835,133.36025)(228.58645,133.18687)(228.74958,133.0173)}%
\IJDraw{\Bezier(228.74958,133.0173)(228.9127,132.84773)(229.08086,132.68198)(229.25377,132.52025)}%
\IJDraw{\Bezier(229.25377,132.52025)(229.42668,132.35852)(229.60436,132.20081)(229.78651,132.04731)}%
\IJDraw{\Bezier(229.78651,132.04731)(229.96866,131.8938)(230.1553,131.7445)(230.34612,131.59956)}%
\IJDraw{\Bezier(230.34612,131.59956)(230.53694,131.45463)(230.73196,131.31407)(230.93087,131.17802)}%
\IJDraw{\Bezier(230.93087,131.17802)(231.12978,131.04198)(231.33259,130.91046)(231.53899,130.78359)}%
\IJDraw{\Bezier(231.53899,130.78359)(231.74539,130.65672)(231.95538,130.53451)(232.16866,130.41707)}%
\IJDraw{\Bezier(232.16866,130.41707)(232.38194,130.29962)(232.5985,130.18695)(232.81804,130.07914)}%
\IJDraw{\Bezier(232.81804,130.07914)(233.03759,129.97134)(233.2601,129.86841)(233.48528,129.77043)}%
\IJDraw{\Bezier(233.48528,129.77043)(233.71046,129.67245)(233.93831,129.57942)(234.1685,129.49142)}%
\IJDraw{\Bezier(234.1685,129.49142)(234.39869,129.40341)(234.63124,129.32043)(234.86581,129.24253)}%
\IJDraw{\Bezier(234.86581,129.24253)(235.10039,129.16462)(235.337,129.09179)(235.57533,129.02408)}%
\IJDraw{\Bezier(235.57533,129.02408)(235.81366,128.95636)(236.05371,128.89376)(236.29517,128.8363)}%
\IJDraw{\Bezier(236.29517,128.8363)(236.53663,128.77884)(236.7795,128.72652)(237.02346,128.67935)}%
\IJDraw{\Bezier(237.02346,128.67935)(237.26743,128.63217)(237.51249,128.59015)(237.75834,128.55328)}%
\IJDraw{\Bezier(237.75834,128.55328)(238.00419,128.5164)(238.25083,128.48467)(238.49796,128.45808)}%
\IJDraw{\Bezier(238.49796,128.45808)(238.74509,128.43149)(238.9927,128.41002)(239.2405,128.39367)}%
\IJDraw{\Bezier(239.2405,128.39367)(239.4883,128.37732)(239.73629,128.36607)(239.98416,128.35989)}%
\IJDraw{\Bezier(239.98416,128.35989)(240.23204,128.35371)(240.47981,128.3526)(240.72718,128.35651)}%
\IJDraw{\Bezier(240.72718,128.35651)(240.97455,128.36043)(241.22152,128.36936)(241.4678,128.38325)}%
\IJDraw{\Bezier(241.4678,128.38325)(241.71409,128.39714)(241.9597,128.416)(242.20434,128.43974)}%
\IJDraw{\Bezier(242.20434,128.43974)(242.44898,128.46349)(242.69266,128.49213)(242.93511,128.52558)}%
\IJDraw{\Bezier(242.93511,128.52558)(243.17756,128.55903)(243.41877,128.5973)(243.65848,128.6403)}%
\IJDraw{\Bezier(243.65848,128.6403)(243.8982,128.68329)(244.13642,128.73102)(244.37287,128.78337)}%
\IJDraw{\Bezier(244.37287,128.78337)(244.60933,128.83573)(244.84403,128.89272)(245.07673,128.95423)}%
\IJDraw{\Bezier(245.07673,128.95423)(245.30942,129.01575)(245.54011,129.0818)(245.76854,129.15227)}%
\IJDraw{\Bezier(245.76854,129.15227)(245.99697,129.22273)(246.22315,129.29762)(246.44685,129.3768)}%
\IJDraw{\Bezier(246.44685,129.3768)(246.67054,129.45599)(246.89174,129.53948)(247.11023,129.62714)}%
\IJDraw{\Bezier(247.11023,129.62714)(247.32872,129.71481)(247.5445,129.80665)(247.75733,129.90254)}%
\IJDraw{\Bezier(247.75733,129.90254)(247.97017,129.99843)(248.18007,130.09837)(248.38683,130.20221)}%
\IJDraw{\Bezier(248.38683,130.20221)(248.59358,130.30606)(248.79719,130.41381)(248.99745,130.52534)}%
\IJDraw{\Bezier(248.99745,130.52534)(249.1977,130.63686)(249.39461,130.75215)(249.58797,130.87107)}%
\IJDraw{\Bezier(249.58797,130.87107)(249.78133,130.98998)(249.97115,131.11251)(250.15723,131.23852)}%
\IJDraw{\Bezier(250.15723,131.23852)(250.34332,131.36452)(250.52567,131.49399)(250.70411,131.62677)}%
\IJDraw{\Bezier(250.70411,131.62677)(250.88256,131.75956)(251.0571,131.89565)(251.22756,132.0349)}%
\IJDraw{\Bezier(251.22756,132.0349)(251.39802,132.17414)(251.5644,132.31654)(251.72655,132.46192)}%
\IJDraw{\Bezier(251.72655,132.46192)(251.8887,132.6073)(252.04661,132.75566)(252.20014,132.90685)}%
\IJDraw{\Bezier(252.20014,132.90685)(252.35367,133.05803)(252.50281,133.21203)(252.64743,133.36867)}%
\IJDraw{\Bezier(252.64743,133.36867)(252.79205,133.52531)(252.93214,133.6846)(253.06757,133.84636)}%
\IJDraw{\Bezier(253.06757,133.84636)(253.20301,134.00811)(253.33378,134.17234)(253.45978,134.33885)}%
\IJDraw{\Bezier(253.45978,134.33885)(253.58578,134.50536)(253.70699,134.67416)(253.82332,134.84507)}%
\IJDraw{\Bezier(253.82332,134.84507)(253.93965,135.01598)(254.05108,135.189)(254.15752,135.36395)}%
\IJDraw{\Bezier(254.15752,135.36395)(254.26396,135.53889)(254.36541,135.71576)(254.46176,135.89436)}%
\IJDraw{\Bezier(254.46176,135.89436)(254.55812,136.07297)(254.64939,136.25331)(254.73549,136.4352)}%
\IJDraw{\Bezier(254.73549,136.4352)(254.82159,136.6171)(254.90251,136.80054)(254.97819,136.98535)}%
\IJDraw{\Bezier(254.97819,136.98535)(255.05388,137.17015)(255.12431,137.35631)(255.18944,137.54365)}%
\IJDraw{\Bezier(255.18944,137.54365)(255.25456,137.73098)(255.31438,137.91948)(255.36884,138.10896)}%
\IJDraw{\Bezier(255.36884,138.10896)(255.42329,138.29844)(255.47238,138.48889)(255.51606,138.68012)}%
\IJDraw{\Bezier(255.51606,138.68012)(255.55974,138.87136)(255.59801,139.06338)(255.63085,139.25598)}%
\IJDraw{\Bezier(255.63085,139.25598)(255.66368,139.44859)(255.69106,139.64178)(255.71298,139.83536)}%
\IJDraw{\Bezier(255.71298,139.83536)(255.7349,140.02895)(255.75136,140.22292)(255.76233,140.41709)}%
\IJDraw{\Bezier(255.76233,140.41709)(255.7733,140.61126)(255.77878,140.80563)(255.77878,141)}%
\IJDraw{\Bezier(255.77878,141)(255.77878,141.19437)(255.7733,141.38874)(255.76233,141.58291)}%
\IJDraw{\Bezier(255.76233,141.58291)(255.75136,141.77708)(255.7349,141.97105)(255.71298,142.16464)}%
\IJDraw{\Bezier(255.71298,142.16464)(255.69106,142.35822)(255.66368,142.55141)(255.63085,142.74402)}%
\IJDraw{\Bezier(255.63085,142.74402)(255.59801,142.93662)(255.55974,143.12864)(255.51606,143.31988)}%
\IJDraw{\Bezier(255.51606,143.31988)(255.47238,143.51111)(255.42329,143.70156)(255.36884,143.89104)}%
\IJDraw{\Bezier(255.36884,143.89104)(255.31438,144.08052)(255.25456,144.26902)(255.18944,144.45635)}%
\IJDraw{\Bezier(255.18944,144.45635)(255.12431,144.64369)(255.05388,144.82985)(254.97819,145.01465)}%
\IJDraw{\Bezier(254.97819,145.01465)(254.90251,145.19946)(254.82159,145.3829)(254.73549,145.5648)}%
\IJDraw{\Bezier(254.73549,145.5648)(254.64939,145.74669)(254.55812,145.92703)(254.46176,146.10564)}%
\IJDraw{\Bezier(254.46176,146.10564)(254.36541,146.28424)(254.26396,146.46111)(254.15752,146.63605)}%
\IJDraw{\Bezier(254.15752,146.63605)(254.05108,146.811)(253.93965,146.98402)(253.82332,147.15493)}%
\IJDraw{\Bezier(253.82332,147.15493)(253.70699,147.32584)(253.58578,147.49464)(253.45978,147.66115)}%
\IJDraw{\Bezier(253.45978,147.66115)(253.33378,147.82766)(253.20301,147.99189)(253.06757,148.15364)}%
\IJDraw{\Bezier(253.06757,148.15364)(252.93214,148.3154)(252.79205,148.47469)(252.64743,148.63133)}%
\IJDraw{\Bezier(252.64743,148.63133)(252.50281,148.78797)(252.35367,148.94197)(252.20014,149.09315)}%
\IJDraw{\Bezier(252.20014,149.09315)(252.04661,149.24434)(251.8887,149.3927)(251.72655,149.53808)}%
\IJDraw{\Bezier(251.72655,149.53808)(251.5644,149.68346)(251.39802,149.82586)(251.22756,149.9651)}%
\IJDraw{\Bezier(251.22756,149.9651)(251.0571,150.10435)(250.88256,150.24044)(250.70411,150.37323)}%
\IJDraw{\Bezier(250.70411,150.37323)(250.52567,150.50601)(250.34332,150.63548)(250.15723,150.76148)}%
\IJDraw{\Bezier(250.15723,150.76148)(249.97115,150.88749)(249.78133,151.01002)(249.58797,151.12893)}%
\IJDraw{\Bezier(249.58797,151.12893)(249.39461,151.24785)(249.1977,151.36314)(248.99745,151.47466)}%
\IJDraw{\Bezier(248.99745,151.47466)(248.79719,151.58619)(248.59358,151.69394)(248.38683,151.79779)}%
\IJDraw{\Bezier(248.38683,151.79779)(248.18007,151.90163)(247.97017,152.00157)(247.75733,152.09746)}%
\IJDraw{\Bezier(247.75733,152.09746)(247.5445,152.19335)(247.32872,152.28519)(247.11023,152.37286)}%
\IJDraw{\Bezier(247.11023,152.37286)(246.89174,152.46052)(246.67054,152.54401)(246.44685,152.6232)}%
\IJDraw{\Bezier(246.44685,152.6232)(246.22315,152.70238)(245.99697,152.77727)(245.76854,152.84773)}%
\IJDraw{\Bezier(245.76854,152.84773)(245.54011,152.9182)(245.30942,152.98425)(245.07673,153.04577)}%
\IJDraw{\Bezier(245.07673,153.04577)(244.84403,153.10728)(244.60933,153.16427)(244.37287,153.21663)}%
\IJDraw{\Bezier(244.37287,153.21663)(244.13642,153.26898)(243.8982,153.31671)(243.65848,153.3597)}%
\IJDraw{\Bezier(243.65848,153.3597)(243.41877,153.4027)(243.17756,153.44097)(242.93511,153.47442)}%
\IJDraw{\Bezier(242.93511,153.47442)(242.69266,153.50787)(242.44898,153.53651)(242.20434,153.56026)}%
\IJDraw{\Bezier(242.20434,153.56026)(241.9597,153.584)(241.71409,153.60286)(241.4678,153.61675)}%
\IJDraw{\Bezier(241.4678,153.61675)(241.22152,153.63064)(240.97455,153.63957)(240.72718,153.64349)}%
\IJDraw{\Bezier(240.72718,153.64349)(240.47981,153.6474)(240.23204,153.64629)(239.98416,153.64011)}%
\IJDraw{\Bezier(239.98416,153.64011)(239.73629,153.63393)(239.4883,153.62268)(239.2405,153.60633)}%
\IJDraw{\Bezier(239.2405,153.60633)(238.9927,153.58998)(238.74509,153.56851)(238.49796,153.54192)}%
\IJDraw{\Bezier(238.49796,153.54192)(238.25083,153.51533)(238.00419,153.4836)(237.75834,153.44672)}%
\IJDraw{\Bezier(237.75834,153.44672)(237.51249,153.40985)(237.26743,153.36783)(237.02346,153.32065)}%
\IJDraw{\Bezier(237.02346,153.32065)(236.7795,153.27348)(236.53663,153.22116)(236.29517,153.1637)}%
\IJDraw{\Bezier(236.29517,153.1637)(236.05371,153.10624)(235.81366,153.04364)(235.57533,152.97592)}%
\IJDraw{\Bezier(235.57533,152.97592)(235.337,152.90821)(235.10039,152.83538)(234.86581,152.75747)}%
\IJDraw{\Bezier(234.86581,152.75747)(234.63124,152.67957)(234.39869,152.59659)(234.1685,152.50858)}%
\IJDraw{\Bezier(234.1685,152.50858)(233.93831,152.42058)(233.71046,152.32755)(233.48528,152.22957)}%
\IJDraw{\Bezier(233.48528,152.22957)(233.2601,152.13159)(233.03759,152.02866)(232.81804,151.92086)}%
\IJDraw{\Bezier(232.81804,151.92086)(232.5985,151.81305)(232.38194,151.70038)(232.16866,151.58293)}%
\IJDraw{\Bezier(232.16866,151.58293)(231.95538,151.46549)(231.74539,151.34328)(231.53899,151.21641)}%
\IJDraw{\Bezier(231.53899,151.21641)(231.33259,151.08954)(231.12978,150.95802)(230.93087,150.82198)}%
\IJDraw{\Bezier(230.93087,150.82198)(230.73196,150.68593)(230.53694,150.54537)(230.34612,150.40044)}%
\IJDraw{\Bezier(230.34612,150.40044)(230.1553,150.2555)(229.96866,150.1062)(229.78651,149.95269)}%
\IJDraw{\Bezier(229.78651,149.95269)(229.60436,149.79919)(229.42668,149.64148)(229.25377,149.47975)}%
\IJDraw{\Bezier(229.25377,149.47975)(229.08086,149.31802)(228.9127,149.15227)(228.74958,148.9827)}%
\IJDraw{\Bezier(228.74958,148.9827)(228.58645,148.81313)(228.42835,148.63975)(228.27554,148.46276)}%
\IJDraw{\Bezier(228.27554,148.46276)(228.12273,148.28577)(227.9752,148.10518)(227.83321,147.92122)}%
\IJDraw{\Bezier(227.83321,147.92122)(227.69121,147.73727)(227.55475,147.54994)(227.42404,147.3595)}%
\IJDraw{\Bezier(227.42404,147.3595)(227.29334,147.16906)(227.16839,146.9755)(227.04942,146.77908)}%
\IJDraw{\Bezier(227.04942,146.77908)(226.93044,146.58267)(226.81744,146.3834)(226.7106,146.18156)}%
\IJDraw{\Bezier(226.7106,146.18156)(226.60376,145.97971)(226.5031,145.77529)(226.40877,145.56859)}%
\IJDraw{\Bezier(226.40877,145.56859)(226.31444,145.36189)(226.22646,145.1529)(226.14497,144.94193)}%
\IJDraw{\Bezier(226.14497,144.94193)(226.06348,144.73096)(225.98849,144.51801)(225.92014,144.30339)}%
\IJDraw{\Bezier(225.92014,144.30339)(225.85178,144.08877)(225.79005,143.87249)(225.73507,143.65486)}%
\IJDraw{\Bezier(225.73507,143.65486)(225.68008,143.43723)(225.63184,143.21825)(225.59043,142.99826)}%
\IJDraw{\Bezier(225.59043,142.99826)(225.54901,142.77827)(225.51443,142.55726)(225.48674,142.33557)}%
\IJDraw{\Bezier(225.48674,142.33557)(225.45904,142.11388)(225.43824,141.89152)(225.42437,141.66881)}%
\IJDraw{\Bezier(225.42437,141.66881)(225.4105,141.4461)(225.40356,141.22305)(225.40356,141)}%
\IJDraw{\ArrowLine[arrowpos=0.5,arrowwidth=5,arrowlength=7](230.00598,130.52349)(241.59254,127.4007)}%
\IJDraw{\ArrowLine[arrowpos=0.5,arrowwidth=5,arrowlength=7](254.48748,148.25526)(244.14,154.33225)}%
\IJSetColor{black}%
\IJDraw{\Vertex(225.40356,141){2.6}}%
\IJDraw{\Vertex(35,141){1.6}}%
\IJDraw{\Text(29,129)[r]{\fontsize{11}{14}\selectfont $0$}}%
\end{picture}%
\endgroup
    }
    \caption{Illustration of how analytic continuation in
the $s_{ij}$ channel produces different winding numbers
for individual symbol letters.
The panels show the contours of
$\ell_6=\bar z-z-x$ (left) and $\ell_7=x+z$ (right),
induced by
$\alpha_{ij}^{\pm}(t)=0.3e^{\pm i\pi t}$,
$0\leq t\leq1$.
We use the solution for $(x,z,\bar z)$ specified in
Eq.~\eqref{eq:x-to-eta}, with starting values
$(\eta,\rho,\tilde\rho)=(-0.01,\,0.02,\,0.3)$.
Each closed contour follows the $+$ path and returns
along the $-$ path.
The contour of $\ell_6$ does not encircle the origin,
whereas that of $\ell_7$ winds around it once,
giving $n_{\ell_6}^{ij}=0$ and $n_{\ell_7}^{ij}=1$.}    \label{fig:letter-map}
\end{figure}

To implement the double discontinuity constraints for the one-mass soft anomalous dimension, we first introduce the $x$, $z$, and $\bar z$ variables used to express the one-mass symbol alphabet in Ref.~\cite{Gardi:2025lws}. These letters are related to $\eta_{ijkL}$, $\rho_{ijkL}$, and $\rho_{iLkj}$ by
\begin{equation}
\eta_{ijkL}=x(x+z-\bar z),
\qquad
\rho_{ijkL}=z\bar z,
\qquad
\rho_{iLkj}=(1-z)(1-\bar z)\,,
\label{eq:ka-rationalization}
\end{equation}
where $L$ represents the massive leg. 
Solving for $x$, $z$, and $\bar z$, and using the notation $\eta\equiv\eta_{ijkL}$, $\rho\equiv\rho_{ijkL}$, and $\tilde\rho\equiv\rho_{iLkj}$, we choose the solution
\begin{subequations}
\label{eq:x-to-eta}
\begin{align}
x&=\frac{\sqrt{(1+\rho-\tilde\rho)^2-4\rho}-\sqrt{(1+\rho-\tilde\rho)^2-4\rho+4\eta}}{2},\\
z&=\frac{1+\rho-\tilde\rho-\sqrt{(1+\rho-\tilde\rho)^2-4\rho}}{2},\\
\bar z&=\frac{1+\rho-\tilde\rho+\sqrt{(1+\rho-\tilde\rho)^2-4\rho}}{2}.
\end{align}
\end{subequations}
The two independent square-root signs give four
solutions for $(x,z,\bar z)$.
We adopt the choice displayed in Eq.~\eqref{eq:x-to-eta}
at the initial Euclidean point and analytically
continue this solution along the prescribed paths.
The contours and winding numbers of individual symbol
letters discussed below  (including those illustrated
in Fig.~\ref{fig:letter-map}) refer to this choice.
Different sign choices can change which letters wind
around the origin.
In terms of these variables, the one-mass symbol
alphabet~\cite{Gardi:2025lws} takes the form
\begin{equation}
 \mathcal{A}_{\mathrm{1m}}=\{\ell_1,\ldots,\ell_{12}\}=
    \left\{\begin{gathered}
        z,\ \bar z,\ 1-z,\ 1-\bar z,\ x,\ \bar z-z-x,\\
        x+z,\ \bar z-x,\ 1-z-x,\ 1-\bar z+x,\\
        x+z-z\bar z,\ \bar z-x-z\bar z
    \end{gathered}\right\}.
 \label{eq:ka-one-mass-alphabet}
\end{equation}
The analytic continuation paths determine the full
channel discontinuities. For the multiple polylogarithms
considered here, these generally contain contributions
beyond the term linear in $i\pi$.
Retaining only this linear term in each channel
discontinuity gives a linear combination of the
coaction operators $\delta_\ell$, as described in
Eq.~\eqref{eq: it_discs=coaction}.
This projection provides a simple and efficient way
to implement the constraints at symbol level, and
will serve as the first stage of the bootstrap
computations below.
The same continuation paths also allow the full
discontinuities to be evaluated as differences of
analytically continued boundary values.
We will use this construction to impose the full
constraints at function level in the one-mass
bootstrap in Sec.~\ref{sec: 1mass_function_bootstrap}.

To determine the appropriate combination of $\delta$ operators (cfr. Eq.~\eqref{coaction discontinuity}) for each Mandelstam channel, we now track how each letter behaves under the analytic  continuation paths derived in the previous subsection. For $\eta$, $\rho$, and $\tilde\rho$, these paths were given in Eq.~\eqref{eq:fv-path-map}, with the relevant exponents given in Table~\ref{tab:winding-numbers}; using Eq.~\eqref{eq:x-to-eta}, we can similarly track how the variables $x$, $z$, and $\bar z$ change, and thus determine how each of the letters in $\mathcal{A}_{1\mathrm m}$ behaves along these paths. In turn, knowing the letters behaviour will allow us to explicitly compute the channel discontinuities as a difference of boundary terms in Sec.~\ref{sec: 1mass_function_bootstrap}.

The analytic continuation of these logarithms is
determined by the winding of their arguments around
zero. For each symbol letter $\ell$, we define the
corresponding winding number by
\begin{equation}
    n_\ell^{AB}
    =\frac{1}{2\pi i}
    \oint_{\Gamma_{AB}}d\log\ell,
\end{equation}
where $\Gamma_{AB}$ denotes the closed contour obtained by concatenating the two analytic continuation paths $\chi_{AB}^\pm(t)$; in particular, we follow the $\chi_{AB}^+(t)$ path, and then return along the $\chi_{AB}^-(t)$ path (in the opposite orientation). 
The discontinuity with respect to $s_{AB}$ satisfies
\begin{align}
    \operatorname{Disc}_{s_{AB}}\mathcal{F}^{(3)}
    =-2\pi i\sum_{\ell\in\mathcal{A}_{1\mathrm m}}
    n_\ell^{AB}\delta_\ell\mathcal{F}^{(3)}+ \mathcal{O}\left((i\pi)^{2} \right)\,,
\end{align}
where we have used the $\delta_\ell$ operators that were introduced in Eq.~\eqref{coaction discontinuity}, which project onto the coaction component that involves $\log\ell$ in its first entry.

As an example, consider the channel $s_{ij}$. From Table~\ref{tab:winding-numbers}, we have $(e_{ij},u_{ij},v_{ij})=(1,1,0)$, so that the continuation induces the paths
\begin{equation}
    \bigl(\eta(t),\rho(t),\tilde\rho(t)\bigr)
    =\bigl(\eta\chi_{ij}(t),
    \rho\chi_{ij}(t),\tilde\rho\bigr),
\end{equation}
where $\chi_{ij}(t)$ denotes the concatenation of the paths $\chi^{\pm}_{ij}(t)$ described above.
Substituting these expressions into the definitions of the symbol letters, we can track their winding numbers. For example, the two letters $\ell_6$ and $\ell_7$ become, respectively, 
\begin{subequations}
    \begin{align}
        \ell_6(t)
        &=\frac{
        \sqrt{[1+\rho(t)-\tilde\rho(t)]^2-4\rho(t)}
        +\sqrt{[1+\rho(t)-\tilde\rho(t)]^2
        -4\rho(t)+4\eta(t)}
        }{2},\\
        \ell_7(t)
        &=\frac{
        1+\rho(t)-\tilde\rho(t)
        -\sqrt{[1+\rho(t)-\tilde\rho(t)]^2
        -4\rho(t)+4\eta(t)}
        }{2},
    \end{align}
\end{subequations}
which lead to the paths shown in Fig.~\ref{fig:letter-map}. There, we see that $\ell_6$ does not encircle zero (nor infinity) so $n_{\ell_6}^{ij}=0$, while $\ell_7$ winds once around the origin, giving $n_{\ell_7}^{ij}=1$. Repeating this analysis for all of the other letters in $\mathcal{A}_{1\mathrm m}$, we find that $\ell_1$, $\ell_5$, and $\ell_{11}$ also have winding number $1$, while all other letters have a winding number equal to zero. Thus, at the level of the coaction, the discontinuity in the \(s_{ij}\) channel is associated with the following combination of coaction entries:
\begin{equation}
    \text{Disc}_{s_{ij}} \quad \leftrightarrow \quad
     \delta_{\ell_1}+\delta_{\ell_5}
    +\delta_{\ell_7}+\delta_{\ell_{11}}.
\end{equation}
Applying the same procedure to the other channels, we find the other operators listed in Table~\ref{tab:ka-one-mass-rows}.

\begin{table}[t]
    \centering
    \small
    \renewcommand{\arraystretch}{1.25}
    \begin{tabular}{c c}
        \toprule
        $AB$ & Linear combination of coaction entries %
        \\
        \midrule
        $ij$ &
        $\delta_{\ell_1}+\delta_{\ell_5}+\delta_{\ell_7}
        +\delta_{\ell_{11}}$\\
        $ik$ &
        $-(\delta_{\ell_1}+\delta_{\ell_3}+\delta_{\ell_5}
        +\delta_{\ell_8}+\delta_{\ell_{10}}
        +\delta_{\ell_{11}}+\delta_{\ell_{12}})$\\
        $iL$ & $\delta_{\ell_3}$\\
        $jk$ &
        $\delta_{\ell_3}+\delta_{\ell_5}
        +\delta_{\ell_9}+\delta_{\ell_{12}}$\\
        $jL$ &
        $-(\delta_{\ell_1}+\delta_{\ell_3}
        +\delta_{\ell_5}+\delta_{\ell_6}
        +\sum_{a=7}^{12}\delta_{\ell_a})$\\
        $kL$ & $\delta_{\ell_1}$\\
        \bottomrule
    \end{tabular}
    \caption{The combinations of coaction entries associated with each channel discontinuity, for the chosen analytic continuations. These combinations apply to general logarithmic entries of the coaction, but can be simplified when acting on the physical first entry.}
    \label{tab:ka-one-mass-rows}
\end{table}

Before constructing our double-discontinuity constraints in terms of these operators, it is worth recalling the first-entry condition satisfied by the three-loop soft anomalous dimension with a massive leg. Namely, although the full symbol alphabet contains twelve letters, only the three functions $\log\eta$, $\log\rho$, and $\log\tilde\rho$ are allowed to appear in the first entry~\cite{Gardi:2025lws}. At three loops, we can therefore write the action of $\Delta_{1,4}$ on each of the quadrupole functions as
\begin{equation}
    \Delta_{1,4} \mathcal{F}^{(3)}_{(ij;kL)}
    =\log\eta\otimes\mathcal{F}_{(ij;kL)}^{(3), \eta}
    +\log\rho\otimes\mathcal{F}_{(ij;kL)}^{(3), \rho}
    +\log\tilde\rho\otimes\mathcal{F}^{(3),\tilde\rho}_{(ij;kL)}.
    \label{eq:ka-physical-first-entry}
\end{equation}
Expressing the corresponding constraints in terms of the $\mathcal{A}_{1\mathrm m}$ symbol alphabet, we have that
\begin{subequations}
\label{eq:ka-first-entry-identities}
    \begin{align}
    \delta_{\ell_5}\mathcal{F}^{(3)}_{(ij;kL)}
    =\delta_{\ell_6}\mathcal{F}^{(3)}_{(ij;kL)}&=\mathcal{F}^{(3),\eta}_{(ij;kL)},\\
    \delta_{\ell_1}\mathcal{F}^{(3)}_{(ij;kL)}
    =\delta_{\ell_2}\mathcal{F}^{(3)}_{(ij;kL)}&=\mathcal{F}^{(3),\rho}_{(ij;kL)},\\
    \delta_{\ell_3}\mathcal{F}^{(3)}_{(ij;kL)}
    =\delta_{\ell_4}\mathcal{F}^{(3)}_{(ij;kL)}&=\mathcal{F}^{(3),\tilde\rho}_{(ij;kL)},\\
    \delta_{\ell_a}\mathcal{F}^{(3)}_{(ij;kL)}&=0,
    \qquad a=7,\ldots,12.
    \end{align}
\end{subequations}
These identities follow directly from the physical
first-entry condition in
Eq.~\eqref{eq:ka-physical-first-entry}, upon expressing
$\eta$, $\rho$, and $\tilde\rho$ through
Eq.~\eqref{eq:ka-rationalization}.
Each allowed first-entry logarithm splits into the
logarithms of a pair of letters with equal coefficients,
while the remaining six letters are absent.
Consequently, either member of each pair in the first
three lines of Eq.~\eqref{eq:ka-first-entry-identities}
can be used as a representative when acting on the
physical first entry.

Analogous identities hold for the other independent permutations of the quadrupole function, $\mathcal{F}^{(3)}_{(ik;jL)}$ and $\mathcal{F}^{(3)}_{(iL;jk)}$, with the corresponding coefficient functions. These identities allow us to simplify the combinations of operators in Table~\ref{tab:ka-one-mass-rows} when they act on the first entry of the physical quadrupole functions. When computing a first discontinuity with respect to $s_{iL}$, $s_{ik}$, or $s_{ij}$, we therefore use the simplified operators
\begin{subequations}
\begin{align}
\text{Disc}_{s_{iL}}\mathcal{F}^{(3)}
&\quad\leftrightarrow\quad
\delta_{\ell_3}\mathcal{F}^{(3)},\\
\text{Disc}_{s_{ik}}\mathcal{F}^{(3)}
&\quad\leftrightarrow\quad
-(\delta_{\ell_1}+\delta_{\ell_3}+\delta_{\ell_5})\mathcal{F}^{(3)},\\
\text{Disc}_{s_{ij}}\mathcal{F}^{(3)}
&\quad\leftrightarrow\quad
(\delta_{\ell_1}+\delta_{\ell_5})\mathcal{F}^{(3)},
\end{align}
\end{subequations}
where $\mathcal{F}$ denotes any of the three independent permutations of the one-mass quadrupole function. 

Using the operators in Table~\ref{tab:ka-one-mass-rows} to compute the second discontinuities in Eq.~\eqref{eq:final_constraints}, we obtain the following constraints on the coaction of the one-mass quadrupole functions:
\begin{subequations}
\begin{align}
\label{eq: 1mass-coaction-1}
\left(
    \delta_{\ell_3}+\delta_{\ell_5}
    +\delta_{\ell_9}+\delta_{\ell_{12}}
\right)
\left(\delta_{\ell_3}\right)
\underbrace{\left(
    \mathcal{F}^{(3)}_{(ij;kL)}
    +\mathcal{F}^{(3)}_{(ik;jL)}
\right)}_{{\Fboot_1}} &= 0,\\
\label{eq: 1mass-coaction-2}
\left(
    \delta_{\ell_1}+\delta_{\ell_3}
    +\delta_{\ell_5}+\delta_{\ell_6}
    +\sum_{a=7}^{12}\delta_{\ell_a}
\right)
\left(
    \delta_{\ell_1}+\delta_{\ell_3}+\delta_{\ell_5}
\right)
\underbrace{\left(
    \mathcal{F}^{(3)}_{(iL;jk)}
    -\mathcal{F}^{(3)}_{(ij;kL)}
\right)}_{{\Fboot_2}} &= 0,\\
\label{eq: 1mass-coaction-3}
\left(\delta_{\ell_1}\right)
\left(\delta_{\ell_1}+\delta_{\ell_5}\right)
\underbrace{\left(
    \mathcal{F}^{(3)}_{(iL;jk)}
    +\mathcal{F}^{(3)}_{(ik;jL)}
\right)}_{{-\Fboot_3}} &= 0.
\end{align}
\end{subequations}
Here the functions $\Fboot_i$ are the one-mass specialisations of Eq.~\eqref{eq:F123Defs}, expressed in the rationalising variables $x$, $z$ and $\bar{z}$
introduced above; the displayed coaction operators therefore act directly on these Jacobi-free combinations.
In these equations, the operator on the right acts first,
in accordance with Eq.~\eqref{eq: iterated_discs}. Its action
on the first entry can therefore be simplified using
Eq.~\eqref{eq:ka-first-entry-identities}; no such simplification
is assumed for the second entry. 

For the channels and kinematic regions considered here,
the full double discontinuities must be independent
of the channel order, as required by
Eq.~\eqref{eq: it_cuts=it_discs}.
The final bootstrap result must therefore give the
same result in both orders, although their coaction
representations take different forms.
When reversing the channel order, the first-entry
simplification must be applied to the operator that
now acts first, while the operator acting second
retains its full form from
Table~\ref{tab:ka-one-mass-rows}.
Both orderings are considered in the bootstrap
of Sec.~\ref{sec: 1mass_symbol}.

We also stress that when we impose these constraints at the level of the coaction, they are insensitive to higher-order contributions in $i\pi$ (as discussed in Sec.~\ref{subsec:coaction}). However, the double discontinuity constraint can be applied at function level by computing the full analytic continuation. In order to do so, we can substitute the endpoint values of the affected logarithms inside the Lyndon basis representation of the function, as done at the end of Sec.~\ref{sec: 1mass_function_bootstrap}. It can indeed be checked that these constraints are satisfied by the quadrupole functions computed in Ref.~\cite{Gardi:2025lws}, and they will be used to bootstrap the one-mass result up to one free parameter in section \ref{sec: 1m3ll-bootstrap}.

To finish this section, we consider the massless soft anomalous dimension. As shown in Eq.~\eqref{eq:fv-sequence}, sending the remaining mass $m_L$ to zero corresponds to taking $\eta\to0$. This sends $x\to0$, leaving us with only the variables $z$ and $\bar z$. The symbol alphabet in this limit reduces to
\begin{equation}
    \mathcal{A}_{0\mathrm m}
    =\{\ell_1, \dots, \ell_4\}=\{z,\bar z,1-z,1-\bar z\},
    \label{eq:ka-massless-alphabet}
\end{equation}
as reported in Ref.~\cite{Almelid:2015jia}.  To obtain the explicit double-discontinuity constraints in this case, we take the massless limit of the one-mass coaction relations, expanding the logarithmic entries in terms of the letters in $\mathcal{A}_{0\mathrm m}$ and collecting the corresponding contributions. In doing so, we use the fact that the spurious letter $z-\bar z$ must drop out of the quadrupole functions in this limit, so we drop any instances of $\delta_{z-\bar{z}}$. In this way, we obtain the constraints
\begin{subequations}
\label{eq: 0mass-coaction}
\begin{align}
\label{eq: 0mass-coaction-1}
\delta_{\ell_3}\delta_{\ell_3}
\left(
    \mathcal{F}^{(3)}_{(ij;kl)}
    +\mathcal{F}^{(3)}_{(ik;jl)}
\right) &={\delta_{\ell_3}^{\,2}\Fboot_1}=0,\\
\label{eq: 0mass-coaction-2}
\left(\delta_{\ell_1}+\delta_{\ell_3}\right)
\left(\delta_{\ell_1}+\delta_{\ell_3}\right)
\left(
    \mathcal{F}^{(3)}_{(il;jk)}
    -\mathcal{F}^{(3)}_{(ij;kl)}
\right) &={\left(\delta_{\ell_1}+\delta_{\ell_3}\right)^2\Fboot_2}=0,\\
\label{eq: 0mass-coaction-3}
\delta_{\ell_1}\delta_{\ell_1}
\left(
    \mathcal{F}^{(3)}_{(il;jk)}
    +\mathcal{F}^{(3)}_{(ik;jl)}
\right) &={-\delta_{\ell_1}^{\,2}\Fboot_3}=0.
\end{align}
\end{subequations}
We will illustrate the power of these constraints in the next section, where we bootstrap the corresponding massless quadrupole functions.


\section{The three-loop massless bootstrap}
\label{sec: 4LLbootstrap}
In this section, we explore the constraining power of our new double-discontinuity constraints by revisiting the bootstrap of the three-loop massless soft anomalous dimension. Starting from the ansatz constructed in Ref.~\cite{Almelid:2017qju}, we show that these new constraints eliminate 8 of the 13 initial free coefficients, substantially reducing the amount of information that must be determined by expanding around special kinematic limits. This suggests that these new analytic constraints may similarly prove valuable when bootstrapping configurations involving massive partons, by reducing our reliance on such limits.

\subsection{Building the ansatz}
We begin by constructing an ansatz for the massless soft anomalous dimension. As discussed in Sec.~\ref{sec:struct}, the allowed colour structures, along with rescaling invariance and Bose symmetry, place strong constraints on the form the three-loop contribution can take; for an amplitude involving $n$ partons, this contribution can be written as~\cite{Almelid:2015jia,Almelid:2017qju}

\begin{eqnarray}
\label{eq: 3loopsSADgeneral}
        \Delta^{(3)}_n \, =& \, 
        &- 16\, C \, \sum_{i=1}^n \sum_{\substack{1 \le j < k \le n \\ j,k \ne i}} f_{a b e} f^e_{\phantom{e} c d}
        \{\mathbf{T}_i^a, \mathbf{T}_i^d\} \mathbf{T}_j^b \mathbf{T}_k^c \\ \nonumber
        &  &+ \, 16 \sum_{1 \le i < j < k < l \le n}
        \mathbf{T}_i^a \mathbf{T}_j^b \mathbf{T}_k^c \mathbf{T}_l^d \,\biggl\{ 
        f_{a b e} f^e_{\phantom{e} c d} \bigg[
        F\left(1-\frac{1}{z},1-\frac{1}{\bar{z}}\right)-F\left(\frac{1}{\bar{z}},\frac{1}{z}\right)
        \bigg]\\ \nonumber
        &&+f_{a c e} f^e_{\phantom{e} b d} \bigg[
        F(1-{\bar{z}},1-z)-F(z,\bar{z})
        \bigg] \\ \nonumber
        &&+ f_{a d e} f^e_{\phantom{e} b c}\bigg[
        F\left(1-\frac{z}{z-1},1-\frac{\bar{z}}{\bar{z}-1}\right)-F\left(\frac{\bar{z}}{\bar{z}-1},\frac{z}{z-1}\right)
        \bigg]
        \biggr\}  \,.
\end{eqnarray}
Here, $C$ is a transcendental constant, while the function $F(z,\bar{z})$ encodes the kinematic dependence on conformal cross-ratios; the variables $z$ and $\bar z$ were introduced in Eq.~\eqref{eq:ka-rationalization} with the massless case corresponding to $x=0$. These variables implicitly depend on the particular choice of four partons $i$, $j$, $k$, and $l$ that appear in each term of the sum.

Ref.~\cite{Almelid:2015jia} has shown that when $\bar z = z^*$ (part of the Euclidean region) the function $F(z,\bar z)$ is single valued, and it can be expressed as a linear combination of single-valued harmonic polylogarithms (SVHPLs) of uniform transcendental weight five~\cite{Brown2004,Dixon_2012}, with the symbol alphabet $\mathcal{A}_{0\mathrm m}$ given in Eq.~\eqref{eq:ka-massless-alphabet}. 
Following Ref.~\cite{Almelid:2017qju}, we assume these
properties and use them as the starting point for bootstrapping
$F(z,\bar z)$. 

When working in the kinematic region where $\bar z = z^*$, we make use of a shorthand notation in which we suppress all $\bar z$ dependence, writing $F(z)$ in place of $F(z,\bar z)$; the transformations that act on $z$ (such as permutations of the parton legs) are then understood to act on $\bar z$ as well. When considering a different kinematic region where $\bar z \neq z^*$, $F(z,\bar z)$ must be analytically continued. Since our chosen parametrisation in Eq.~\eqref{eq:ka-rationalization} induces a Galois symmetry $z\leftrightarrow\bar{z}$, our initial ansatz for $F(z)$ should also reflect this property.

Following Ref.~\cite{Almelid:2017qju}, we construct
a pure, uniform-weight-five SVHPL ansatz with the
physical first-entry structure, imposing Bose symmetry
and invariance under $z\leftrightarrow\bar z$.
We also remove combinations that cancel from the
soft anomalous dimension, taking account of functional
identities and the colour Jacobi identity.
This gives the following parametrisation of $F(z)$
and $C$:
\begin{subequations}
\label{eq:massless_ansatz}
    \begin{align}
    \label{eq: ansatzF}
    \begin{split}
        F (z) &= a_1\mathcal{L}_{00000} + a_2\mathcal{L}_{00100} + a_3\mathcal{L}_{10001} + a_4\mathcal{L}_{10101} + a_5 (\mathcal{L}_{01001} + \mathcal{L}_{10010})\\
        &+ a_6 [\mathcal{L}_{00101} +\mathcal{L}_{10100} +2(\mathcal{L}_{00011} +\mathcal{L}_{11000})]+a_7 [\mathcal{L}_{11010} +\mathcal{L}_{01011} \\
        &+3(\mathcal{L}_{00011} +\mathcal{L}_{11000})]+ a_8 \zeta_2\mathcal{L}_{000} + a_9 \zeta_2 (\mathcal{L}_{001} + \mathcal{L}_{100}) + a_{10} \zeta_3 \mathcal{L}_{00} + a_{11} \zeta_2^2\mathcal{L}_0,
    \end{split}\\
    \label{eq:ansatzC}
        C\,\,&=\,\,a_{12}\zeta_5 + a_{13}\zeta_2\zeta_3,
\end{align}
\end{subequations}
where $\mathcal{L}_w$ denotes the SVHPL associated with the word $w$, and $a_1$ through $a_{13}$ represent undetermined rational coefficients. We recall from Sec.~\ref{sec:SADlevel_constr} that the Jacobi identity implies that different choices for $F(z)$ can give the same physical soft anomalous dimension; as such, the ansatz above should be understood as one representative of an equivalence class. While we could carry out the bootstrap directly in terms of the physical combinations of $F(z)$, we prefer to stick with the above parametrisation as it gives rise to more compact expressions.

\subsection{The double discontinuity constraint}
\label{sec:doubleDisc}

We now impose the double discontinuity constraints of Eq.~(\ref{eq:alpha_discs}) on our ansatz. Recall that these constraints only apply to physical combinations built by appropriately combining pairs of the three quadrupole functions 
\begin{subequations}
\begin{align}
    \mathcal{F}^{(3)}_{(ij;kl)}&=16\Bigl[F\left(1-\frac{1}{z},1-\frac{1}{\bar{z}}\right)-F\left(\frac{1}{\bar{z}},\frac{1}{z}\right)\Bigr],\\
    \mathcal{F}^{(3)}_{(ik;jl)}&=16\Bigl[F(1-{\bar{z}},1-z)-F(z,\bar{z})\Bigr],\\
    \mathcal{F}^{(3)}_{(il;jk)}&=16\Bigl[F\left(1-\frac{z}{z-1},1-\frac{\bar{z}}{\bar{z}-1}\right)-F\left(\frac{\bar{z}}{\bar{z}-1},\frac{z}{z-1}\right)\Bigr],
\end{align}
\end{subequations}
rather than to each of them individually. These physical combinations are precisely the Jacobi-free functions $\Fboot_i$ appearing in Eq.~\eqref{eq: 0mass-coaction}. Here the channel functions $\mathcal F^{(3)}$, and hence the Jacobi-free functions $\Fboot_1,\Fboot_2,\Fboot_3$ defined in Eq.~\eqref{eq:F123Defs}, retain the full coefficient normalisation of Eq.~\eqref{eq:Gamma}, with the coupling expanded in $\alpha_s/(4\pi)$. The function $F(z,\bar z)$ and the constant $C$ use the conventions of Refs.~\cite{Almelid:2015jia,Almelid:2017qju,Gardi:2025lws}.
At the level of the coaction, the constraints are given in Eq.~\eqref{eq: 0mass-coaction}. For example, expressed in terms of $F(z,\bar{z})$, the constraint in Eq.~\eqref{eq: 0mass-coaction-1} becomes
\begin{equation}
\delta_{\ell_3}\delta_{\ell_3}
\left[
F\left(1-\frac{1}{z},1-\frac{1}{\bar{z}}\right)-F\left(\frac{1}{\bar{z}},\frac{1}{z}\right)
+F(1-{\bar{z}},1-z)-F(z,\bar{z})
\right]=0.
\end{equation}
Imposing the constraints in Eq.~\eqref{eq: 0mass-coaction} gives a system of linear equations for the coefficients in Eq.~\eqref{eq: ansatzF}, which jointly require 8 of them to vanish:
\begin{equation} \label{eq:massless_double_disc_constraints}
a_1=a_2=a_3=a_5=a_6=a_7=a_8=a_{10}=0.
\end{equation} 
After imposing the double-discontinuity constraints,
the ansatz for $F(z)$ reduces to
\begin{equation}
    F(z)=a_4\mathcal{L}_{10101}
    +a_9\zeta_2(\mathcal{L}_{001}+\mathcal{L}_{100})
    +a_{11}\zeta_2^2\mathcal{L}_0.
    \label{eq:after-dd}
\end{equation}
The two coefficients in $C$ remain unaffected, since these contributions involve no kinematic dependence. We are therefore left with 5 free coefficients, without having imposed any information about the Regge or collinear limits. We now turn to these limits to fix the remaining coefficients.

\subsection{The Regge and collinear constraints}
\label{sec:massless_regge_collinear}

We now supplement the double-discontinuity constraints with
information about the Regge and collinear limits of the massless
soft anomalous dimension.

\medskip
\noindent\textbf{Regge limit.}
We consider the four-particle Regge limit $s\gg -t$.
At leading power in $(-t)/s$, the amplitude develops
logarithmically enhanced terms of the form
\begin{equation}
    \alpha_s^p\log^m\left(\frac{s}{-t}\right),
    \qquad m\leq p,
\end{equation}
which dress the Born amplitude and can be resummed to all
orders in perturbation theory.
Much of the expected behaviour of the soft anomalous
dimension in this limit is already matched by the dipole
formula~\cite{DelDuca:2011ae,DelDuca:2011wkl,
DelDuca:2013ara,DelDuca:2014cya}.
Consequently, at three loops,
the quadrupole terms cannot
contain real contributions with $m\geq1$ or imaginary
contributions with $m\geq2$~\cite{Caron-Huot:2013fea,
Caron-Huot:2017fxr,Almelid:2017qju}.\footnote{
Further constraints on the four-loop soft anomalous
dimension have been obtained by combining explicit
high-energy calculations with an understanding of
Regge factorisation and exponentiation, including
the separation of Regge-pole and Regge-cut
contributions~\cite{Falcioni:2020lvv,
falcioni_scattering_2022}.
}

To obtain the physical Regge expansions, we first analytically continue from the Euclidean region.
Writing
\begin{equation}
    u\equiv\rho_{ijkl}=z\bar z,
    \qquad
    v\equiv\rho_{ilkj}=(1-z)(1-\bar z),
\end{equation}
representatives of the three physical channels are
reached along the paths
\begin{equation}
\begin{array}{ccc}
    \toprule
    \text{Incoming pair}
    &u(\tau)&v(\tau)\\
    \hline
    ij&u\,e^{-2\pi i\tau}&v\\
    ik&u\,e^{+2\pi i\tau}&v\,e^{+2\pi i\tau}\\
    il&u&v\,e^{-2\pi i\tau}\\
    \bottomrule
\end{array}
\label{eq:massless-regge-continuations}
\end{equation}
where in each case $0\leq\tau\leq1$.
The analytic continuation must be retained when taking
the subsequent Regge limit: the final values of the
cross-ratios alone do not specify the branches of the
functions.

For incoming partons $i$ and $j$, imposing four-particle
momentum conservation gives the boundary values
\begin{equation}
    z\to\frac{s_{ij}}{s_{ij}+s_{jk}}-i\varepsilon,
    \qquad
    \bar z\to\frac{s_{ij}}{s_{ij}+s_{jk}}+i\varepsilon.
\end{equation}
Thus, in the limit $s_{ij}\gg -s_{jk}$, both variables
approach $1$ from above on the real line, with the
monodromy inherited from
Eq.~\eqref{eq:massless-regge-continuations}.
Defining
\begin{equation}
    L\equiv\log\frac{s_{ij}}{-s_{jk}},
\end{equation}
the reduced ansatz in Eq.~\eqref{eq:after-dd} gives,
at leading power,
\begin{subequations}
\label{eq:Regge-Re-a4a9a11}
\begin{align}
\mathrm{Re}(\Fboot_1)
&\xrightarrow{s_{ij}\gg -s_{jk}}
16\Bigl[
(-4a_{11}-48a_4+24a_9)\zeta_2^2L
+(48a_4-8a_9)\zeta_2\zeta_3
-8a_4\zeta_5
\Bigr],
\\
\mathrm{Re}(\Fboot_2)
&\xrightarrow{s_{ij}\gg -s_{jk}}
16\Bigl[
(2a_{11}+48a_4-24a_9)\zeta_2^2L
+(-48a_4+4a_9)\zeta_2\zeta_3
+4a_4\zeta_5
\Bigr],
\\
\mathrm{Re}(\Fboot_3)
&\xrightarrow{s_{ij}\gg -s_{jk}}
16\Bigl[
2a_{11}\zeta_2^2L
+4a_9\zeta_2\zeta_3
+4a_4\zeta_5
\Bigr],
\end{align}
\end{subequations}
and
\begin{subequations}
\label{eq:Regge-Im-a4a9a11}
\begin{align}
\frac{\mathrm{Im}(\Fboot_1)}{\pi}
&\xrightarrow{s_{ij}\gg -s_{jk}}
16\Bigl[
(2a_{11}+24a_4-12a_9)\zeta_2^2
\Bigr],
\\
\frac{\mathrm{Im}(\Fboot_2)}{\pi}
&\xrightarrow{s_{ij}\gg -s_{jk}}
16\Bigl[
(4a_4-2a_9)\zeta_2L^2
-8a_4\zeta_3L
+\left(2a_{11}+\frac{64a_4}{5}-24a_9\right)
\zeta_2^2
\Bigr],
\\
\frac{\mathrm{Im}(\Fboot_3)}{\pi}
&\xrightarrow{s_{ij}\gg -s_{jk}}
16\Bigl[
(-4a_4+2a_9)\zeta_2L^2
+8a_4\zeta_3L
-\left(4a_{11}+\frac{184a_4}{5}-36a_9\right)
\zeta_2^2
\Bigr].
\end{align}
\end{subequations}
Requiring the coefficients of $L^m$ to vanish for
$m\geq1$ in the real parts and $m\geq2$ in the imaginary
parts gives
\begin{equation}
\label{eq:massless_regge_constraints}
    a_9=2a_4,
    \qquad
    a_{11}=0.
\end{equation}
This uniquely determines the ansatz for $F(z)$, up to an overall normalisation.

Repeating the analysis in the other two physical channels gives the same conditions.
This follows from the permutation symmetries already incorporated in the ansatz; namely, when the functions and
their analytic continuation paths are transformed
consistently, the Regge expansions are related by
the corresponding permutations.
These other channels therefore provide no additional constraints.

To compare the information supplied by the Regge and double-discontinuity constraints, we return temporarily to the original ansatz in Eq.~\eqref{eq:massless_ansatz}, before imposing either set.
The homogeneous Regge conditions give six independent linear constraints on the eleven coefficients in $F(z)$, leaving five undetermined.\footnote{Ref.~\cite{Almelid:2017qju} also found 6 independent constraints in each individual Regge limit. An error in the analytic continuation in some channels, however, made different channels appear to supply additional independent conditions, leading to a combined count of 8. With the continuation corrected, all channels yield the same 6 constraints. The final result of that paper remains unchanged.}
Each physical Regge channel yields the same six constraints.
The two coefficients in the kinematic constant $C$ remain unconstrained, since $C$ contributes no Regge logarithms.

By comparison, the double-discontinuity constraints impose eight independent conditions on $F(z)$, leaving three coefficients undetermined.
Four of the six Regge conditions are already implied by these constraints, so imposing Regge behaviour supplies precisely the two additional conditions in Eq.~\eqref{eq:massless_regge_constraints}.
Together, the two sets therefore provide ten independent constraints on the eleven coefficients in $F(z)$, fixing its functional form up to an overall normalisation.
Neither set alone achieves this.

\medskip
\noindent\textbf{Collinear limit.}
We next consider the timelike collinear limit, in which
two final-state partons become
collinear~\cite{Becher:2009qa,Dixon:2009ur,
Almelid:2017qju,Becher:2019avh,Duhr:2025cye}.
In this limit, scattering amplitudes factorise into
a universal splitting operator times an on-shell
amplitude with one fewer external parton,
\begin{equation}
    \mathcal{M}_n
    \xrightarrow{p_i\parallel p_j}
    \mathbf{Sp}_{ij}\,\mathcal{M}_{n-1}.
\end{equation}
Strict collinear factorisation requires the splitting
operator to depend only on the kinematics and quantum
numbers of the collinear
partons~\cite{BERENDS1989595,Bern:1995ix,
Kosower:1999xi,Mangano:1990by,Catani:2011st,Feige_2014}.
Matching the infrared factors of the $n$- and
$(n-1)$-parton amplitudes therefore gives a splitting
anomalous dimension that must be independent of the
kinematics and colour charges of the non-collinear
partons~\cite{Becher:2009qa,Dixon:2009ur}.  

For the quadrupole functions considered here, we may
evaluate the collinear limit directly in the Euclidean
region.
Continuing to the physical region in which all coloured
partons are outgoing gives the two-particle invariants
a common phase, which cancels in the conformal
cross-ratios throughout the continuation.
The Euclidean collinear limit therefore yields the
same constraints as the physical timelike limit.

For massless amplitudes, the constraints on the soft
anomalous dimension from multi-particle timelike
collinear limits already follow from those imposed
by two-particle collinear
limits~\cite{Duhr:2025cye}.
It therefore suffices here to consider two final-state
partons becoming collinear, parametrised by
\begin{equation}
    p_i\to\xi P,
    \qquad
    p_j\to(1-\xi)P,
    \qquad
    P^2=0.
\end{equation}
In this limit, we have $s_{ij}\to0$ and
\begin{equation}
    \rho_{ijkl}\to0,
    \qquad
    \rho_{ilkj}\to1,
\end{equation}
corresponding to $z,\bar z\to0$.
As derived in Ref.~\cite{Almelid:2017qju},
strict collinear factorisation implies, in the
normalisation of Eq.~\eqref{eq:F123Defs},
\begin{equation}
\label{eq:collinear}
    \left.(\Fboot_1-\Fboot_2)\right|_{p_i\parallel p_j}=0,
    \qquad
    \left.(\Fboot_1+\Fboot_2)\right|_{p_i\parallel p_j}=128C.
\end{equation}
Here the Jacobi-free functions are the massless
representatives of Eq.~\eqref{eq:F123Defs},
\begin{subequations}
\begin{align}
    \Fboot_1
    &\equiv
    \mathcal{F}^{(3)}_{(ij;kl)}
    +\mathcal{F}^{(3)}_{(ik;jl)},
    \\
    \Fboot_2
    &\equiv
    \mathcal{F}^{(3)}_{(il;jk)}
    -\mathcal{F}^{(3)}_{(ij;kl)}.
    \label{F2comb}
\end{align}
\end{subequations}

The double-discontinuity and Regge constraints automatically enforce
the first condition in Eq.~\eqref{eq:collinear}.
For the second, we find
\begin{equation}
\label{eq:collinear_equation}
    \left.(\Fboot_1+\Fboot_2)\right|_{p_i\parallel p_j}
    =16\left(
    8a_4\zeta_5+16a_4\zeta_2\zeta_3
    \right)
    =128C.
\end{equation}
Using the ansatz for $C$ in Eq.~\eqref{eq:ansatzC},
this gives two further conditions,
\begin{equation}
    a_{12}=a_4,
    \qquad
    a_{13}=2a_4.
\end{equation}
Thus, combining the double-discontinuity, Regge, and collinear
constraints, we are left with
\begin{subequations}
\begin{align}
\label{eq:massless-F-final}
    F(z)
    &=a_4\left[
    \mathcal{L}_{10101}
    +2\zeta_2\left(
    \mathcal{L}_{001}+\mathcal{L}_{100}
    \right)
    \right],
    \\
\label{eq:massless-C-final}
    C
    &=a_4\left(
    \zeta_5+2\zeta_2\zeta_3
    \right).
\end{align}
\end{subequations}
This reproduces the known three-loop massless
result~\cite{Almelid:2015jia} up to an overall
normalisation, whose correct value is $a_4=1$.
(Since all the constraints imposed here are homogeneous,
determining this overall coefficient requires additional input.)

The double-discontinuity constraints are essential to
determining the functional form from the homogeneous
Regge and collinear conditions.
Without them, these conditions leave additional
undetermined coefficients, whose determination requires
input from an explicit amplitude or soft anomalous
dimension calculation.
Such input can, for example, be supplied by matching
non-vanishing terms in the Regge expansion.
With the new analytic constraints, only the overall
normalisation remains to be determined.


\section{The three-loop one-mass bootstrap}
\label{sec: 1m3ll-bootstrap}

Having illustrated how the double-discontinuity constraint significantly strengthens the massless soft anomalous dimension bootstrap, we now turn our attention to the case with one massive parton which was recently determined by an explicit diagrammatic calculation~\cite{Gardi:2025lws}. Our bootstrap begins from the assumption that the quadrupole functions can be expressed in terms of pure multiple polylogarithms of uniform transcendental weight five, which draw on the twelve-letter alphabet $\mathcal{A}_{1\mathrm m}$ from Eq.~\eqref{eq:ka-one-mass-alphabet}. As in the massless case, we exclude kinematics-dependent algebraic prefactors.

For clarity we stress that the purpose of this section is to demonstrate the constraining power of the various properties of the soft anomalous dimension, in particular the new double discontinuity constraint. Hence we make no attempt to independently derive the alphabet or the purity properties of its  function space. These are taken as prior assumptions. 

Since the space of multiple polylogarithms defined by the one-mass symbol alphabet is significantly larger than what we encounter in the massless case, we carry out the bootstrap in two stages.\footnote{The bootstrap procedure we employ broadly  follows those employed in planar $\mathcal{N}=4$ supersymmetric Yang-Mills theory; see for instance~\cite{Dixon:2011pw,Caron-Huot:2016owq,Caron-Huot:2019vjl,Caron-Huot:2020bkp,Arkani-Hamed:2022rwr}.} First, working at the level of the symbol, we impose integrability, the physical first-entry condition, Bose and Galois symmetries, the Jacobi identity, and the known massless and collinear limits. This leaves us with 74 free coefficients, which the double discontinuity constraints reduce to just 9. We then lift our remaining ansatz to function level, including terms proportional to Riemann zeta values. Requiring that this new ansatz also exhibit the expected symmetries, branch cut constraints, and known collinear and massless limits leaves a one-parameter family that contains the known result of Ref.~\cite{Gardi:2025lws}. We fix the final coefficient by matching the leading
logarithm in a simple Euclidean limit to the explicit result.

\subsection{One-mass quadrupole function}
For one massive parton $L$ and three massless partons $i,j,k$, the quadrupole contribution to the three-loop soft anomalous dimension takes the form
\begin{equation}
	\begin{aligned}
	\Gamma^{(3)}\Big|_{\mathrm{quadrupole}}
	={}\mathbf Q_{(ij;kL)}\,\mathcal F_{(ij;kL)}^{(3)}+\mathbf Q_{(ik;jL)}\,\mathcal F_{(ik;jL)}^{(3)}
	+\mathbf Q_{(iL;jk)}\,\mathcal F_{(iL;jk)}^{(3)},
	\end{aligned}
	\label{eq:one-massive-quadrupole}
\end{equation}
where $\Gamma^{(3)}$ multiplies $(\alpha_s/4\pi)^3$, and the colour factors are defined by 
\begin{equation}
	\mathbf Q_{(ij;kl)}=f^{abe}f^{cde}
	\mathbf T_i^a\mathbf T_j^b\mathbf T_k^c\mathbf T_l^d.
\end{equation}
In this configuration, Bose symmetry allows us to relate all three channels by permutations of the massless legs $i,j,k$, keeping the massive leg $L$ fixed. We can therefore express their kinematic coefficients in terms of a single function $\mathcal F_{1,3}^{(3)}$ evaluated at different permutations of its arguments,
\begin{subequations}
\begin{align}
	\mathcal F_{(ij;kL)}^{(3)}&\equiv\mathcal F_{1,3}^{(3)}( r_{ikL},r_{jkL},r_{ijL})\\
	\mathcal F_{(ik;jL)}^{(3)}&\equiv\mathcal F_{1,3}^{(3)}(r_{ijL},r_{jkL} ,r_{ikL})\\
	\mathcal F_{(iL;jk)}^{(3)}&\equiv\mathcal F_{1,3}^{(3)}(r_{ijL} ,r_{ikL},r_{jkL}),
\end{align}
\end{subequations}
where the variables $r_{ijL}$ are the $r$-ratios defined in Eq.~\eqref{eq: r-ratios}, which respect both rescaling and Lorentz invariance.\footnote{To see why there are precisely three independent Lorentz- and
rescaling-invariant variables, we count the velocity components.
The three lightlike velocities each have three independent components,
while the timelike velocity has four. Removing six Lorentz degrees
of freedom and four independent rescalings leaves $3\times3+4-6-4=3$ independent variables (note that we have not imposed momentum conservation).}

The antisymmetry of the colour factors fixes the corresponding
symmetry of the kinematic coefficients. For example, exchanging
the massless legs $i$ and $k$, with $j$ and $L$ held fixed, gives
\begin{equation}
    \mathbf Q_{(ki;jL)}=-\mathbf Q_{(ik;jL)}.
    \label{eq:one-mass-colour-antisymmetry}
\end{equation}
Bose symmetry of the full colour-dressed expression therefore
requires
\begin{equation}
    \mathcal F_{1,3}^{(3)}(r_{ijL},r_{jkL},r_{ikL})
    =-\mathcal F_{1,3}^{(3)}(r_{jkL},r_{ijL},r_{ikL}),
    \label{eq:channel-bose-antisymmetry}
\end{equation}
so $\mathcal F_{1,3}^{(3)}$ is antisymmetric under exchange of
its first two arguments.

Combining the colour Jacobi identity
\begin{equation}
	\mathbf Q_{(ij;kL)}-\mathbf Q_{(ik;jL)}+\mathbf Q_{(iL;jk)}=0
    \label{1Mass-Color-Jacobi}
\end{equation}
with Eq.~\eqref{eq:channel-bose-antisymmetry}, we eliminate
$\mathbf Q_{(iL;jk)}$ and obtain
\begin{equation}
	\begin{aligned}
	\Gamma^{(3)}\Big|_{\mathrm{quadrupole}}
	={}&-\mathbf Q_{(ij;kL)}
	\underbrace{\bigl[\mathcal F_{1,3}^{(3)}(r_{ijL},r_{ikL},r_{jkL})
	-\mathcal F_{1,3}^{(3)}(r_{ikL},r_{jkL},r_{ijL})\bigr]}_{\Fboot_2(r_{ijL},r_{ikL},r_{jkL})}\\
	&-\mathbf Q_{(ik;jL)
    }
	\underbrace{\bigl[\mathcal F_{1,3}^{(3)}(r_{ikL},r_{ijL},r_{jkL})
	-\mathcal F_{1,3}^{(3)}(r_{ijL},r_{jkL}, r_{ikL})\bigr]}_{\Fboot_2(r_{ikL},r_{ijL},r_{jkL})}.
	\end{aligned}
	\label{eq:one-massive-jacobi-free}
\end{equation}
The combination in the first square bracket is precisely the Jacobi-free function $\Fboot_2$ defined in Eq.~\eqref{eq:jacobi-free-F2-definition}, specialised to the one-mass configuration. The second line is the corresponding combination with $j$ and $k$ exchanged; for this mass configuration, this simply interchanges $r_{ijL}$ and $r_{ikL}$. Thus, the two functions in Eq.~\eqref{eq:one-massive-jacobi-free} correspond to the same function evaluated on different arguments. In what follows, we express this function in terms of the variables $x$, $z$, and~$\bar z$ that have been introduced in Ref.~\cite{Gardi:2025lws} to rationalise the alphabet. According to Eq.~(\ref{eq:ka-rationalization}) these are related to the $r$-ratios by
\begin{equation}
\frac{r_{ijL}r_{jkL}}{r_{ikL}}=x(x+z-\bar z),\qquad    \frac{r_{ijL}}{r_{ikL}}=z\bar z,\qquad
    \frac{r_{jkL}}{r_{ikL}}=(1-z)(1-\bar z)\,.
    \label{eq:one-mass-coordinates}
\end{equation}
Implementing this change of variables we define
\begin{equation}
    \widetilde{\Fboot}_2(z,\bar z,x)\equiv \Fboot_2(r_{ijL},r_{ikL},r_{jkL})\,.
\end{equation}
The explicit diagrammatic result for this function,
used for comparison below, is provided in the ancillary
material accompanying Ref.~\cite{Gardi:2025lws} and is
publicly available in the
\href{https://github.com/zzh118/SoftAD}{\texttt{SoftAD} repository}. 
We bootstrap this function in two stages: first constraining
its symbol in Sec.~\ref{sec: 1mass_symbol}, then completing
the bootstrap at function level in
Sec.~\ref{sec: 1mass_function_bootstrap}, including terms
invisible to the symbol.

\subsection{Symbol bootstrap}
\label{sec: 1mass_symbol}

\begin{table}[!t]
	\centering
	\renewcommand{\arraystretch}{1.2}
	\begin{tabularx}{\textwidth}{@{}Xr@{}}
		\toprule
		Symbol-level condition & \shortstack{Remaining free\\coefficients}\\
		\midrule
		Integrability and physical first entry & 642\\
		Bose symmetry & 373\\
		Cyclic Jacobi & 211\\
		Galois symmetry & 117\\
		Massless limit & 100\\
		Collinear limit & 74\\
		Double discontinuity & 9\\
		\bottomrule
	\end{tabularx}
	\caption{Number of free coefficients that remain in the symbol of $\widetilde{\Fboot}_2(z,\bar{z},x)$ after each constraint (as well as the ones above it) are imposed.}
	\label{tab:one-mass-symbol-counts}
\end{table}

We begin by recalling the one-mass symbol alphabet $\mathcal{A}_{1\mathrm m}$ given in Eq.~\eqref{eq:ka-one-mass-alphabet} above,  following  Ref.~\cite{Gardi:2025lws}:
\begin{equation}
    \mathcal{A}_{\mathrm{1m}}=\{\ell_1,\ldots,\ell_{12}\}=
    \left\{\begin{gathered}
        z,\ \bar z,\ 1-z,\ 1-\bar z,\ x,\ \bar z-z-x,\\
        x+z,\ \bar z-x,\ 1-z-x,\ 1-\bar z+x,\\
        x+z-z\bar z,\ \bar z-x-z\bar z
    \end{gathered}\right\}\,.
    \label{eq:one-mass-alphabet}
\end{equation}
We assume that $\widetilde{\Fboot}_2(z,\bar z,x)$ is a pure multiple polylogarithm of uniform transcendental weight five, and thus start from a general symbol ansatz constructed of all words of the form
\begin{equation}
    \ell_{i_1}\otimes \ell_{i_2}\otimes \ell_{i_3}\otimes \ell_{i_4}\otimes \ell_{i_5},
    \qquad i_1,\ldots,i_5\in\{1,\ldots,12\}.
\end{equation}
This means we have $12^5$ free coefficients before imposing any constraints.

\paragraph*{Integrability and first entry.}
We first require that our symbol ansatz is integrable, namely that partial derivatives applied to this ansatz commute. That is, for each pair of neighbouring entries, the terms involving $d\log\ell_{i_k}\wedge d\log\ell_{i_{k+1}}$ must sum to zero, with all remaining entries held fixed. These conditions impose linear relations among the coefficients of our ansatz.

We also impose the physical first-entry condition, which requires the first entry of the symbol to involve only combinations of letters associated with physical branch points on the principal Riemann sheet~\cite{Gaiotto:2011dt,Abreu:2015zaa}.\footnote{The first-entry condition need not by itself exclude all unphysical branch cuts, even at symbol level; see, for example, Ref.~\cite{Dixon:2020bbt}. In the present bootstrap, the full physical branch cut condition is imposed at function level in Sec.~\ref{sec: 1mass_function_bootstrap} below, including constraints on terms proportional to transcendental constants that are invisible to the symbol.} For the one-mass quadrupole function, the allowed first-entry combinations in the $z$, $\bar z$, and $x$ variables are $z\bar z$, $(1-z)(1-\bar z)$, and $x(x+z-\bar z)$~\cite{Gardi:2025lws}. Together with integrability, the first-entry condition reduces our ansatz to $642$ free coefficients, as indicated in the first row of Table~\ref{tab:one-mass-symbol-counts}.

\paragraph*{Bose symmetry.}
We use the colour and kinematic antisymmetry relations in
Eqs.~\eqref{eq:one-mass-colour-antisymmetry} and
\eqref{eq:channel-bose-antisymmetry} to determine how the
Jacobi-free function $\Fboot_2$ transforms under $i\leftrightarrow k$.
Consider its definition in the first line of
Eq.~\eqref{eq:one-massive-jacobi-free}.
Under this exchange, the first term in the square brackets
becomes the negative of the second, while the second becomes
the negative of the first. Thus $\Fboot_2$ is symmetric under
this exchange:
\begin{equation}
    \Fboot_2(r_{ijL},r_{ikL},r_{jkL})=\Fboot_2(r_{jkL},r_{ikL},r_{ijL}).
    \label{eq:F2-bose}
\end{equation}
According to Eq.~(\ref{eq:one-mass-coordinates}),  in the $z$, $\bar{z}$, $x$ variables this permutation corresponds to $(z,\bar z,x)\mapsto(1-\bar z,1-z,x)$. Altogether, this gives us the constraint
\begin{equation}
    \widetilde{\Fboot}_2(z,\bar z,x)=\widetilde{\Fboot}_2(1-\bar z,1-z,x).
    \label{eq:F2-bose-coordinates}
\end{equation}
Imposing this symmetry on the symbol ansatz leaves $373$ free coefficients.

\paragraph*{Jacobi relation.}
Next, we impose the cyclic relation between the three functions introduced in Eq.~\eqref{eq:F123Defs}, namely
\begin{equation}
    \Fboot_1+\Fboot_2+\Fboot_3=0.
    \label{eq:F123-cyclic}
\end{equation}
In the one-mass configuration, the three functions are related by cyclic permutations of the massless legs, allowing us to express the Jacobi relation of Eq.~(\ref{eq:F123-cyclic}) entirely in terms of~$\widetilde{\Fboot}_2$:
\begin{equation}
\underbrace{\mathsf P_{(ikj)}\widetilde{\Fboot}_2}_{\widetilde{\Fboot}_1}
+\widetilde{\Fboot}_2
+\underbrace{\mathsf P_{(ikj)}^2\widetilde{\Fboot}_2}_{\widetilde{\Fboot}_3}=0,
\label{eq:F2-cyclic}
\end{equation}
where $\mathsf P_{(ikj)}$ cyclically permutes the legs $i,j ,k$, namely  
$\{
i\mapsto k,\quad k\mapsto j,\quad j\mapsto i\}$ and thus 
\begin{equation}
\label{eq:Pikjdef}
\mathsf P_{(ikj)}:\quad (r_{ijL},r_{ikL},r_{jkL})
\longmapsto (r_{ikL},r_{jkL},r_{ijL}).
\end{equation}
In the variables appearing in the alphabet, this mapping becomes
\begin{equation}
\mathsf P_{(ikj)}:\quad (z,\bar z,x)
\longmapsto
\left(\frac{1}{1-z},\frac{1}{1-\bar z},
\frac{x}{(1-z)(1-\bar z)}\right),
\label{eq:jacobi-xzzb}
\end{equation}
while 
\begin{equation}
\mathsf P_{(ikj)}^2:\quad (z,\bar z,x)
\longmapsto
\left(1-\frac{1}{z},1-\frac{1}{\bar z},
\frac{x}{z\bar z}\right).
\label{eq:jacobi-xzzbTrans2}
\end{equation}
Requiring that the cyclic sum in Eq.~\eqref{eq:F2-cyclic} vanishes at symbol level reduces the number of free coefficients to $211$.

\paragraph*{Galois symmetry.}
When expressed in terms of $r_{ijL}$, $r_{ikL}$, and $r_{jkL}$, the symbol alphabet involves two square roots:
\begin{equation}
	\begin{gathered}
		x=\frac{-\sqrt{\Delta}-\sqrt{\Delta+4r_{ijL}r_{ikL}r_{jkL}}}{2r_{ikL}},\\
		z=\frac{r_{ijL}+r_{ikL}-r_{jkL}+\sqrt{\Delta}}{2r_{ikL}},\qquad
		\bar z=\frac{r_{ijL}+r_{ikL}-r_{jkL}-\sqrt{\Delta}}{2r_{ikL}},
	\end{gathered}
	\label{eq:one-mass-inverse-coordinates}
\end{equation}
where
\begin{equation}
	\Delta=r_{ijL}^2+r_{ikL}^2+r_{jkL}^2-2r_{ijL}r_{ikL}-2r_{ijL}r_{jkL}-2r_{ikL}r_{jkL}.
\end{equation}
The quantities under the two square roots admit a
geometric interpretation. 
In the rest frame of the massive parton $L$, write the
three massless directions as $(1,\boldsymbol n_a)$,
with $\boldsymbol n_a^2=1$. Then
\begin{equation}
    r_{abL}=\frac{1-\boldsymbol n_a\cdot\boldsymbol n_b}{2}.
\end{equation}
The Källén function $\Delta$ and the quantity under
the second square root can be expressed as
\begin{align}
    \Delta
    &=-\frac14\left|
    (\boldsymbol n_i-\boldsymbol n_j)
    \times(\boldsymbol n_i-\boldsymbol n_k)
    \right|^2,
    \\
    \Delta+4r_{ijL}r_{ikL}r_{jkL}
    &=-\frac14\left[
    \boldsymbol n_i\cdot
    (\boldsymbol n_j\times\boldsymbol n_k)
    \right]^2.
\end{align}
The first expression is minus the squared area of the
triangle with vertices $\boldsymbol n_i$,
$\boldsymbol n_j$, and $\boldsymbol n_k$.
The second is minus one quarter of the squared volume
of the parallelepiped spanned by these three vectors.

The quadrupole function is naturally a function of the
three ratios $r_{ijL}$, $r_{ikL}$, and $r_{jkL}$.
The rationalising variables $(x,z,\bar z)$ provide
multiple representations of the same values of these
ratios, related by changes in the signs of the two
square roots. The choice of representation cannot
affect the quadrupole function on a fixed analytic
branch. 
We therefore require invariance under the corresponding
Galois transformations, which are obeyed by the explicit
one-mass result~\cite{Gardi:2025lws}:
\begin{subequations}
\begin{align}
	\Fboot_2(x,z,\bar z)&=\Fboot_2(\bar z-z-x,z,\bar z),
	\label{eq:F2-first-galois}\\
	\Fboot_2(x,z,\bar z)&=\Fboot_2(-x,\bar z,z)\,.
	\label{eq:F2-second-galois}
\end{align}
\end{subequations}
The first of these follows from flipping the sign of
$\sqrt{\Delta+4r_{ijL}r_{ikL}r_{jkL}}$, and the second from simultaneously flipping the
signs of both square roots. Imposing these two conditions leaves $117$ free coefficients in the symbol.
These identities compare coordinate representations of
the same analytic branch. Analytic continuation around
a singularity can instead change that branch; for
discussions of this analytic structure and its geometric
interpretation, see
Refs.~\cite{Hannesdottir:2022xki,Hannesdottir:2024hke}.

\paragraph*{Lightlike limit.}
Next, we impose boundary conditions in the lightlike limit. Namely, as
$\beta_L^2\to0$, the function should recover the known massless result.
This limit leaves the ratios $r_{ijL}/r_{ikL}$ and $r_{jkL}/r_{ikL}$ fixed, while $r_{ijL}r_{jkL}/r_{ikL}\propto\beta_L^2\to0$; the variables from Eq.~\eqref{eq:one-mass-coordinates} realise this limit as $x\to0$ for fixed values of $z$ and $\bar z$. Thus, we require that in the lightlike limit the $\Fboot_2$ combination is given by (\ref{F2comb}) 
\begin{align}
	\lim_{x\to0}\widetilde{\Fboot}_2(z,\bar z,x)=16\Bigg[&
	F\!\left(\frac1{\bar z},\frac1z\right)
	-F\!\left(1-\frac1z,1-\frac1{\bar z}\right)\notag\\
	&+F\!\left(\frac1{1-z},\frac1{1-\bar z}\right)
	-F\!\left(\frac{\bar z}{\bar z-1},\frac z{z-1}\right)
	\Bigg],
	\label{eq:F2-massless-boundary}
\end{align}
where $F(z,\bar z)$ is the function from Eq.~\eqref{eq:massless-F-final}, with $a_4=1$.

The same boundary condition follows from strict collinear
factorisation in the timelike triple-collinear limit
$\smash{p_i\parallel p_j\parallel p_k}$~\cite{Duhr:2025cye}.
In this limit, the three ratios $r_{ijL}$, $r_{ikL}$,
and $r_{jkL}$ vanish at fixed relative ratios, reproducing
the kinematic limit reached when the massive parton
becomes lightlike.

Considering our alphabet  $\mathcal{A}_{1\mathrm m}$ in the $x\to0$ limit with fixed values of $z$ and $\bar z$ we obtain
\begin{equation}
	\{\ell_1,\ldots,\ell_{12}\}\longrightarrow
	\left\{\begin{gathered}
		z,\ \bar z,\ 1-z,\ 1-\bar z,\ x,\ \bar z-z,\\
		z,\ \bar z,\ 1-z,\ 1-\bar z,\ z(1-\bar z),\ \bar z(1-z)
	\end{gathered}\right\}.
	\label{eq:massless-leading-alphabet}
\end{equation}
Notice that this limit contains two spurious letters, namely $x$ and $\bar z - z$, which do not occur in the lightlike result. All instances of these letters must cancel in the strict $x\to0$ limit of $\widetilde{\Fboot}_2(z,\bar z,x)$. Imposing Eq.~\eqref{eq:F2-massless-boundary} at symbol level leaves $100$ free coefficients.

\paragraph*{Collinear limit.}
Next, we impose boundary conditions in the limit where the two velocities $\beta_j$ and $\beta_k$ become (timelike) collinear, corresponding to $(r_{ijL},r_{ikL},r_{jkL})\to(r,r,0)$~\cite{Liu:2022elt,Gardi:2025lws}. 
In the $z$, $\bar z$, and $x$ variables, this limit
can be approached using an equivalent form of the
parametrisation given in the Supplemental Material
of Ref.~\cite{Gardi:2025lws}:
\begin{equation}
    x=\tau,\qquad z=1-y\tau,\qquad
    \bar z=1-\bar y\tau,\qquad \tau\to0,
    \label{eq:F2-collinear-path}
\end{equation}
for fixed values of $y$ and $\bar y$. 
Strict collinear factorisation with a massive spectator
relates this limit to the same constant $C$ that appears
in the massless soft anomalous
dimension~\cite{Liu:2022elt,Duhr:2025cye}.
For the Jacobi-free combination considered here, this
requires $\widetilde{\Fboot}_2\to64C$, consistently with
the explicit result of Ref.~\cite{Gardi:2025lws}.
At symbol level, this condition simply requires the
limiting symbol to vanish; the value of the constant
is imposed at function level in
Sec.~\ref{sec: 1mass_function_bootstrap}.

Imposing this requirement at leading order in $\tau$ reduces the number of free coefficients to $74$. The remaining two-particle collinear limits are related
to this one by Bose symmetry and the cyclic Jacobi
relation, and therefore impose no additional constraints.

\paragraph*{Double-discontinuity constraint.}
Finally, we impose the double-discontinuity constraint
\begin{equation}
    \operatorname{Disc}_{s_{ik}}\operatorname{Disc}_{s_{jL}-m_L^2}
    \widetilde{\Fboot}_2
    =\operatorname{Disc}_{s_{jL}-m_L^2}\operatorname{Disc}_{s_{ik}}\widetilde{\Fboot}_2=0,
    \label{eq:F2-double-discontinuity}
\end{equation}
where $s_{ij}=(p_i+p_j)^2$ and $m_L^2=p_L^2$. The analogous constraints on $\Fboot_1$ and $\Fboot_3$ follow
from those on $\Fboot_2$ by cyclically relabelling
the massless legs. Indeed,
$\Fboot_1=\mathsf P_{(ikj)}\Fboot_2$ and
$\Fboot_3=\mathsf P_{(ikj)}^2\Fboot_2$, as used in
Eq.~\eqref{eq:F2-cyclic}. These permutations map the
channel pair $(s_{ik},s_{jL}-m_L^2)$ to
$(s_{jk},s_{iL}-m_L^2)$ and
$(s_{ij},s_{kL}-m_L^2)$, respectively.
Relabelling the continuation paths together with
the channels therefore recovers the constraints
on all three functions. This relabelling preserves
the order of the two discontinuities.

To implement the constraint on $\Fboot_2$ at symbol
level, we use Eq.~\eqref{eq: 1mass-coaction-2}
and its counterpart with the channel order reversed,
applying the first-entry simplification to the
operator that acts first in each case.
On the integrable symbol ansatz with the physical
first-entry condition, we verify that both orderings
give identical constraints, despite their different
coaction representations. Reversing the order therefore
imposes no additional conditions.
The double-discontinuity constraint fixes $65$
coefficients, reducing the number of free coefficients
from $74$ to $9$. This reduction demonstrates the
constraining power of the new analytic constraints.
The bootstrap must still be completed at function
level, including terms invisible to the symbol.

\subsection{Function bootstrap}
\label{sec: 1mass_function_bootstrap}

We next lift the 9-parameter ansatz that we derived in Sec.~\ref{sec: 1mass_symbol} to a function. This involves adding contributions that are proportional to transcendental constants, which are annihilated by the symbol; doing so introduces an additional 793 free coefficients in the ansatz (bringing the total to 802). However, we find that imposing all the constraints in
Table~\ref{tab:function-counts} fixes nearly all of these free coefficients, and leaves just 1 undetermined. We note that the double-discontinuity constraint does not provide additional constraints at this stage, in part because we have already imposed these constraints at symbol level, where it fixed $65$ free coefficients. 
Let us now discuss the function-level bootstrap step by step.
\begin{table}[!t]
\centering
\renewcommand{\arraystretch}{1.15}
\begin{tabularx}{\textwidth}{@{}Xr@{}}
\toprule
Function-level constraint & \shortstack{Remaining free\\coefficients}\\
\midrule
function-level ansatz & 802\\
Bose symmetry & 450\\
Massless limit & 379\\
Cyclic Jacobi & 210\\
Collinear limit & 130\\
Galois symmetries & 44\\
Branch cuts on principal Riemann sheet& 1\\
Double discontinuity & 1\\
Euclidean boundary matching & 0\\
\bottomrule
\end{tabularx}
\caption{Number of free coefficients in the
function ansatz for $\widetilde{\Fboot}_2$ after each constraint (as well as the ones above it) are imposed.}
\label{tab:function-counts}
\end{table}

\paragraph*{Euclidean region.}
We work in the Euclidean region where 
\begin{gather}
r_{ijL},\, r_{ikL},\, r_{jkL}<0.
\label{eq:one-mass-euclidean-invariants}
\end{gather}
In $(x,z,\bar z)$ space, this Euclidean region has seven distinct
domains (see the supplementary material of Ref.~\cite{Gardi:2025lws}).
We focus on $R_1^E$, defined by
\begin{equation}
 R_1^E:\qquad 0<x<\bar z-z\qquad 0<z<\bar z<1,
 \label{eq:euclidean-R1E}
\end{equation}
and impose all of the constraints that follow in this domain, unless stated otherwise.

\paragraph*{Polylogarithmic ansatz.}
As at symbol level, we assume that $\widetilde{\Fboot}_2(z,\bar z,x)$ is a pure multiple-polylogarithmic function of uniform transcendental weight five, with no rational or algebraic prefactors with kinematic dependence. By inspecting various limits of the symbol alphabet, one can convince oneself that only Riemann zeta values should be expected to appear in this function. We therefore formulate our function-level ansatz as
\begin{equation}
    \widetilde{\Fboot}_2(z,\bar z,x)
    =P_5+\zeta_2 P_3+\zeta_3 P_2+\zeta_4 P_1
    +c_1\zeta_5+c_2\zeta_2\zeta_3,
    \label{eq:function-ansatz}
\end{equation}
where each $P_n$ is a linear combination of weight-$n$ multiple polylogarithms that draw on the twelve-letter alphabet $\mathcal{A}_{1\mathrm m}$ from Eq.~\eqref{eq:one-mass-alphabet}. We additionally restrict the symbol of $P_5$ to the span of the 9-parameter ansatz bootstrapped in the last subsection. Note that we do not include odd powers of $i\pi$, as the soft anomalous dimension is manifestly real in the Euclidean region; terms proportional to $i \pi$ will be generated when one analytically continues out of the Euclidean region.

\paragraph*{Lower-weight function spaces.}
We construct $P_1$, $P_2$, and $P_3$ as the spaces of multiple polylogarithms that draw on the alphabet $\mathcal{A}_{1\mathrm m}$ at the corresponding transcendental weight. The dimensions of these function spaces are as follows:
\begin{equation}
    \begin{array}{c | c c c}
    \toprule
        \text{Function space} & P_1 & P_2 & P_3 \\ \hline
        \text{Number of free coefficients} & 12 & 98 & 681\\
        \bottomrule
    \end{array}
    \label{eq:lower-weight-dimensions}
\end{equation}
Note that, at this stage, the functions in these spaces may give rise to unphysical branch cuts; we require such branch cuts to be absent as a separate constraint below. Since $P_1$, $P_2$, and $P_3$ each multiply a unique constant in Eq.~\eqref{eq:function-ansatz}, the total number of new function-level coefficients can be read off of the sum of the entries in~\eqref{eq:lower-weight-dimensions} after one adds 2 parameters to account for $\zeta_5$ and $\zeta_2 \zeta_3$.

\paragraph*{Bose symmetry.}
We recall from Eq.~\eqref{eq:F2-bose-coordinates} that Bose symmetry under the exchange $i\leftrightarrow k$ requires
\begin{equation}
    \widetilde{\Fboot}_2(z,\bar z,x)
    =\widetilde{\Fboot}_2(1-\bar z,1-z,x).
    \label{eq:F2-bose-function}
\end{equation}
Note that this transformation preserves the Euclidean domain $R_1^E$ defined in Eq.~\eqref{eq:euclidean-R1E}. To impose this symmetry on the function ansatz, we make use of \texttt{HyperInt}~\cite{Panzer_2015} to express the transformed function $\widetilde{\Fboot}_2(1-\bar z,1-z,x)$ in the same fibration basis as $\widetilde{\Fboot}_2(z,\bar z,x)$. This requires careful treatment of analytic continuation, since individual GPLs may acquire imaginary contributions even though the complete function is real in the Euclidean region. 
To specify the branches of the individual GPLs when
rewriting them in a common fibration basis, we choose
the prescription\footnote{For the treatment of branches in
\texttt{HyperInt}, see Ref.~\cite{Panzer_2015}.
Our prescription corresponds to the choice
$(\delta_x,\delta_z,\delta_{\bar z})=(+1,+1,-1)$.
The prescriptions for transformed GPL arguments
are inherited from this choice and must be
propagated through the changes of variables.}
\begin{equation}
    x\longrightarrow x+i\epsilon,\qquad
    z\longrightarrow z+i\epsilon,\qquad
    \bar z\longrightarrow\bar z-i\epsilon,\qquad \epsilon>0,
    \label{eq:cyclic-continuation-prescription}
\end{equation}
This prescription is preserved by the Bose transformation
$(x,z,\bar z)\mapsto(x,1-\bar z,1-z)$ and is used
consistently in the subsequent analytic continuations.

Having expressed both functions in a common fibration basis, we expand their difference in a basis of linearly independent GPL products and require each coefficient to vanish. This yields $352$ independent linear constraints on the ansatz in Eq.~\eqref{eq:function-ansatz}, leaving $450$ free parameters.

\paragraph*{Massless limit.}
We next impose the massless boundary condition from Eq.~\eqref{eq:F2-massless-boundary} by taking $x\to0^+$ for fixed values of $z$ and $\bar z$. By going to a fibration basis, we can push all dependence on $x$ into the upper integration endpoint of the GPLs in our ansatz (rather than their indices). Moreover, by putting all the indices into a Lyndon ordering, we can ensure that all of the GPLs that depend on $x$ take the form~\cite{Dixon:2013eka}
\begin{equation}
    G(\cdots,a;x) \qquad \text{or} \qquad G(0;x)
    =\log x,
\end{equation}
where $a \neq 0$ is some rational function of $z$ and $\bar z$. At leading power in the $x \to 0$ limit, all of the former functions vanish, and we are just left with large logs (multiplying GPLs that depend on $z$ and $\bar z$). Since the massless limit is finite, we require all positive powers of $\log x$ to drop out in the $x \to 0$ limit; we additionally require that the remaining finite function reproduces the known massless result. Together, these conditions impose $71$ independent linear constraints, reducing the number of free coefficients to $379$.

\paragraph*{Cyclic Jacobi.}
We next impose the cyclic Jacobi identity from Eq.~\eqref{eq:F2-cyclic} at function level. Although these permutations preserve the Euclidean conditions $r_{ijL},r_{ikL},r_{jkL}<0$, they do not preserve our reference domain $R_1^E$. Instead, $\mathsf P_{(ikj)}$
of Eq.~(\ref{eq:Pikjdef}) maps $R_1^E$ to the domain $R_2^E$, defined in the notation of Ref.~\cite{Gardi:2025lws} by
\begin{equation}
    R_2^E:\qquad 1<z<\bar z,\qquad 0<x<\bar z-z,
    \label{eq:euclidean-R2E}
\end{equation}
while $\mathsf P_{(ikj)}^2$ of Eq.~(\ref{eq:jacobi-xzzbTrans2}) maps it to $R_3^E$, defined by
\begin{equation}
    R_3^E:\qquad z<\bar z<0,\qquad 0<x<\bar z-z.
    \label{eq:euclidean-R3E}
\end{equation}
To impose the Jacobi identity, we therefore analytically continue all functions back to the region $R_1^E$. Requiring the sum in Eq.~\eqref{eq:F2-cyclic} to vanish yields $169$ additional independent linear constraints, reducing the number of free coefficients from $379$ to $210$.

\paragraph*{Collinear limit.}
We next impose the collinear boundary condition at function level. To approach this limit within our reference domain $R_1^E$, we take
\begin{equation}
    x=wt,\qquad z=1-t,\qquad \bar z=1-yt,\qquad t\to0^+,
    \label{eq:F2-function-collinear-path}
\end{equation}
with $0<y<1$ and $0<w<1-y$ held fixed. We then express the ansatz in a fibration basis so that all dependence on $t$ resides in the upper integration endpoint of GPLs. As in the massless limit, we use a Lyndon basis to make all powers of $\log t$ explicit. To impose the two-particle  collinear-limit constraints according to  Refs.~\cite{Liu:2022elt,Gardi:2025lws,Duhr:2025cye}
we require that all these large logarithms of $t$ drop out, and that the finite part of $\widetilde{\Fboot}_2(z,\bar z,x)$ approaches $64 C=64(2\zeta_2\zeta_3+\zeta_5)$. This imposes $80$ independent linear constraints,  reducing the number of free coefficients to $130$.

\paragraph*{Galois symmetries.}
As we saw in Eq.~\eqref{eq:one-mass-inverse-coordinates}, the variables $z$, $\bar z$, and $x$ involve square roots. We next impose the requirement that the $\widetilde{\Fboot}_2$  function is Galois symmetric under flipping the sign of either of these square roots. To apply these symmetries, we apply each transformation to our ansatz and express the resulting expression in a common fibration basis with the original function, and require the difference between these functions to vanish. As in the case of Bose symmetry, we must retain all $i\pi$ and zeta terms generated by analytic continuation. Imposing the reflection $x\mapsto\bar z-z-x$ in Eq.~\eqref{eq:F2-first-galois}, with $z$ and $\bar z$ fixed, yields $71$ independent constraints, reducing the number of free coefficients from $130$ to $59$. Subsequently imposing the transformation $(z,\bar z,x)\mapsto(\bar z,z,-x)$ in Eq.~\eqref{eq:F2-second-galois} yields another $15$ independent constraints, leaving $44$ free coefficients.

\paragraph*{Branch cut conditions.}
We next require that $\widetilde{\Fboot}_2(z,\bar z,x)$ has no unphysical branch cuts on the principal Riemann sheet. The only allowed branch points on this sheet occur at $z\bar z=0$, $(1-z)(1-\bar z)=0$, and $x(\bar z-z-x)=0$, corresponding to zeros of the Mandelstam invariants. The physical first-entry condition was imposed at symbol level in Sec.~\ref{sec: 1mass_symbol}; as explained there, it does not by itself enforce the full physical branch cut condition. At function level, additional constraints are needed because terms proportional to transcendental constants can introduce unphysical branch cuts that are invisible to the symbol.

To ensure that no further branch points arise, we go into a basis that exposes each disallowed singularity. For instance, consider the letter $\ell_7=x+z$. If we first introduce a new variable $u=x+z$, we can express our ansatz in a fibration and Lyndon basis such that all dependence on $u$ resides in the upper integration endpoint of GPLs. This allows us to require that linear combinations of terms that give rise to large logs as $u \to 0$ not be present, after analytically continuing back to $R_1^E$ (using the prescription from Eq.~\eqref{eq:cyclic-continuation-prescription}). Already, this condition at $x+z=0$ yields $43$ independent linear constraints, which reduces the number of free coefficients down to $1$. We find that all of the other branch cut constraints are automatically satisfied once these 43 constraints are imposed.

\paragraph*{Double discontinuity.}
Finally, we check the double-discontinuity constraints at function level. It is convenient to work with $\Fboot_3$ from Eq.~\eqref{eq:F123Defs}, which can be obtained from our ansatz for $\widetilde{\Fboot}_2$ using the cyclic permutation $\mathsf P_{(ikj)}^2$ in Eq.~\eqref{eq:jacobi-xzzbTrans2}. As described by the coproduct relation in Eq.~\eqref{eq: 1mass-coaction-3}, the discontinuity in $s_{ij}$ induces paths around $\ell_1=z=0$ and $\ell_5=x=0$, while the subsequent discontinuity in $s_{kL}$ induces a path around $\ell_1=z=0$. Here we use the fact that the letters $\ell_7,\ldots,\ell_{12}$ do not give rise to branch points on the principal Riemann sheet. To separate these discontinuities, we parametrise the kinematics by
\begin{equation}
    x=t,\qquad z=st,
    \label{eq:F2-function-dd-coordinates}
\end{equation}
for fixed $\bar z$, and denote the cyclically transformed function by $\widetilde{\Fboot}_3(z,\bar z,x)\equiv\mathsf P_{(ikj)}^2\widetilde{\Fboot}_2(z,\bar z,x)$. The two orders of the double-discontinuity constraint then become
\begin{equation}
    \begin{aligned}
    \operatorname{Disc}_{s=0}\operatorname{Disc}_{t=0}
    \widetilde{\Fboot}_3(st,\bar z,t)&=0,\\
    \operatorname{Disc}_{t=0}\operatorname{Disc}_{s=0}
    \widetilde{\Fboot}_3(st,\bar z,t)&=0.
    \end{aligned}
    \label{eq:st-DoubleDisc-F3}
\end{equation}
The corresponding constraints on $\Fboot_1$ and $\Fboot_2$ follow by cyclic relabelling.\footnote{Indeed, the Bose and cyclic relations already imposed at function level, Eqs.~\eqref{eq:F2-bose-function} and \eqref{eq:F2-cyclic}, imply $\mathsf P_{(ikj)}\widetilde{\Fboot}_3=\widetilde{\Fboot}_2$ and $\mathsf P_{(ikj)}^2\widetilde{\Fboot}_3=\widetilde{\Fboot}_1$. These permutations map the channel pair $(s_{ij},s_{kL})$ to $(s_{ik},s_{jL})$ and $(s_{jk},s_{iL})$, respectively. Relabelling the continuation paths together with the channels therefore recovers all three constraints in Eq.~\eqref{eq:final_constraints}, in both orders.} 
To impose these constraints, we express the transformed ansatz in a fibration and Lyndon basis that makes all powers of $\log t$ explicit. We evaluate the first discontinuity by shifting $\log t\to\log t+2\pi i$ and subtracting the original expression. We then apply the same procedure to $\log s$ and require the coefficients of all surviving GPL products to vanish. We also evaluate the two discontinuities in the opposite order. Both double discontinuities vanish for the entire one-parameter family, so no additional coefficients are fixed at this stage.

\paragraph*{Euclidean boundary matching.}
To fix the remaining coefficient, we consider a particularly simple
boundary limit within the Euclidean domain: we set $z=t$ and
$\bar z=1-t$, keeping $x\in(0,1)$ fixed as $t\to0^+$.
This path lies in $R_1^E$ for $0<t<(1-x)/2$.
It approaches the simultaneous zeros of
$\beta_i\cdot\beta_L$ and $\beta_k\cdot\beta_L$, with the
massless pairwise scalar products held finite and nonzero.
Such a configuration requires complex external momenta for a
timelike massive leg, and hence does not represent a physical
scattering threshold. Expanding the explicit result of
Ref.~\cite{Gardi:2025lws} along this path, we find
\begin{equation}
    \widetilde{\Fboot}_2(t,1-t,x)=A(x)\log(1/t)+\mathcal O(1),
\end{equation}
With \(\ell_0=\log x\) and \(\ell_1=\log(1-x)\), the coefficient becomes
\[
\begin{aligned}
A(x)=32\,\Bigg\{&
4\bigl[\operatorname{Li}_4(x)+\operatorname{Li}_4(1-x)\bigr]\\
&-2(\ell_0+\ell_1)
 \bigl[\operatorname{Li}_3(x)+\operatorname{Li}_3(1-x)\bigr]\\
&+\frac{(\ell_0-\ell_1)^4}{24}-\ell_0^2\ell_1^2
+\frac{5\pi^2}{12}(\ell_0-\ell_1)^2
+\frac{\pi^2}{3}\ell_0\ell_1+\frac{131\pi^4}{360}
\Bigg\}.
\end{aligned}
\]
In particular, all powers $\log^k(1/t)$ with $k\geq2$ cancel.
The corresponding expansion of our one-parameter ansatz is
\begin{equation}
    \widetilde{\Fboot}_2(t,1-t,x)
    =\frac{c+96}{72}\log^5(1/t)
    +\mathcal O\!\left(\log^4(1/t)\right),
    \label{eq:one-mass-fifth-log}
\end{equation}
where $c$ is the remaining free coefficient. Matching the vanishing fifth-logarithm coefficient fixes $c=-96$.
Although this final matching uses only a single leading-logarithmic
coefficient, the resulting function reproduces the full analytic
result of Ref.~\cite{Gardi:2025lws}, including all transcendental
constant terms.



\section{Conclusions}
\label{sec: concl}

In this paper, we have derived new analytic constraints
on the multileg soft anomalous dimension at three loops by analysing
how the topology of the contributing webs
constrains their unitarity cuts. These constraints restrict the double discontinuities that can appear in the quadrupole functions correlating four coloured partons, for all configurations of massive and massless partons. This elucidates an important property of these functions at three loops and provides useful input for the bootstrap of the QCD soft anomalous dimension. We have checked these constraints against the known massless~\cite{Almelid:2015jia} and one-mass~\cite{Gardi:2025lws} cases, and found agreement.

Interestingly, these constraints hold separately for each web that contributes to the quadrupole functions; this is particularly nontrivial for the connected webs that couple all four Wilson lines, where every individual two-particle unitarity cut is non-vanishing. Nevertheless, certain pairs of non-overlapping two-particle channels do not admit compatible cuts. As illustrated in Fig.~\ref{fig: crossed cut}, the corresponding cuts cross and as such cannot be imposed simultaneously, implying that the associated double discontinuities vanish~\cite{Abreu:2014cla}.

To illustrate the constraining power of these vanishing double discontinuities, we first revisited the bootstrap of the three-loop massless soft anomalous dimension. Starting from the 13-parameter ansatz of Ref.~\cite{Almelid:2017qju}, we found that, remarkably,  the double-discontinuity constraints eliminate 8 free coefficients. The remaining coefficients can then be fixed, up to an overall normalisation, using only the homogeneous constraints associated with the Regge and collinear limits, which can be deduced from the relevant factorisation and exponentiation properties in these limits. 

We next applied the same strategy to the more involved case of one massive and three massless partons, which was only recently computed~\cite{Gardi:2025lws}. Here we took as input the known twelve-letter symbol alphabet. After imposing integrability, symmetries, and the known massless and collinear limits, we obtained a symbol-level ansatz with $74$ free coefficients, which the double-discontinuity constraints reduce to just $9$. Lifting this ansatz to function level introduces additional terms proportional to transcendental constants, bringing the total number of free coefficients to $802$. Imposing the corresponding function-level constraints, including the absence of unphysical branch cuts, leaves a single undetermined parameter. 
We then fixed the remaining parameter by matching the
leading-logarithmic coefficient in a simple Euclidean
limit to its vanishing value in the existing diagrammatic
result~\cite{Gardi:2025lws}.
This reproduces the complete one-mass quadrupole function.

As illustrated by these two applications, the new
double-discontinuity constraints provide useful input
to bootstrap calculations, beyond what is captured by
symmetries, factorisation properties, and known kinematic
limits. They substantially restrict the ansatz and reduce
the boundary information needed to determine the result.
Since these constraints apply independently of the masses
of the external partons, we expect them to prove valuable
for bootstrapping configurations with two or more massive
partons.

Extending the bootstrap to these configurations may
require more complicated function spaces, including
elliptic multiple
polylogarithms~\cite{Milloy:2020hzi,Bourjaily:2022bwx,Henn:2023pqn}.
Another direction for investigation is whether the
hierarchical
principle~\cite{pham,Hannesdottir:2022xki,McLeod:2026kaf,Crisanti:2026cuf}
yields additional constraints on the relevant web
integrals when more partons are massive.
Establishing whether such constraints complement those
derived here would be an interesting step towards
bootstrapping the unknown multi-mass soft anomalous
dimension.

The analysis of compatible cuts could also be extended
to higher loop orders, possibly leading to all-orders constraints
such as already exist for Feynman integrals and the wavefunction
of the universe~\cite{Benincasa:2020aoj,Benincasa:2021qcb,Hannesdottir:2025bss}.
Beyond restrictions on double discontinuities, it would
be interesting to investigate whether constraints can
be derived on the letters appearing near the end of
the symbol~\cite{Hannesdottir:2021kpd}, or on all pairs
of adjacent entries, as has been found in the context
of massless scattering
amplitudes~\cite{Drummond:2017ssj,Caron-Huot:2019bsq}.
Such investigations could identify further analytic
constraints and thereby reduce the boundary information
needed to bootstrap the soft anomalous dimension
with additional massive partons or at higher loop orders.

The Steinmann relations have provided powerful constraints
in the bootstrap of planar $\mathcal{N}=4$ super-Yang--Mills
amplitudes~\cite{Caron-Huot:2016owq,Dixon:2016nkn}. They require sequential discontinuities in
partially overlapping momentum channels to vanish.
The constraints derived here concern non-overlapping
two-particle channels, for which the topology of the webs
contributing to the three-loop quadrupole functions
excludes the corresponding cuts.
Despite this difference in channel configurations,
the two sets of constraints share a connection between
vanishing sequential discontinuities and the absence
of a compatible set of unitarity cuts.
This common mechanism reflects the role of causality
in both sets of constraints.
In time-ordered perturbation theory, this connection
is explicit: the required set of cuts cannot be
accommodated within a consistent causal ordering
of the interaction vertices.


\section*{Acknowledgements}
E.G. thanks Andries Waelkens, Niamh Maher,   Samuel Abreu and German Rodrigo for very useful discussions regarding cuts and sequential discontinuities of webs. 
We are grateful to the organisers and participants of the workshop \emph{Jumpstarting Elliptic Bootstrap Methods for Scattering Amplitudes} at the Niels Bohr Institute (supported by the Carlsberg Foundation), as well as the \emph{Symbology@15} conference at the Max Planck Institute for Physics in Garching, for hospitality and valuable feedback on preliminary versions of these results.
This research
has been supported by the STFC Consolidated Grant “Particle Physics at the Higgs Centre.” A.J.M.~and L.L.~acknowledge support from the Royal Society grant
URF{\textbackslash}R1{\textbackslash}221233, and A.J.M.~additionally acknowledges support from the European Research Council (ERC) under the European Union’s Horizon Europe research and innovation program grant agreement 101163627 (ERC Starting Grant ``AmpBoot'').
E.G. would like to thank the CERN theory department for hospitality  as a Scientific Associate during the final stages of writing this paper. 
 We acknowledge the use of LLM models from OpenAI (including Astra 6) and Anthropic (including Opus 5.5) during the writing of this paper. Models were used to check calculations, correct grammar and check for typos. We retain full responsibility for the content of this paper. 

\clearpage

\appendix

\section{Mandelstam invariants and the \texorpdfstring{$\alpha_{AB}$}{alpha} variables}
\label{sec:alpha_details}
In this appendix, we explain the relation between the two-particle Mandelstam invariant $s_{AB}$ and the $\alpha_{AB}$ variable, including the branch convention underlying
Eq.~\eqref{eq:boundary-alpha}. 
The $\alpha$ variables were introduced in Ref.~\cite{Correa:2012nk}, and have since been used in a number of calculations; see, for example, Refs.~\cite{Gardi:2013saa,Grozin:2015kna,Grozin:2014hna,Henn:2023pqn, Henn:2019swt}.
Throughout this appendix, $A$ and $B$ label massive
lines with $m_A,m_B>0$, and all other kinematic variables $s_{CD}$ are held fixed at values away from the threshold and pseudothreshold branch cuts.

\subsection{Definition and rationalisation of the Källén square root}
\label{app:alpha-definition}
First, we return to the definition of $\alpha_{AB}$ from
Eq.~\eqref{eq: alpha} and, inserting the definitions $s_{AB}=(p_A+p_B)^2$ and $p_A^2=m_A^2$, derive the relationship
\begin{equation}
    -\left(\alpha_{AB}+\frac{1}{\alpha_{AB}}\right)=\frac{s_{AB}-m_A^2-m_B^2}{m_Am_B},
\label{eq:am-definition}
\end{equation}
which can be inverted to give
\begin{equation}
\label{eq:s-as-a-function-of-alpha}
    s_{AB}=m_A^2+m_B^2-m_Am_B\left(\alpha_{AB}+\frac{1}{\alpha_{AB}}\right).
\end{equation}
We can also rewrite Eq.~\eqref{eq:am-definition} to obtain
\begin{equation}
    \alpha_{AB}^2+\frac{s_{AB}-m_A^2-m_B^2}{m_A m_B}\alpha_{AB}+1=0,
\label{eq:am-quadratic}
\end{equation}
which admits a pair of solutions $\alpha_{AB}^{(1)}$ and $\alpha_{AB}^{(2)}$, given by
\begin{subequations}
\label{eq:am-roots}
\begin{align}
 \alpha_{AB}^{(1)}
 &=\frac{m_A^2+ m_B^2-s_{AB} + \sqrt{\lambda(s_{AB}, m_A^2, m_B^2)}}{2m_Am_B}, \\
 \alpha_{AB}^{(2)}
 &=\frac{m_A^2+ m_B^2-s_{AB} - \sqrt{\lambda(s_{AB}, m_A^2, m_B^2)}}{2m_Am_B},
\end{align}
\end{subequations}
where the argument of the square root is the K\"all\'en polynomial
\begin{align}
\label{eq:am-kallen}
    \begin{split}
\lambda(s_{AB},m_A^2,m_B^2)&=\left(s_{AB}-m_A^2-m_B^2\right)^2-4m^2_Am_B^2\\
    &=\big(s_{AB}-(m_A+m_B)^2\big)\big(s_{AB}-(m_A-m_B)^2\big).
    \end{split}
\end{align}
Note that these solutions satisfy $\alpha_{AB}^{(1)}=1/{\alpha_{AB}^{(2)}}$. 

The benefit of working in terms of $\alpha_{AB}$ becomes evident when we notice that 
\begin{equation}
    \lambda(s_{AB},m_A^2,m_B^2)=m^2_Am_B^2\left(\alpha_{AB}-\frac{1}{\alpha_{AB}}\right)^2.
\label{eq:am-rational-root}
\end{equation}
Thus, both \(s_{AB}\) and the square root of the Källén function are rational functions of \(\alpha_{AB}\); the latter variable provides a rational uniformisation, parametrising both Riemann sheets associated with \(\sqrt{\lambda}\) with a single complex coordinate.

\subsection{A branch choice and the map of the complex planes}
\label{app:alpha-map}
Let us now investigate how the complex $s_{AB}$ plane gets mapped into the complex $\alpha_{AB}$ plane. The threshold value $(m_A+m_B)^2$ that appears in Eq.~\eqref{eq:am-kallen} is the minimum value of $s_{AB}$ that allows the production of two positive-energy on-shell particles. By contrast, the pseudo-threshold at $(m_A-m_B)^2$ does not correspond to the production of positive-energy particles; however, while it does not appear as a singularity on the first Riemann sheet, it can arise on the second sheet. To determine their locations in the $\alpha_{AB}$-plane,
we use Eqs.~\eqref{eq:am-kallen} and
\eqref{eq:am-rational-root} to obtain the exact factorisations
\begin{subequations}
\begin{align}
 s_{AB}-(m_A+m_B)^2
 &=-m_Am_B\frac{(1+\alpha_{AB})^2}{\alpha_{AB}}, \\
 s_{AB}-(m_A-m_B)^2
 &=-m_Am_B\frac{(1-\alpha_{AB})^2}{\alpha_{AB}}.
 \label{eq:am-threshold-factors}
\end{align}
\end{subequations}
These equations imply that the threshold occurs at $\alpha_{AB} = -1$, while the pseudo-threshold appears at $\alpha_{AB} = 1$.

\begin{figure}[tbp]
  \centering
  \resizebox{\textwidth}{!}{%
%
\begingroup%
\definecolor{AMBlue}{HTML}{2667A8}%
\definecolor{AMGreen}{HTML}{16826C}%
\definecolor{AMGold}{HTML}{A66B12}%
\definecolor{AMAxis}{HTML}{737B84}%
\definecolor{AMCut}{HTML}{4D555E}%
\definecolor{AMPaleBlue}{HTML}{EBF1F7}%
\definecolor{AMPaleGreen}{HTML}{E1EEEC}%
\ifcsname ver@axodraw4j.sty\endcsname%
  \def\AMSetRGB#1,#2,#3\relax{\SetColor{#1 #2 #3 setrgbcolor}}%
  \def\AMSetColor#1{%
    \color{#1}%
    \convertcolorspec{named}{#1}{rgb}{\AMRGB}%
    \expandafter\AMSetRGB\AMRGB\relax}%
\else%
  \let\AMSetColor\SetColor%
\fi%
\def\AMFillBox(#1,#2)(#3,#4){%
  \put(#1,#2){\rule{\dimexpr#3pt-#1pt\relax}%
    {\dimexpr#4pt-#2pt\relax}}}%
\def\AMDraw#1{\put(0,0){#1}\ignorespaces}%
\setlength{\unitlength}{1pt}%
\ifcsname ver@axodraw4j.sty\endcsname%
  \SetScale{0.99626400996}
\else%
  \SetScale{1}%
\fi%
\SetOffset(0,0)\unskip%
\SetScaledOffset(0,0)\unskip%
\SetArrowStroke{0}%
\fontsize{13}{15}\selectfont%
\begin{picture}(650,430)%
%
%
\AMSetColor{AMPaleBlue}%
\AMDraw{\AMFillBox(8,213)(205,277)}%
\AMSetColor{AMPaleGreen}%
\AMDraw{\AMFillBox(8,149)(205,213)}%
\AMSetColor{AMAxis}%
\SetWidth{0.65}%
\AMDraw{\LongArrow[arrowwidth=4,arrowlength=5](8,213)(217,213)}%
\AMDraw{\LongArrow[arrowwidth=4,arrowlength=5](31,143)(31,286)}%
\AMDraw{\Text(219,213)[l]{\fontsize{11.5}{13}\selectfont $\operatorname{Re}s_{AB}$}}%
\AMDraw{\Text(34,293)[l]{$\operatorname{Im}s_{AB}$}}%
\AMDraw{\ZigZag(136,213)(205,213){1.7}{8}}%
\AMSetColor{AMGold}%
\SetWidth{2.1}%
\AMDraw{\Line(64,213)(136,213)}%
\AMDraw{\LongArrow[arrowwidth=5.5,arrowlength=6.5](88,213)(116,213)}%
%
\SetWidth{1.1}%
\AMSetColor{AMBlue}%
\AMDraw{\LongArrow[arrowwidth=4,arrowlength=5](143,220)(198,220)}%
\AMDraw{\Text(175,230)[c]{\fontsize{11.5}{13}\selectfont $s_{AB}+i\varepsilon$}}%
\AMDraw{\Text(123,257)[c]{$\operatorname{Im}s_{AB}>0$}}%
\AMSetColor{AMGreen}%
\AMDraw{\LongArrow[arrowwidth=4,arrowlength=5](143,206)(198,206)}%
\AMDraw{\Text(175,195)[c]{\fontsize{11.5}{13}\selectfont $s_{AB}-i\varepsilon$}}%
\AMDraw{\Text(123,167)[c]{$\operatorname{Im}s_{AB}<0$}}%
\AMSetColor{black}%
\AMDraw{\Vertex(64,213){2.5}}%
\AMDraw{\Vertex(136,213){2.5}}%
\AMDraw{\Text(70,200)[c]{$s_{AB}^{\mathrm{ps.}}$}}%
\AMDraw{\Text(130,200)[c]{$s_{AB}^{\mathrm{th.}}$}}%
\AMDraw{\Text(107,316)[c]{\fontsize{17}{19}\selectfont $s_{AB}$ plane}}%
%
\AMSetColor{AMAxis}%
\SetWidth{1.15}%
\AMDraw{\LongArrow[arrowwidth=6,arrowlength=8](230,245)(330,340)}%
\AMDraw{\LongArrow[arrowwidth=6,arrowlength=8](230,179)(330,110)}%
\AMSetColor{black}%
\AMDraw{\Text(260,305)[c]{$\alpha_{AB}^{(1)}(s_{AB})$}}%
\AMDraw{\Text(274,127)[c]{$\alpha_{AB}^{(2)}(s_{AB})$}}%
%
\AMSetColor{AMPaleGreen}%
\AMDraw{\AMFillBox(345,290)(635,392)}%
\AMSetColor{AMPaleBlue}%
\AMDraw{\Vertex(490,290){62}}%
\AMSetColor{white}%
\AMDraw{\AMFillBox(427,227)(553,290)}%
\AMSetColor{AMAxis}%
\SetWidth{0.65}%
\AMDraw{\LongArrow[arrowwidth=4,arrowlength=5](343,290)(643,290)}%
\AMDraw{\LongArrow[arrowwidth=4,arrowlength=5](490,272)(490,400)}%
\AMDraw{\Text(640,276)[r]{$\operatorname{Re}\alpha_{AB}$}}%
\AMDraw{\Text(499,397)[l]{$\operatorname{Im}\alpha_{AB}$}}%
%
\AMSetColor{AMCut}%
\SetWidth{0.9}%
\AMDraw{\ZigZag(345,290)(425,290){1.7}{9}}%
\AMDraw{\ZigZag(431,290)(486,290){1.7}{7}}%
\AMSetColor{AMGold}%
\SetWidth{2.1}%
\AMDraw{\Arc[arrow,arrowpos=0.5,arrowwidth=5.5,arrowlength=6.5](490,290)(62,0,180)}%
%
\SetWidth{1.1}%
\AMSetColor{AMBlue}%
\AMDraw{\LongArrow[arrowwidth=4,arrowlength=5](434,297)(482,297)}%
\AMDraw{\Text(458,310)[c]{\fontsize{11.5}{13}\selectfont $-\tau+i\varepsilon$}}%
\AMDraw{\Text(495,335)[c]{\fontsize{11.5}{13}\selectfont $\operatorname{Im}s_{AB}>0$}}%
\AMSetColor{AMGreen}%
\AMDraw{\LongArrow[arrowwidth=4,arrowlength=5](419,297)(353,297)}%
\AMDraw{\Text(383,310)[c]{\fontsize{11.5}{13}\selectfont $-\tau^{-1}+i\varepsilon$}}%
\AMDraw{\Text(591,363)[c]{\fontsize{11.5}{13}\selectfont $\operatorname{Im}s_{AB}<0$}}%
\AMSetColor{black}%
\AMDraw{\Vertex(428,290){2.5}}%
\AMDraw{\Vertex(552,290){2.5}}%
\SetWidth{0.8}%
\AMDraw{\BCirc(490,290){2.1}}%
\AMDraw{\Text(428,276)[c]{$s_{AB}^{\mathrm{th.}}=-1$}}%
\AMDraw{\Text(552,276)[c]{$s_{AB}^{\mathrm{ps.}}=1$}}%
\AMDraw{\Text(490,421)[c]{\fontsize{17}{19}\selectfont $\alpha_{AB}^{(1)}$ plane}}%
%
\AMSetColor{AMPaleBlue}%
\AMDraw{\AMFillBox(345,41)(635,143)}%
\AMSetColor{AMPaleGreen}%
\AMDraw{\Vertex(490,143){62}}%
\AMSetColor{white}%
\AMDraw{\AMFillBox(427,143)(553,206)}%
\AMSetColor{AMAxis}%
\SetWidth{0.65}%
\AMDraw{\LongArrow[arrowwidth=4,arrowlength=5](343,143)(643,143)}%
\AMDraw{\LongArrow[arrowwidth=4,arrowlength=5](490,33)(490,185)}%
\AMDraw{\Text(640,158)[r]{$\operatorname{Re}\alpha_{AB}$}}%
\AMDraw{\Text(499,187)[l]{$\operatorname{Im}\alpha_{AB}$}}%
\AMSetColor{AMCut}%
\SetWidth{0.9}%
\AMDraw{\ZigZag(345,143)(425,143){1.7}{9}}%
\AMDraw{\ZigZag(431,143)(486,143){1.7}{7}}%
\AMSetColor{AMGold}%
\SetWidth{2.1}%
\AMDraw{\Arc[clock,arrow,arrowpos=0.5,arrowwidth=5.5,arrowlength=6.5](490,143)(62,0,180)}%
\SetWidth{1.1}%
\AMSetColor{AMGreen}%
\AMDraw{\LongArrow[arrowwidth=4,arrowlength=5](434,136)(482,136)}%
\AMDraw{\Text(458,123)[c]{\fontsize{11.5}{13}\selectfont $-\tau-i\varepsilon$}}%
\AMDraw{\Text(495,100)[c]{\fontsize{11.5}{13}\selectfont $\operatorname{Im}s_{AB}<0$}}%
\AMSetColor{AMBlue}%
\AMDraw{\LongArrow[arrowwidth=4,arrowlength=5](419,136)(353,136)}%
\AMDraw{\Text(383,123)[c]{\fontsize{11.5}{13}\selectfont $-\tau^{-1}-i\varepsilon$}}%
\AMDraw{\Text(591,70)[c]{\fontsize{11.5}{13}\selectfont $\operatorname{Im}s_{AB}>0$}}%
\AMSetColor{black}%
\AMDraw{\Vertex(428,143){2.5}}%
\AMDraw{\Vertex(552,143){2.5}}%
\SetWidth{0.8}%
\AMDraw{\BCirc(490,143){2.1}}%
\AMDraw{\Text(428,158)[c]{$s_{AB}^{\mathrm{th.}}=-1$}}%
\AMDraw{\Text(552,158)[c]{$s_{AB}^{\mathrm{ps.}}=1$}}%
\AMDraw{\Text(490,215)[c]{\fontsize{17}{19}\selectfont $\alpha_{AB}^{(2)}$ plane}}%
\end{picture}%
\endgroup
  }
  \caption{The $s_{AB}$ complex plane and its images under
    the two roots $\alpha_{AB}^{(1)}$ and $\alpha_{AB}^{(2)}$. The upper half of the $\alpha_{AB}$ plane is the image of the first algebraic root, while the lower half is the image of the second. This picture makes evident why, using $\alpha_{AB}^{(1)}$, the boundary value $\alpha_{AB}^+$ ends up inside the unit circle, while $\alpha_{AB}^-$ sits outside. Inversion symmetry exchanges the two roots and connects $\alpha_{AB}^-$ to its equivalent representation inside the unit circle. On the second Riemann sheet, a further branch cut appear between the pseudo-threshold and infinity. The standard routing of the cut along the negative real axis is mapped onto the positive real axis in the $\alpha$ plane, between $\alpha_{AB}=0$ and $\alpha_{AB}=1$ and between $\alpha_{AB}=1$ and infinity along the real axis.}
  \label{fig:alpha-map}
\end{figure}
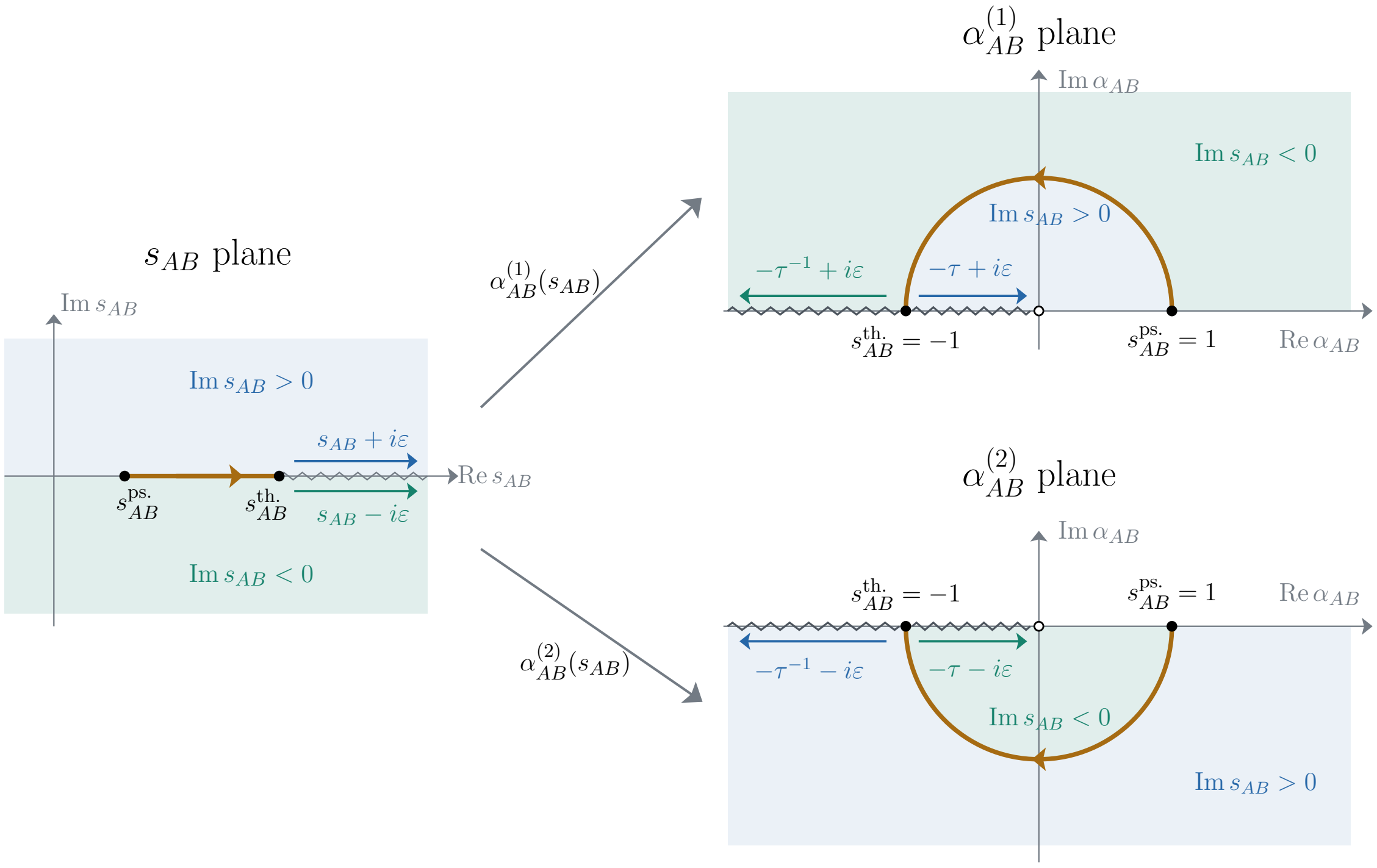

More generally, to obtain precisely the convention used in Eq.~\eqref{eq:boundary-alpha}, we start in the region where $ (m_A-m_B)^2 <s_{AB}< (m_A+m_B)^2$, and pick the root $\alpha_{AB}^{(1)}$. Then we can write 
\begin{equation}
    \alpha_{AB}^{(1)}(s_{AB})
    =\frac{m_A^2+m_B^2-s_{AB}
    +i\sqrt{s_{AB}-(m_A-m_B)^2}
    \sqrt{(m_A+m_B)^2-s_{AB}}}{2m_Am_B},
    \label{eq:am-chosen-branch}
\end{equation}
where each square root is on its principal sheet, with the corresponding branch cut on the negative real axis of its argument. Between the threshold and pseudothreshold, we can parametrise $\alpha_{AB}^{(1)}$ in terms of a new variable $\theta_{AB}$ as
\begin{equation}
    \alpha_{AB}^{(1)}=e^{i\theta_{AB}},\qquad
    \cos\theta_{AB}=\frac{m^2_A+m^2_B-s_{AB}}{2m_Am_B}, %
    \qquad 0<\theta_{AB}<\pi.
 \label{eq:am-unit-semicircle}
\end{equation}
Increasing $s_{AB}$ along the real axis from the pseudo-threshold to the threshold therefore
traces the upper unit semicircle from $\alpha_{AB}=1$ to $\alpha_{AB}=-1$, as shown in Fig.~\ref{fig:alpha-map}. More generally, the branch choice in Eq.~\eqref{eq:am-chosen-branch} picks out a one-to-one map to the upper-half complex $\alpha_{AB}$ plane:
\begin{equation}
    \mathbb C\setminus\bigl(( -\infty,s_{AB}^\text{ps.}]\cup[s_{AB}^\text{th.},\infty)\bigr)
    \quad\longrightarrow\quad \{\alpha_{AB}:\operatorname{Im}\alpha_{AB}>0\}.
    \label{eq:am-slit-domain}
\end{equation}
To see which region lies inside the unit circle, we write $\alpha_{AB}=\operatorname{Re}(\alpha_{AB})+i\operatorname{Im}(\alpha_{AB})$ and take the
imaginary part of Eq.~\eqref{eq:s-as-a-function-of-alpha}:
\begin{equation}
 \operatorname{Im}s_{AB}
 =-m_Am_B\,\operatorname{Im}(\alpha_{AB})\left(1-|\alpha_{AB}|^{-2}\right).
 \label{eq:am-imaginary-map}
\end{equation}
Since $\operatorname{Im}\alpha_{AB}^{(1)}>0$ in Eq.~\eqref{eq:am-unit-semicircle}, the upper half $s_{AB}$ plane maps to $|\alpha_{AB}^{(1)}|<1$, while the lower half $s_{AB}$ plane maps to $|\alpha_{AB}^{(1)}|>1$ (in both cases, staying in the upper half of the $\alpha_{AB}^{(1)}$ plane). Real values $s_{AB}< (m_A-m_B)^2$ have positive reciprocal images, one in $(0,1)$ and the other in $(1,\infty)$, while real values $s_{AB}>(m_A+m_B)^2$ have negative reciprocal images, one in $(-1,0)$ and the other in $(-\infty,-1)$. 

In order to understand the inversion symmetry introduced by the definition of $\alpha_{AB}$ in Eq.~\eqref{eq: alpha}, it is interesting to see what happens if we choose the other root, $\alpha_{AB}^{(2)}$, when carrying out this analysis. Following the same argument, and keeping in mind that choosing $(m_A-m_B)^2<s_{AB}<(m_A+m_B)^2$ means that $\alpha_{AB}^{(2)}=e^{-i\theta_{AB}}$ for $0<\theta_{AB}<\pi$, we find the results depicted in Fig.~\ref{fig:alpha-map}.
Considering both roots together, we recover the full complex \(\alpha_{AB}\) plane. Since the two roots are related by inversion, \(\alpha_{AB}^{(2)}=1/\alpha_{AB}^{(1)}\), every value of \(s_{AB}\) can be represented by a point inside or on the unit circle, allowing us to restrict our choice of root to \(|\alpha_{AB}|\leq 1\).

\subsection{Derivation of Eq.~\eqref{eq:boundary-alpha}}
\label{app:alpha-boundaries}
We now derive the boundary values used in Eq.~\eqref{eq:boundary-alpha}, to show that these boundary values can be mapped to points inside the unit circle just above and below the branch cut connecting $\alpha_{AB}=0$ and $\alpha_{AB}=-1$. First, take $s_{AB}>(m_A+m_B)^2$ and let
\begin{align}
        \tau=\frac{s_{AB}-m^2_A-m^2_B-\sqrt{\lambda(s_{AB},m_A^2,m_B^2)}}{2m_Am_B}\in(0,1).
 \label{eq:am-tau}
\end{align}
for $\lambda(s_{AB},m_A^2,m_B^2) > 0$ as given in Eq.~\eqref{eq:am-kallen}. This shows that $\tau\to1$ at threshold and $\tau\to0$ as $s_{AB}\to+\infty$. On the branch chosen in Eq.~\eqref{eq:am-chosen-branch},
the square roots obey
\begin{align}
 \label{eq:am-root-boundaries}
\begin{split}
    \sqrt{(m_A+m_B)^2-(s_{AB}\pm i0)} &= \sqrt{\big(s_{AB}-(m_A+m_B)^2\big)e^{\mp i\pi}} \\
    &=\mp i\sqrt{s_{AB}-(m_A+m_B)^2}.
    \end{split}
\end{align}
On the two sides of the branch cut, we thus have
\begin{subequations}
    \begin{align}
        \operatorname{Re}\left(\alpha_{AB}^{(1)}(s_{AB}+i\varepsilon)\right)&=\frac{m^2_A+m^2_B-s_{AB}+\sqrt{\lambda(s_{AB},m_A^2,m_B^2)}}{2m_Am_B}=-\tau,\\
        \operatorname{Re}\left(\alpha_{AB}^{(1)}(s_{AB}-i\varepsilon)\right)&=\frac{m^2_A+m^2_B-s_{AB}-\sqrt{\lambda(s_{AB},m_A^2,m_B^2)}}{2m_Am_B}=-\tau^{-1}.
    \end{align}
\end{subequations}
To determine the imaginary part, we differentiate the inverse map
\begin{equation}
 \frac{ \partial\alpha_{AB}^{(1)}}{ \partial s}
 =\frac{\alpha_{AB}^2}{m_Am_B(1-\alpha_{AB}^2)},
 \label{eq:am-derivative}
\end{equation}
then evaluate this derivative at the real boundary value (either $\tau$ or $\tau^{-1}$) and multiply by the infinitesimal displacement $i\varepsilon$. For fixed $s_{AB}>(m_A+m_B)^2$ this yields
\begin{equation}
    \begin{aligned}
        \alpha_{AB}^{(1)}(s_{AB}+i\varepsilon)&=-\tau+\frac{i\varepsilon\tau^2}{m_Am_B(1-\tau^2)}+O(\varepsilon^2),\\
        \alpha_{AB}^{(1)}(s_{AB}-i\varepsilon)&=-\tau^{-1}+\frac{i\varepsilon}{m_Am_B(1-\tau^2)}+O(\varepsilon^2).
    \end{aligned}
    \label{eq:am-finite-regulator}
\end{equation}
Both coefficients multiplying $+i\varepsilon$ are positive. Removing the root specification therefore gives exactly Eq.~\eqref{eq:boundary-alpha}:
\begin{equation}
 \alpha_{AB}^{+}\equiv\alpha_{AB}(s_{AB}+i\varepsilon)=-\tau+i\varepsilon,
 \qquad
 \alpha_{AB}^{-}\equiv\alpha_{AB}(s_{AB}-i\varepsilon)=-\tau^{-1}+i\varepsilon.
 \label{eq:am-main-boundaries}
\end{equation}
Finally, inversion symmetry maps the point $\alpha_{AB}^{-}$ to its equivalent representative inside the unit circle, completing the argument.

\section{Graphical representation of the colour structures}
\label{app:color}
In this appendix, we provide a graphical representation of each of the colour structures that appears in Table~\ref{tab:col_struct}. Several webs can contribute to each colour structure through colour conservation \cite{Almelid} and the non-abelian exponentiation theorem \cite{Gardi:2013ita}. Here we only show the connected web that defines each colour structure.

We start from the dipoles, using thicker double lines to stress the presence of massive Wilson lines. From left to right, the completely massless ($\mathbf{D}_{(ij)}$), mixed ($\mathbf{D}_{(Ij)}$), and massive ($\mathbf{D}_{(IJ)}$) dipoles are represented by the connected webs:
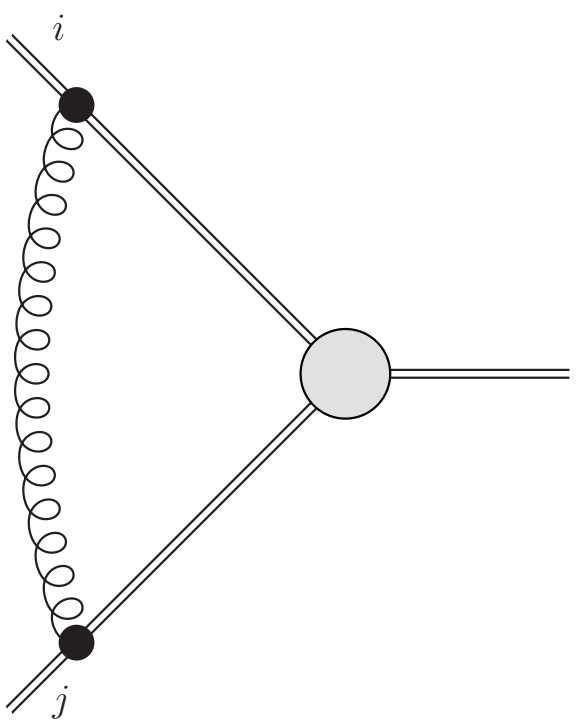
\begin{center}
\begin{tabular}{c | c | c}
    \scalebox{0.42}{
    \begin{picture}(300,325) (200,-69)
    \SetWidth{1}
    \SetColor{Black}
    \Line[double,sep=3.5](210,234)(360,84)
    \Line[double,sep=3.5](210,-66)(360,84)
    \SetWidth{1.0}
    \GluonArc(590,84)(370,161.08,198.92){7.5}{15}
    \Vertex(240,-36){8}
    \Vertex(240,204){8}
    \Text(230,234)[lb]{\LARGE{\Black{$i$}}}
    \Text(230,-70)[lb]{\LARGE{\Black{$j$}}}
    \SetWidth{1}
    \Line[double,sep=3.5](360,84)(460,84)
    \SetWidth{1.0}
    \GOval(360,84)(20,20)(0){0.882}
\end{picture}}&
    \scalebox{0.42}
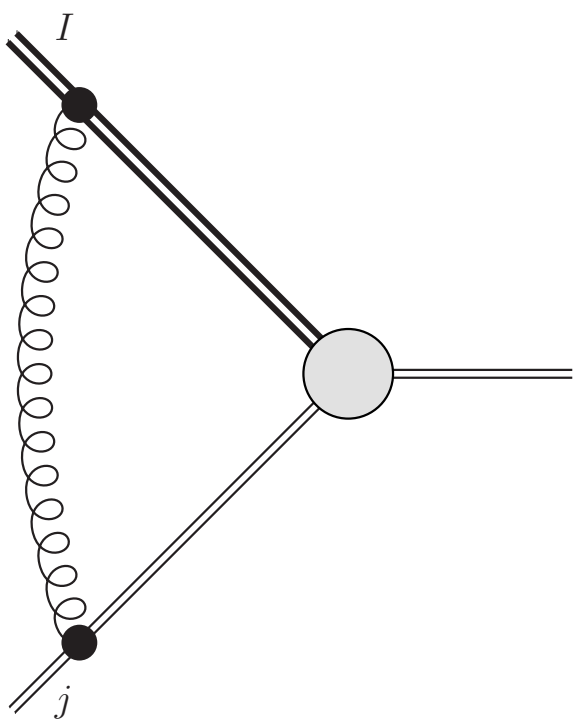
{
    \begin{picture}(300,325)(200,-69)
    \SetWidth{2.8}
    \SetColor{Black}
    \Line[double,sep=5.5](210,234)(360,84)
    \SetWidth{1}
    \Line[double,sep=3.5](210,-66)(360,84)
    \SetWidth{1.0}
    \GluonArc(590,84)(370,161.08,198.92){7.5}{15}
    \Vertex(240,-36){8}
    \Vertex(240,204){8}
    \Text(230,234)[lb]{\LARGE{\Black{$I$}}}
    \Text(230,-70)[lb]{\LARGE{\Black{$j$}}}
    \SetWidth{1}
    \Line[double,sep=3.5](360,84)(460,84)
    \SetWidth{1.0}
    \GOval(360,84)(20,20)(0){0.882}
\end{picture}}
    &
    \scalebox{0.42}
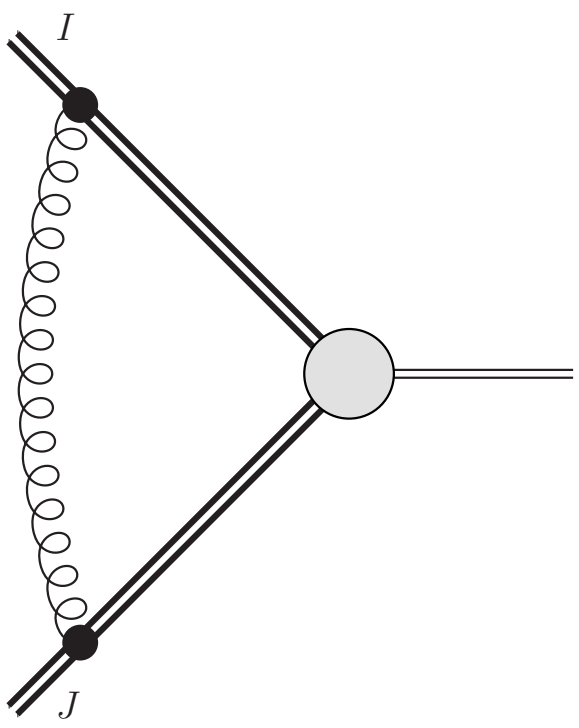
{
    \begin{picture}(300,325)(200,-69)
    \SetWidth{2.8}
    \SetColor{Black}
    \Line[double,sep=5.5](210,234)(360,84)
    \Line[double,sep=5.5](210,-66)(360,84)
    \SetWidth{1.0}
    \GluonArc(590,84)(370,161.08,198.92){7.5}{15}
    \Vertex(240,-36){8}
    \Vertex(240,204){8}
    \Text(230,234)[lb]{\LARGE{\Black{$I$}}}
    \Text(230,-70)[lb]{\LARGE{\Black{$J$}}}
    \SetWidth{1}
    \Line[double,sep=3.5](360,84)(460,84)
    \SetWidth{1.0}
    \GOval(360,84)(20,20)(0){0.882}
\end{picture}}\\
    $\mathbf{D}_{(ij)}=\mathbf{T}_i\cdot\mathbf{T}_j$ & $\mathbf{D}_{(Ij)}=\mathbf{T}_I\cdot\mathbf{T}_j$ & $\mathbf{D}_{(IJ)}=\mathbf{T}_I\cdot\mathbf{T}_J$
\end{tabular}    
\end{center}
Next, we show (on the left) the connected web defining the mixed tripole $\mathbf{T}_{(IJk)}$ and (on the right) the one defining the massive tripole $\mathbf{T}_{(IJK)}$:
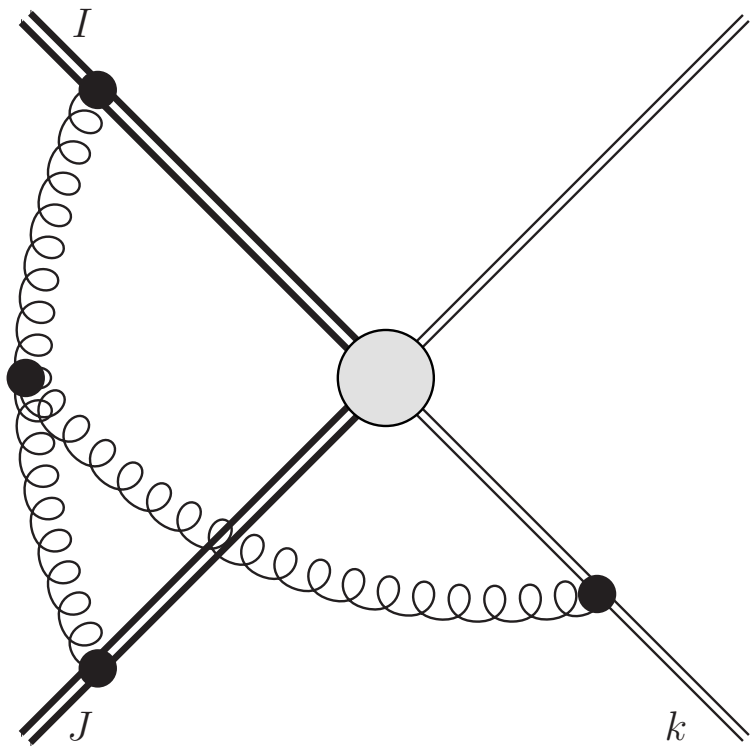
\begin{center}
\begin{tabular}{c|c}
    \scalebox{0.45}{
    \begin{picture}(315,306) (174,-87)
            \SetWidth{2.8}
            \SetColor{Black}
            \Line[double,sep=5.5](180,216)(330,66)
            \Line[double,sep=5.5](180,-84)(330,66)
            \SetWidth{1}
            \Line[double,sep=3.5](330,66)(480,216)
            \Line[double,sep=3.5](330,66)(480,-84)
            \SetWidth{1.0}
            \GluonArc(463.167,66)(280.167,154.639,205.361){7.5}{17}
            \GluonArc(375,221)(249.098,-141.52,-79.592){7.5}{17}
            \Vertex(210,186){8}
            \Vertex(210,-55){8}
            \Vertex(418,-24){8}
            \Vertex(180,66){8}
            \GOval(330,66)(20,20)(0){0.882}
            \Text(200,209)[lb]{\LARGE{\Black{$I$}}}
            \Text(448,-85)[lb]{\LARGE{\Black{$k$}}}
            \Text(198,-85)[lb]{\LARGE{\Black{$J$}}}
        \end{picture}}\hspace{20pt}
    &
    \hspace{30pt}\scalebox{0.45}
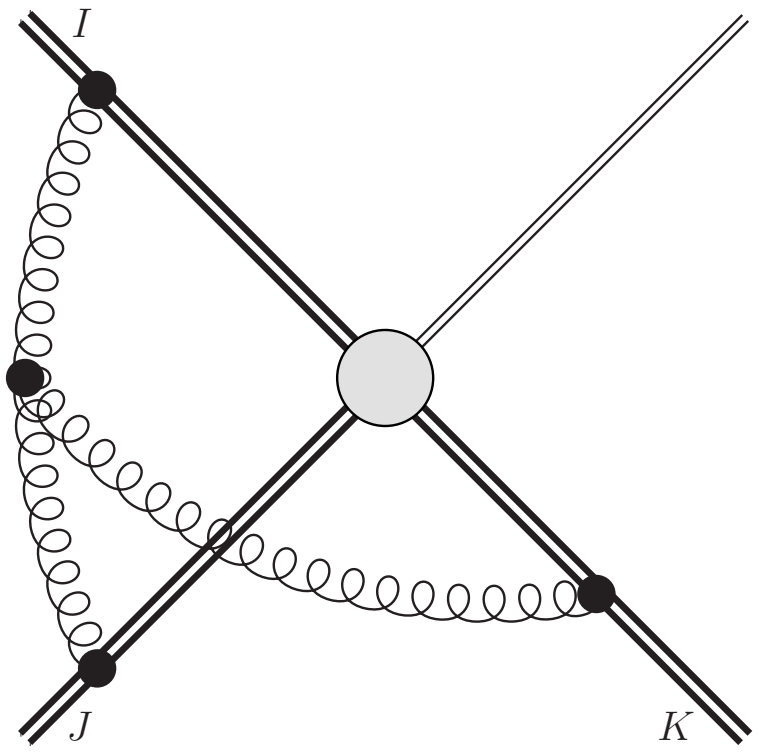
{
        \begin{picture}(315,306) (174,-87)
            \SetWidth{2.8}
            \SetColor{Black}
            \Line[double,sep=5.5](180,216)(480,-84)
            \Line[double,sep=5.5](180,-84)(330,66)
            \SetWidth{1}
            \Line[double,sep=3.5](330,66)(480,216)
            \SetWidth{1.0}
            \GluonArc(463.167,66)(280.167,154.639,205.361){7.5}{17}
            \GluonArc(375,221)(249.098,-141.52,-79.592){7.5}{17}
            \Vertex(210,186){8}
            \Vertex(210,-55){8}
            \Vertex(418,-24){8}
            \Vertex(180,66){8}
            \GOval(330,66)(20,20)(0){0.882}
            \Text(200,209)[lb]{\LARGE{\Black{$I$}}}
            \Text(445,-85)[lb]{\LARGE{\Black{$K$}}}
            \Text(198,-85)[lb]{\LARGE{\Black{$J$}}}
        \end{picture}}\\
    $\mathbf{T}_{(IJk)}=if_{abc}\mathbf{T}_I^a\mathbf{T}_J^b\mathbf{T}_k^c$\hspace{20pt} & \hspace{30pt}$\mathbf{T}_{(IJK)}=if_{abc}\mathbf{T}_I^a\mathbf{T}_J^b\mathbf{T}_K^c$
\end{tabular}    
\end{center}
We then have the quadrupoles with a repeated leg, starting from $\mathbf{Q}_{(ij;ik)}$ on the left and $\mathbf{Q}_{(Ij;Ik)}$ on the right:
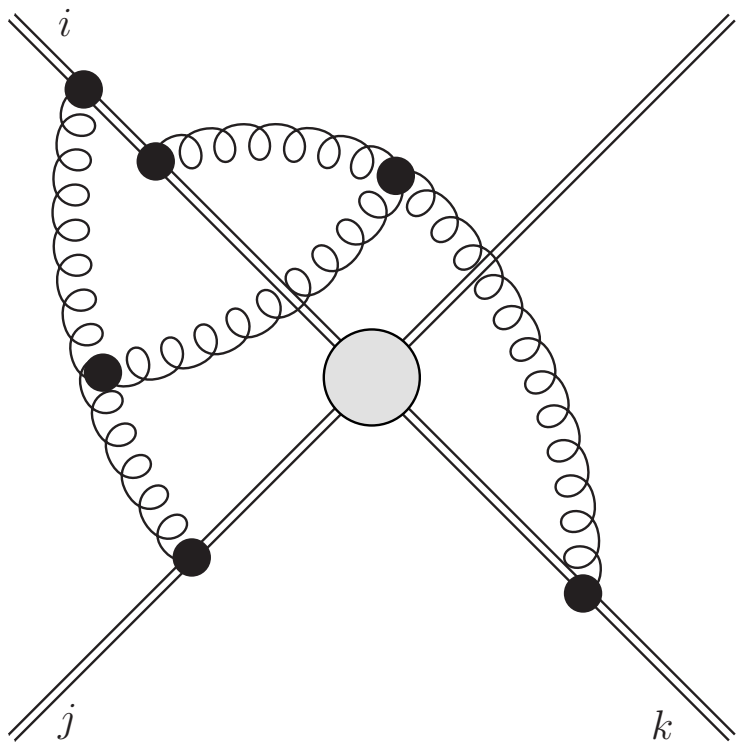
\begin{center}
\begin{tabular}{c|c}
    \scalebox{0.45}{
    \begin{picture}(315,306) (174,-87)
        \SetWidth{1}
        \SetColor{Black}
        \Line[double,sep=3.5](180,216)(330,66)
        \Line[double,sep=3.5](180,-84)(330,66)
        \SetWidth{1}
        \Line[double,sep=3.5](330,66)(480,216)
        \Line[double,sep=3.5](330,66)(480,-84)
        \SetWidth{1.0}
        \GluonArc(437.190,135.736)(232.684,167.525,218.464){7.5}{13}   
        \GluonArc(190.328,240.926)(175.126,279.091,328.721){7.5}{9}   
        \GluonArc(283.51,44.86)(119.35,61.75,111.38){7.5}{6}    
        \GluonArc(190.84,-21.35)(227.17,359.33,408.96){7.5}{12}  
        \Vertex(210,186){8}
        \Vertex(240,156){8}
        \Vertex(340,150){8}
        \Vertex(255,-9){8}
        \Vertex(418,-24){8}
        \Vertex(218,68){8}
        \GOval(330,66)(20,20)(0){0.882}
        \Text(200,209)[lb]{\LARGE{\Black{$i$}}}
        \Text(448,-85)[lb]{\LARGE{\Black{$k$}}}
        \Text(200,-85)[lb]{\LARGE{\Black{$j$}}}
        \end{picture}}\hspace{20pt}
    &
    \hspace{20pt}\scalebox{0.45}{
        \begin{picture}(315,306) (174,-87)
            
            \SetColor{Black}
            \SetWidth{2.8}
            \Line[double,sep=5.5](180,216)(330,66)
            \SetWidth{1}
            \Line[double,sep=3.5](180,-84)(330,66)
            \SetWidth{1}
            \Line[double,sep=3.5](330,66)(480,216)
            \Line[double,sep=3.5](330,66)(480,-84)
            \SetWidth{1.0}
            \GluonArc(437.190,135.736)(232.684,167.525,218.464){7.5}{13}   
            \GluonArc(190.328,240.926)(175.126,279.091,328.721){7.5}{9}   
            \GluonArc(283.51,44.86)(119.35,61.75,111.38){7.5}{6}          
            \GluonArc(190.84,-21.35)(227.17,359.33,408.96){7.5}{12}       
            \Vertex(210,186){8}
            \Vertex(240,156){8}
            \Vertex(340,150){8}
            \Vertex(255,-9){8}
            \Vertex(418,-24){8}
            \Vertex(218,68){8}
            \GOval(330,66)(20,20)(0){0.882}
            \Text(200,209)[lb]{\LARGE{\Black{$I$}}}
            \Text(448,-85)[lb]{\LARGE{\Black{$k$}}}
            \Text(200,-85)[lb]{\LARGE{\Black{$j$}}}
        \end{picture}}\\
    $\mathbf{Q}_{(ij;ik)}=\frac{1}{2}f_{abe}f^{e}_{\phantom{e}cd}\left\{\mathbf{T}^a_i,\mathbf{T}^c_i\right\}\mathbf{T}^b_j\mathbf{T}^d_k$\hspace{20pt} & \hspace{30pt}$\mathbf{Q}_{(Ij;Ik)}=\frac{1}{2}f_{abe}f^{e}_{\phantom{e}cd}\left\{\mathbf{T}^a_I,\mathbf{T}^c_I\right\}\mathbf{T}^b_j\mathbf{T}^d_k$
\end{tabular}
\end{center}
Then, on the left, we show the quadrupole with two massive legs, one of which is repeated $\mathbf{Q}_{(IJ;Ik)}$ and, on the right, the fully massive quadrupole with one repeated leg $\mathbf{Q}_{(IJ;IK)}$:
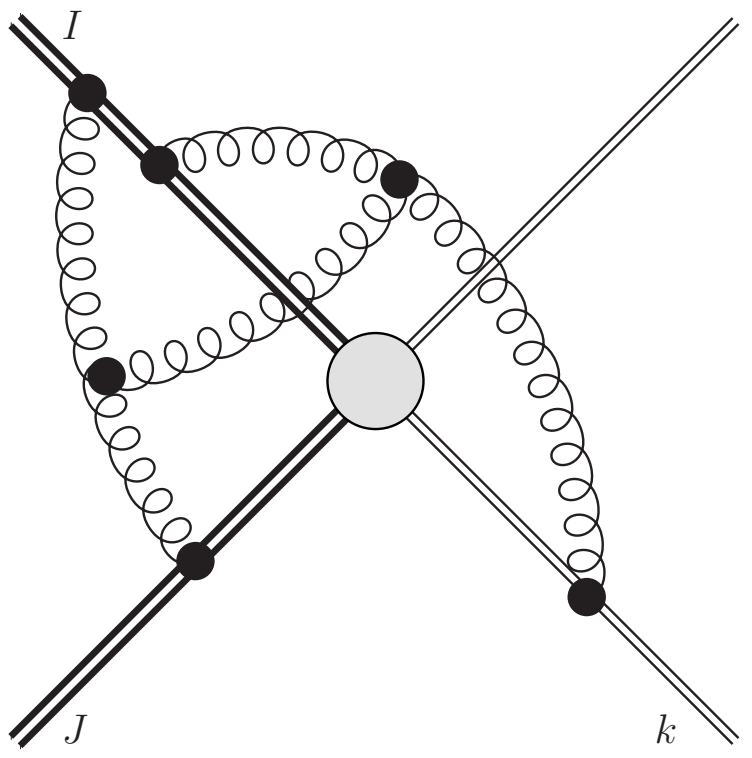
\begin{center}
 \begin{tabular}{c|c}
    \scalebox{0.45}{
    \begin{picture}(315,306) (174,-87)
        \SetWidth{2.8}
        \SetColor{Black}
        \Line[double,sep=5.5](180,216)(330,66)
        \Line[double,sep=5.5](180,-84)(330,66)
        \SetWidth{1}
        \Line[double,sep=3.5](330,66)(480,216)
        \Line[double,sep=3.5](330,66)(480,-84)
        \SetWidth{1.0}
        \GluonArc(437.190,135.736)(232.684,167.525,218.464){7.5}{13}   
        \GluonArc(190.328,240.926)(175.126,279.091,328.721){7.5}{9}   
        \GluonArc(283.51,44.86)(119.35,61.75,111.38){7.5}{6}    
        \GluonArc(190.84,-21.35)(227.17,359.33,408.96){7.5}{12}  
        \Vertex(210,186){8}
        \Vertex(240,156){8}
        \Vertex(340,150){8}
        \Vertex(255,-9){8}
        \Vertex(418,-24){8}
        \Vertex(218,68){8}
        \GOval(330,66)(20,20)(0){0.882}
        \Text(200,209)[lb]{\LARGE{\Black{$I$}}}
        \Text(448,-85)[lb]{\LARGE{\Black{$k$}}}
        \Text(200,-85)[lb]{\LARGE{\Black{$J$}}}
        \end{picture}}\hspace{20pt}
    &
    \hspace{20pt}\scalebox{0.45}{
        \begin{picture}(315,306) (174,-87)
            
            \SetColor{Black}
            \SetWidth{2.8}
            \Line[double,sep=5.5](180,216)(330,66)
            \Line[double,sep=5.5](180,-84)(330,66)
            \Line[double,sep=5.5](330,66)(480,-84)
            \SetWidth{1.0}
            \Line[double,sep=3.5](330,66)(480,216)  
            \SetWidth{1.0}
            \GluonArc(437.190,135.736)(232.684,167.525,218.464){7.5}{13}   
            \GluonArc(190.328,240.926)(175.126,279.091,328.721){7.5}{9}   
            \GluonArc(283.51,44.86)(119.35,61.75,111.38){7.5}{6}          
            \GluonArc(190.84,-21.35)(227.17,359.33,408.96){7.5}{12}       
            \Vertex(210,186){8}
            \Vertex(240,156){8}
            \Vertex(340,150){8}
            \Vertex(255,-9){8}
            \Vertex(418,-24){8}
            \Vertex(218,68){8}
            \GOval(330,66)(20,20)(0){0.882}
            \Text(200,209)[lb]{\LARGE{\Black{$I$}}}
            \Text(445,-85)[lb]{\LARGE{\Black{$K$}}}
            \Text(203,-85)[lb]{\LARGE{\Black{$J$}}}
        \end{picture}}\\
    $\mathbf{Q}_{(IJ;Ik)}=\frac{1}{2}f_{abe}f^{e}_{\phantom{e}cd}\left\{\mathbf{T}^a_I,\mathbf{T}^c_I\right\}\mathbf{T}^b_J\mathbf{T}^d_k$\hspace{20pt} & \hspace{30pt}$\mathbf{Q}_{(IJ;IK)}=\frac{1}{2}f_{abe}f^{e}_{\phantom{e}cd}\left\{\mathbf{T}^a_I,\mathbf{T}^c_I\right\}\mathbf{T}^b_J\mathbf{T}^d_K$
 \end{tabular}
\end{center}
We then continue showing quadrupoles involving four distinct legs. In the line below, we show the webs defining the colour structures $\mathbf{Q}_{(ij;kl)}$ (on the left) and  $\mathbf{Q}_{(Ij;kl)}$ (on the right):
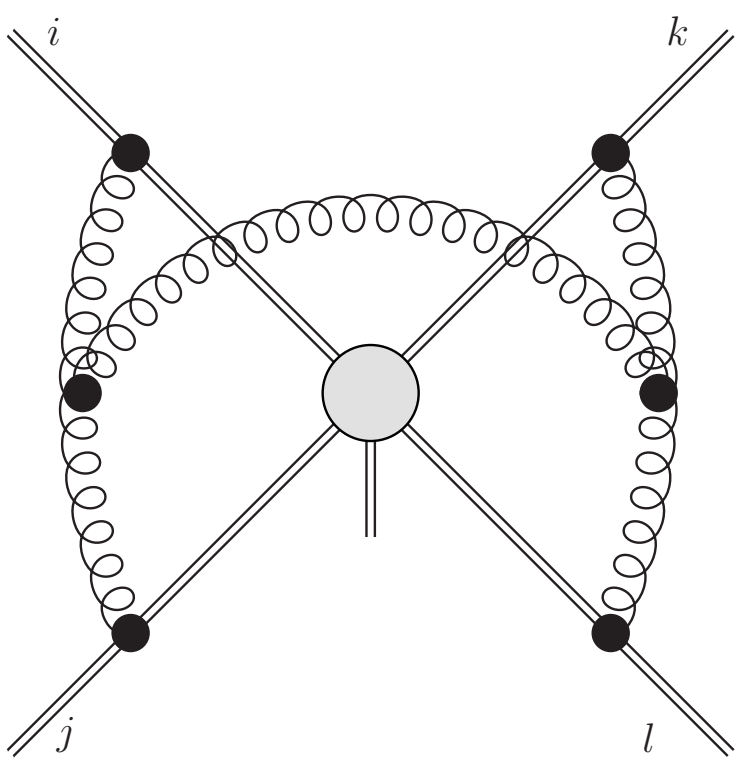
\begin{center}
\begin{tabular}{c|c}
    \scalebox{0.45}{
        \begin{picture}(308,324) (207,-69)
            \SetWidth{1}
            \SetColor{Black}
            \Line[double,sep=3.5](210,234)(510,-66)
            \Line[double,sep=3.5](210,-66)(510,234)
            \Line[double,sep=3.5](360,84)(360,24)
            \Vertex(240,84){8}
            \Vertex(480,84){8}
            \Vertex(260,-16){8}
            \Vertex(260,184){8}
            \Vertex(460,184){8}
            \Vertex(460,-16){8}
            \SetWidth{1.0}
            \GOval(360,84)(20,20)(0){0.882}
            \GluonArc(478,84)(240,155.3757,204.6243){7.5}{13}
            \GluonArc(240,84)(240,-24.6243,24.6243){7.5}{13}
            \GluonArc[clock](360,25.5)(133.5,154.011,25.989){7.5}{20}
            \Text(226,230)[lb]{\LARGE{\Black{$i$}}}
            \Text(485,230)[lb]{\LARGE{\Black{$k$}}}
            \Text(230,-66)[lb]{\LARGE{\Black{$j$}}}
            \Text(475,-66)[lb]{\LARGE{\Black{$l$}}}
        \end{picture}}\hspace{30pt}
    &
    \hspace{30pt}\scalebox{0.45}{
        \begin{picture}(308,324) (207,-69)
            \SetColor{Black}
            \SetWidth{2.8}
            \Line[double,sep=5.5](210,234)(360,84)
            \SetWidth{1}
            \Line[double,sep=3.5](360,84)(510,-66)
            \Line[double,sep=3.5](210,-66)(510,234)
            \Line[double,sep=3.5](360,84)(360,24)
            \Vertex(240,84){8}
            \Vertex(480,84){8}
            \Vertex(260,-16){8}
            \Vertex(260,184){8}
            \Vertex(460,184){8}
            \Vertex(460,-16){8}
            \SetWidth{1.0}
            \GOval(360,84)(20,20)(0){0.882}
            \GluonArc(478,84)(240,155.3757,204.6243){7.5}{13}
            \GluonArc(240,84)(240,-24.6243,24.6243){7.5}{13}
            \GluonArc[clock](360,25.5)(133.5,154.011,25.989){7.5}{20}
            \Text(226,230)[lb]{\LARGE{\Black{$I$}}}
            \Text(485,230)[lb]{\LARGE{\Black{$k$}}}
            \Text(230,-66)[lb]{\LARGE{\Black{$j$}}}
            \Text(475,-66)[lb]{\LARGE{\Black{$l$}}}
        \end{picture}}\\
    $\mathbf{Q}_{(ij;kl)}=f_{abe}f^{e}_{\phantom{e}cd}\mathbf{T}^a_i\mathbf{T}^b_j\mathbf{T}^c_k\mathbf{T}^d_l$\hspace{30pt} & \hspace{30pt}$\mathbf{Q}_{(Ij;kl)}=f_{abe}f^{e}_{\phantom{e}cd}\mathbf{T}^a_I\mathbf{T}^b_j\mathbf{T}^c_k\mathbf{T}^d_l$
\end{tabular}
\end{center}
Then, $\mathbf{Q}_{(Ij;Kl)}$ (on the left) and $\mathbf{Q}_{(IJ;kl}$ (on the right):
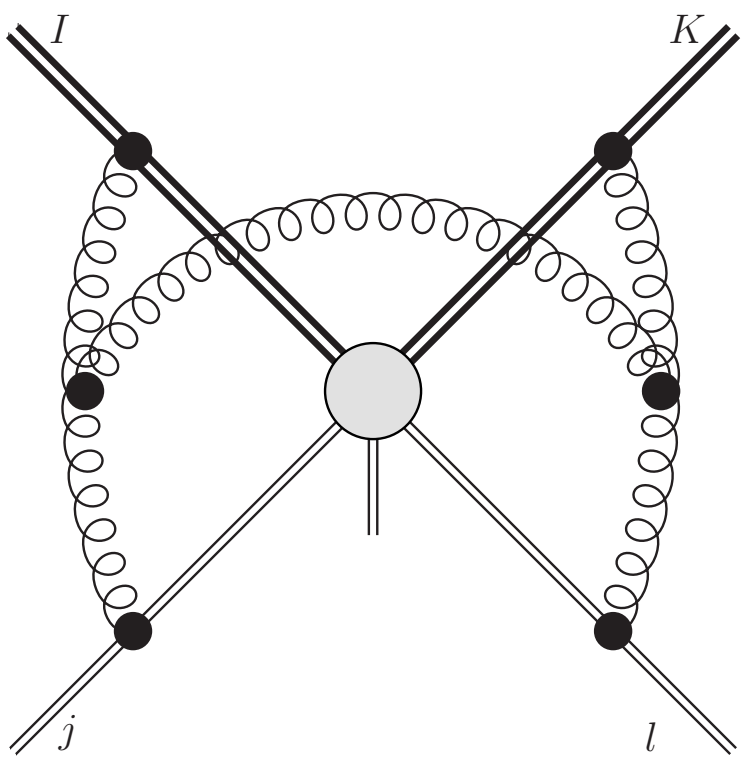
\begin{center}
\begin{tabular}{c|c}
    \scalebox{0.45}{
        \begin{picture}(308,324) (207,-69)
            \SetColor{Black}
            \SetWidth{2.8}
            \Line[double,sep=5.5](210,234)(360,84)
            \Line[double,sep=5.5](360,84)(510,234)
            \SetWidth{1}
            \Line[double,sep=3.5](360,84)(510,-66)
            \Line[double,sep=3.5](210,-66)(360,84)
            \Line[double,sep=3.5](360,84)(360,24)
            \Vertex(240,84){8}
            \Vertex(480,84){8}
            \Vertex(260,-16){8}
            \Vertex(260,184){8}
            \Vertex(460,184){8}
            \Vertex(460,-16){8}
            \SetWidth{1.0}
            \GOval(360,84)(20,20)(0){0.882}
            \GluonArc(478,84)(240,155.3757,204.6243){7.5}{13}
            \GluonArc(240,84)(240,-24.6243,24.6243){7.5}{13}
            \GluonArc[clock](360,25.5)(133.5,154.011,25.989){7.5}{20}
            \Text(226,230)[lb]{\LARGE{\Black{$I$}}}
            \Text(485,230)[lb]{\LARGE{\Black{$K$}}}
            \Text(230,-66)[lb]{\LARGE{\Black{$j$}}}
            \Text(475,-66)[lb]{\LARGE{\Black{$l$}}}
        \end{picture}
    }\hspace{30pt}
    &
    \hspace{30pt}\scalebox{0.45}{
        \begin{picture}(308,324) (207,-69)
            \SetColor{Black}
            \SetWidth{2.8}
            \Line[double,sep=5.5](210,-66)(360,84)
            \Line[double,sep=5.5](210,234)(360,84)
            \SetWidth{1}
            \Line[double,sep=3.5](360,84)(510,234)
            \Line[double,sep=3.5](360,84)(510,-66)
            \Line[double,sep=3.5](360,84)(360,24)
            \Vertex(240,84){8}
            \Vertex(480,84){8}
            \Vertex(260,-16){8}
            \Vertex(260,184){8}
            \Vertex(460,184){8}
            \Vertex(460,-16){8}
            \SetWidth{1.0}
            \GOval(360,84)(20,20)(0){0.882}
            \GluonArc(478,84)(240,155.3757,204.6243){7.5}{13}
            \GluonArc(240,84)(240,-24.6243,24.6243){7.5}{13}
            \GluonArc[clock](360,25.5)(133.5,154.011,25.989){7.5}{20}
            \Text(226,230)[lb]{\LARGE{\Black{$I$}}}
            \Text(485,230)[lb]{\LARGE{\Black{$k$}}}
            \Text(230,-66)[lb]{\LARGE{\Black{$J$}}}
            \Text(475,-66)[lb]{\LARGE{\Black{$l$}}}
        \end{picture}}\\ $\mathbf{Q}_{(Ij;Kl)}=f_{abe}f^{e}_{\phantom{e}cd}\mathbf{T}^a_I\mathbf{T}^b_j\mathbf{T}^c_K\mathbf{T}^d_l$ \hspace{30pt}& \hspace{30pt}$\mathbf{Q}_{(IJ;kl)}=f_{abe}f^{e}_{\phantom{e}cd}\mathbf{T}^a_I\mathbf{T}^b_J\mathbf{T}^c_k\mathbf{T}^d_l$
\end{tabular}
\end{center}
Finally, we have $\mathbf{Q}_{(IJ;Kl)}$ (on the left) and $\mathbf{Q}_{(IJ;KL)}$ (on the right):
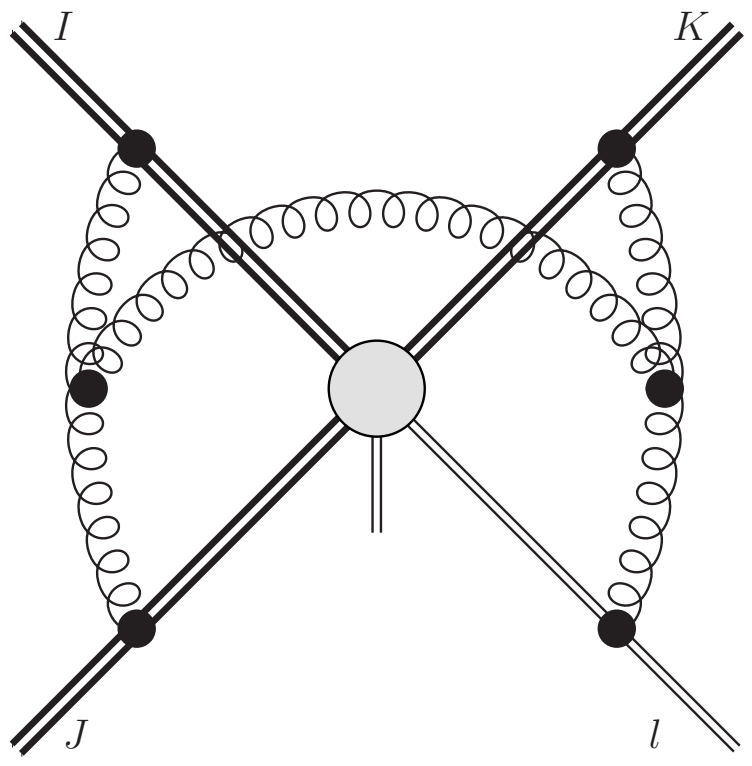
\begin{center}
\begin{tabular}{c|c}
    \scalebox{0.45}{
        \begin{picture}(308,324) (207,-69)
            \SetColor{Black}
            \SetWidth{2.8}
            \Line[double,sep=5.5](210,-66)(360,84)
            \Line[double,sep=5.5](210,234)(360,84)
            \Line[double,sep=5.5](360,84)(510,234)
            \SetWidth{1}
            \Line[double,sep=3.5](360,84)(510,-66)
            \Line[double,sep=3.5](360,84)(360,24)
            \Vertex(240,84){8}
            \Vertex(480,84){8}
            \Vertex(260,-16){8}
            \Vertex(260,184){8}
            \Vertex(460,184){8}
            \Vertex(460,-16){8}
            \SetWidth{1.0}
            \GOval(360,84)(20,20)(0){0.882}
            \GluonArc(478,84)(240,155.3757,204.6243){7.5}{13}
            \GluonArc(240,84)(240,-24.6243,24.6243){7.5}{13}
            \GluonArc[clock](360,25.5)(133.5,154.011,25.989){7.5}{20}
            \Text(226,230)[lb]{\LARGE{\Black{$I$}}}
            \Text(485,230)[lb]{\LARGE{\Black{$K$}}}
            \Text(230,-66)[lb]{\LARGE{\Black{$J$}}}
            \Text(475,-66)[lb]{\LARGE{\Black{$l$}}}
        \end{picture}}\hspace{30pt}
    &
    \hspace{30pt}\scalebox{0.45}{
        \begin{picture}(308,324) (207,-69)
            \SetWidth{2.8}
            \SetColor{Black}
            \Line[double,sep=5.5](210,234)(510,-66)
            \Line[double,sep=5.5](210,-66)(510,234)
            \SetWidth{1}
            \Line[double,sep=5.5](360,84)(360,24)
            \Vertex(240,84){8}
            \Vertex(480,84){8}
            \Vertex(260,-16){8}
            \Vertex(260,184){8}
            \Vertex(460,184){8}
            \Vertex(460,-16){8}
            \SetWidth{1.0}
            \GOval(360,84)(20,20)(0){0.882}
            \GluonArc(478,84)(240,155.3757,204.6243){7.5}{13}
            \GluonArc(240,84)(240,-24.6243,24.6243){7.5}{13}
            \GluonArc[clock](360,25.5)(133.5,154.011,25.989){7.5}{20}
            \Text(226,230)[lb]{\LARGE{\Black{$I$}}}
            \Text(485,230)[lb]{\LARGE{\Black{$K$}}}
            \Text(230,-66)[lb]{\LARGE{\Black{$J$}}}
            \Text(475,-66)[lb]{\LARGE{\Black{$L$}}}
        \end{picture}
    }\\
    $\mathbf{Q}_{(IJ;Kl)}=f_{abe}f^{e}_{\phantom{e}cd}\mathbf{T}^a_I\mathbf{T}^b_J\mathbf{T}^c_K\mathbf{T}^d_l$\hspace{30pt} & \hspace{30pt}$\mathbf{Q}_{(IJ;KL)}=f_{abe}f^{e}_{\phantom{e}cd}\mathbf{T}^a_I\mathbf{T}^b_J\mathbf{T}^c_K\mathbf{T}^d_L$
\end{tabular}
\end{center}

\clearpage

\bibliographystyle{JHEP}
\bibliography{biblio.bib}

\end{document}